\documentclass[final,1p,times,authoryear]{elsarticle}
\usepackage{bm}
\usepackage{amsthm}
\usepackage{amssymb}
\usepackage{mathtools, cuted}
\usepackage{xcolor}
\newcommand{\modif}[1]{\textcolor{black}{#1}}
\newcommand\scalemath[2]{\scalebox{#1}{\mbox{\ensuremath{\displaystyle #2}}}}

\newcommand{\sigmac}{\sigma_\mathrm{c}}
\newcommand{\deltac}{D_\mathrm{c}}
\newcommand{\Cw}{C_\mathrm{w}}
\newcommand{\fw}{f_\mathrm{w}}
\newcommand{\dfw}{f'_\mathrm{w}}
\newcommand{\kw}{\vert k \vert \omega}
\newcommand{\KI}{K_\mathrm{I}}
\newcommand{\KIc}{K_\mathrm{Ic}}
\newcommand{\Gc}{G_\mathrm{c}}
\newcommand{\abs}[1]{\vert {#1} \vert}

\usepackage{lineno}

\journal{Journal of the Mechanics and Physics of Solids}

\begin{document}
% \linenumbers

\begin{frontmatter}


\title{Quasi-static crack front deformations in cohesive materials}
\author[add1]{Mathias Lebihain}
\author[add2]{Thibault Roch}
\author[add2]{Jean-Fran\c{c}ois Molinari}
\cortext[cor1]{Corresponding author : mathias.lebihain@enpc.fr}
\address[add1]{Laboratoire Navier, CNRS (UMR 8205), Ecole des Ponts ParisTech, Universit{\'e} Gustave Eiffel, , 6-8 avenue Blaise Pascal, 77455 Marne-la-Vall{\'e}e, France}
\address[add2]{Computational Solid Mechanics Laboratory, Civil Engineering Institute, Materials Science and Engineering Institute, {\'E}cole Polytechnique F{\'e}d{\'e}rale de Lausanne, Station 18, CH-1015 Lausanne, Switzerland}

\begin{abstract}
When a crack interacts with material heterogeneities, its front distorts and adopts complex tortuous configurations that are reminiscent of the energy barriers encountered during crack propagation. As such, the study of crack front deformations is key to rationalize the effective failure properties of micro-structured solids and interfaces. Yet, the impact of a localized dissipation in a finite region behind the crack front, called the \emph{process zone}, has often been overlooked. In this work, we derive the equation ruling 3D coplanar crack propagation in heterogeneous \emph{cohesive} materials where the opening of the crack is resisted by some traction in its wake. 
We show that the presence of a process zone results in two \emph{competing} effects on the deformation of crack fronts: (i) it makes the front \emph{more compliant} to small-wavelength perturbations, and (ii) it \emph{smooths out} local fluctuations of strength and process zone size, from which emerge heterogeneities of fracture energy. Their respective influence on front deformations is shown to strongly impact the stability of perturbed crack fronts, as well as their stationary shapes when interacting with arrays of tough obstacles. Overall, our theory provides a unified framework to predict the variety of front profiles observed in experiments, even when the small-scale yielding hypothesis of linear elastic fracture mechanics breaks down.
\end{abstract}

\end{frontmatter}

\section{Introduction}
Significant efforts have been made in the past decades to unravel the influence of material heterogeneities on the failure behavior of composites. Understanding how crack fronts deform may seem a rather anecdotal subject matter in this regard. Yet, the front deformations are reminiscent of the interaction between a crack and material disorder, and therefore conceal a wealth of information on the disorder intensity and its structure. Their study has provided invaluable insights on the spatio-temporal dynamics of propagating cracks, and on the effective toughness of composite brittle materials (see \cite{lazarus_review_2011}, \cite{bonamy_failure_2011} and references therein).

During its interaction with heterogeneities of material properties, the crack front distorts and adopts complex tortuous configurations. Predicting its propagation path requires considering all possible geometric configurations, as well as a suitable criterion that selects the ``most favorable'' one. This lead to a predominance of perturbative approaches in fracture mechanics, where quantities of interest such as the stress intensity factors can be linked to perturbations of the crack front geometry \citep{gao_shear_1986} and that of its surface \citep{movchan_perturbations_1998}. In particular, the first-order theory of \cite{rice_first-order_1985} based on \cite{bueckner_weight_1987}'s weight functions has been quite successful in capturing experimental observations of coplanar crack propagation along weak interfaces. It provides a convincing framework to describe quantitatively the deformations of a crack front interacting with a single tough defect  
 \citep{chopin_crack_2011, patinet_pinning_2013, vasoya_experimental_2016} of sometimes complex shape \citep{xia_toughening_2012}, periodic arrays of obstacles \citep{dalmas_pinning_2009}, or disorderly-placed heterogeneities \citep{delaplace_high_1999}. It also helped in rationalizing the micro-dynamics of depinning of a crack exiting an obstacle \citep{chopin_depinning_2018}, or the intermittent macro-dynamics of crack propagation in disordered media \citep{maloy_local_2006, ponson_crack_2010, bares_aftershock_2018}. The same theory has been extensively applied to predict the effective toughness arising from periodic \citep{gao_trapping_1989}, asymmetric \citep{xia_adhesion_2015}, or even disordered \citep{patinet_quantitative_2013, demery_microstructural_2014, lebihain_towards_2021} distributions of fracture properties. Higher-order theories also provided valuable insights on bridging mechanisms \citep{bower_bridging_1991}, on fingering instabilities \citep{vasoya_fingering_2016}, or on the microbranching transition at higher crack velocities \citep{kolvin_nonlinear_2017}.
 
A major pitfall of these models is that they arguably overlook the influence of the spatially localized weakening dynamics near the crack front. Indeed, the LEFM theory is based on the so-called \emph{small-scale yielding} assumption, which states that all the dissipation occurs in an infinitesimally small region in the vicinity of the crack front. As such, LEFM does not provide any meaningful dissipation length scale, and treats all asperity scales indifferently. Yet, as we zoom in on the front of a propagating crack, one ultimately finds a region of finite size where the material behaves inelastically and the validity of the LEFM framework breaks down. It is crucial to understand how heterogeneities smaller or bigger than this dissipative region may influence the overall failure behavior of a composite. Cohesive-zone models \citep{dugdale_yielding_1960, barenblatt_processzone_1962} provide a way to do that, as they assume that the material does not weaken instantly but in a region of finite size -- called ``\emph{process zone}'' -- located at the crack front where the crack opening is resisted by a distribution of \emph{cohesive} stresses. In these models, the fracture properties are not solely characterized by the fracture energy $\Gc$, but rather by (i) the \emph{strength} $\sigmac$ of the material, and (ii) its \emph{process zone size} $\omega$, two quantities from which emerge the fracture energy \citep{palmer_growth_1973}. The development of a unified framework that describes the failure behavior of \emph{cohesive} and \emph{heterogeneous} materials may only take place through a preliminary study of the front deformations \emph{in presence of a finite process zone}.\\

In this work, we extend the LEFM perturbative theory of \cite{rice_first-order_1985} to mode I coplanar crack propagation in \emph{cohesive} materials. This is performed by deriving first \cite{bueckner_weight_1987}'s crack face weight function for a semi-infinite coplanar crack perturbed within its plane. As a result, we can explore for the first time the influence of a finite process zone size on the front deformations of a crack propagating in tensile mode I. We show that a cohesive crack accommodates perturbations differently depending on the size of the perturbation wavelength with respect to that of the process zone. We also emphasize that the presence of cohesive stress behind the crack front may strongly modify the fluctuations of fracture energy the crack actually perceives during its propagation. This theory successfully predicts the front profiles observed in the peeling experiments of \cite{chopin_crack_2011}, for which the small-scale yielding hypothesis of LEFM is suspected to break down.

The paper is organized as follows: in Section~\ref{sec:Model}, we derive the expression of the crack face weight function perturbations at first-order in the front deformation, following a method proposed by  \cite{rice_weight_1989}. It allows us to calculate the stress intensity factor arising from cohesive stresses acting behind the front for materials translationally invariant in the propagation direction. Building on this model, we revisit in Section~\ref{sec:StabilityAnalysis} the stability problem of a perturbed crack front treated by \cite{rice_first-order_1985} for perfectly brittle materials. We highlight here the influence of the finite process zone size on the relative ``stiffness'' of the crack front to some modal perturbations. Next, we investigate in Section~\ref{sec:FrontDeformations} the stationary shape of a crack front interacting with periodic arrays of tough obstacles. We show here that the presence of cohesive stresses behind the crack front does not only affect the stiffness of the crack front, but also influences the energy landscape the crack actually experiences during its propagation. The front profiles predicted by our theory are successfully compared to those obtained by \cite{chopin_crack_2011} during adhesive peeling experiments of an elastomer block from a patterned glass substrate.

\section{First-order variations of the mode I stress intensity factor for a semi-infinite coplanar cohesive crack}
\label{sec:Model}

\subsection{Cohesive approach to three-dimensional coplanar crack propagation}
\label{subsec:Model_Cohesive}

\begin{figure}[h]
\centering \includegraphics[width=\textwidth]{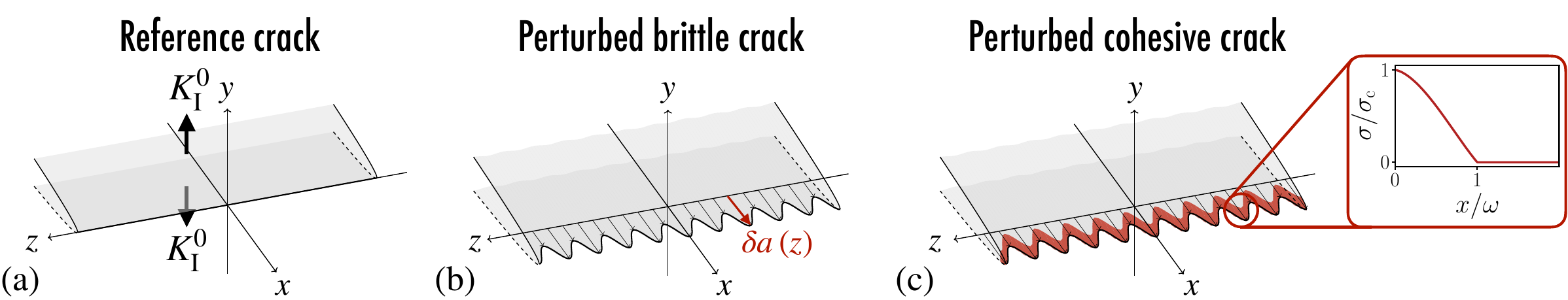}
\caption{(a) A semi-infinite planar crack $\Gamma$ with a straight crack front $\mathcal{F}$ is loaded under pure mode I through some system of forces, giving rise to a stress intensity factor $\KI^0$. (b) Due to the presence of microscopic heterogeneities, the crack front $\mathcal{F}^*$ of the perturbed crack $\Gamma^*$ distorts by a quantity $\delta a(z)$ along the direction of propagation $\left(Ox\right)$. (c) The material does not weaken instantly in the crack wake, but progressively in a closed region behind the rupture front (in red) called \emph{process zone}. Inset: In this region, the cohesive stress $\sigma$ decay from its peak value $\sigmac$, referred to as the \emph{strength} of the material, down to zero along a typical distance $\omega$.}
\label{fig:PerturbedCrackFront}
\end{figure}

We consider a semi-infinite crack embedded in an infinite body made of an isotropic linear elastic material. In the initial reference configuration $\Gamma$, the crack is planar and its front is straight (see Fig.~\ref{fig:PerturbedCrackFront}a). We adopt the usual convention of LEFM, and thus use a Cartesian frame $Oxyz$ with origin $O$ chosen arbitrarily within the crack plane, axis $(Ox)$ parallel to the direction of crack propagation, $(Oy)$ parallel to the direction orthogonal to the crack plane, and $(Oz)$ parallel to the crack front. The associated unit vectors are denoted $\left(e_{x}, e_{y}, e_{z}\right)$. The crack $\Gamma$ is loaded in pure mode I through some system of forces independent of the coordinate $z$, so that the unperturbed stress intensity factor (SIF) $\KI^0$ is independent of the point of observation along the crack front. The influence of the loading conditions and the specimen geometry may be accounted for in the evolution of $\KI^0$ with crack advance.\\

Now, let the front $\mathcal{F}$ undergo some infinitesimal coplanar perturbation, and note $\delta a$ the local orthogonal distance between the perturbed front $\mathcal{F}^*$ with respect to the reference one (see Fig.~\ref{fig:PerturbedCrackFront}b). The LEFM perturbed stress intensity factor $K_\mathrm{lefm}$ along the front reads \citep{rice_first-order_1985}:
\begin{equation}
\label{eq:LEFM_SIF_Variations}
K_\mathrm{lefm}\left(z\right) = \KI^0 + \delta K_\mathrm{lefm}\left(z\right) = \KI^0 \left[1 + \frac{1}{\KI^0} \dfrac{\partial \KI^0}{\partial a} \delta a(z) - \frac{1}{2\pi} \mathrm{PV} \int_{-\infty}^{+\infty} \dfrac{\delta a\left(z\right) - \delta a(z')}{\left(z-z'\right)^2} dz'\right]
\end{equation}
where $\delta K_\mathrm{lefm}$ corresponds to the first-order variations of $K_\mathrm{lefm}$ due to the front deformations $\delta a$, and the symbol $\mathrm{PV}$ denotes a Cauchy principal value. Equation~\eqref{eq:LEFM_SIF_Variations} is non-local, since the rupture behavior at a given point is affected by the position of the other points due to long-range elastic interactions.\\

In the following, use will be made of Fourier transforms in the direction $(Oz)$ of the crack front. The definition adopted here for the Fourier transform $\hat{\chi}(k)$ of an arbitrary function $\chi(z)$ is:
\begin{equation}
\label{eq:Fourier_Definition}
\hat{\chi}(k)= \frac{1}{2\pi}\int_{-\infty}^{+\infty} \chi(z) e^{-ikz} dz \,\Leftrightarrow\, \chi(z) = \int_{-\infty}^{+\infty} \hat{\chi}(k) e^{ikz} dk
\end{equation}
Then Eq.~\eqref{eq:LEFM_SIF_Variations} reads in the Fourier space:
\begin{equation}
\dfrac{\widehat{\delta K_\mathrm{lefm}}\left(k\right)}{\KI^0 } = \left(\dfrac{1}{\KI^0} \dfrac{\partial \KI^0}{\partial a} -\dfrac{\vert k\vert}{2}\right) \widehat{\delta a}\left(k\right)
\label{eq:LEFM_SIF_Fourier}
\end{equation}

The LEFM approach leads to a ``non-physical'' situation where the stress is found singular at the crack front. Cohesive zone models (CZMs) address this issue by assuming that the opening of the crack is \emph{resisted} in its wake by a tensile stress $\sigma\left(z,x\right)$, called \emph{cohesive stress}, acting behind the crack front (see Fig.~\ref{fig:PerturbedCrackFront}c). It typically evolves from a peak value $\sigmac$, called \emph{strength} of the material, down to negligible values along a characteristic distance $\omega$, called \emph{process zone size} (see inset of Fig.~\ref{fig:PerturbedCrackFront}c). \modif{The evolution of $\sigma$ behind the crack front usually results from a traction-separation law $\sigma = f(\delta)$ describing material degradation with the local opening displacement $\delta$ \citep{dugdale_yielding_1960, barenblatt_processzone_1962}.} This cohesive stress $\sigma$ gives rise to a \emph{negative} stress intensity factor $- K_\mathrm{czm}\left(z\right)$ that balances $K_\mathrm{lefm}\left(z\right)$, ensuring then that the stress is non-singular at each point along the crack front:
\begin{equation}
\label{eq:CohesiveApproach}
K_\mathrm{total}\left(z\right) = K_\mathrm{lefm}\left(z\right) - K_\mathrm{czm}\left(z\right) = 0
\end{equation}
Building on Bueckner-Rice's weight function theory \citep{bueckner_weight_1987, rice_weight_1989}, one may express the cohesive stress intensity factor $K_\mathrm{czm}$ as:
\begin{equation}
\label{eq:Cohesive_SIF_General}
K_\mathrm{czm}\left(z\right) = \int_{\Gamma^*} k^*\left(\Gamma^*; z, z', x\right) \sigma\left(\mathbf{x}\right) dz' dx
\end{equation}
where $k^*\left(\Gamma^*; z, z', x\right)$ is the mode I crack face weight function (CFWF). It corresponds to the stress intensity factor generated at point $z$ by a pair of unitary tensile forces applied at a point $(z', x)$ located on the faces of $\Gamma^*$.

\subsection{First-order variations of the mode I crack face weight function}
\label{subsec:Model_CFWF}

In order to compute the cohesive stress intensity factor of Eq.~\eqref{eq:Cohesive_SIF_General}, we need a closed form for the crack face weight function $k^*$ of the perturbed front $\mathcal{F}^*$. We follow here the ideas developed by \cite{rice_weight_1989} and \cite{favier_coplanar_2006}, upon which \modif{\cite{leblond_second_2012} built to derive the first-order variations of the fundamental kernel $Z\left(\Gamma^*; z; z'\right) = \sqrt{\pi}/8 \lim\limits_{x \rightarrow 0} k\left(\Gamma^*; z; z', x\right)/\sqrt{x}$}. We build upon them here to derive the \emph{variations of the mode I crack face weight function} for a semi-infinite coplanar crack at \emph{first order} in the perturbation $\delta a$. \modif{Our results generalize the calculations of \cite{leblond_second_2012} that only looked at the asymptotic behavior of  $k^*$ near the crack front ($x \rightarrow 0$).}\\

\begin{figure}[h]
\centering \includegraphics[width=0.6\textwidth]{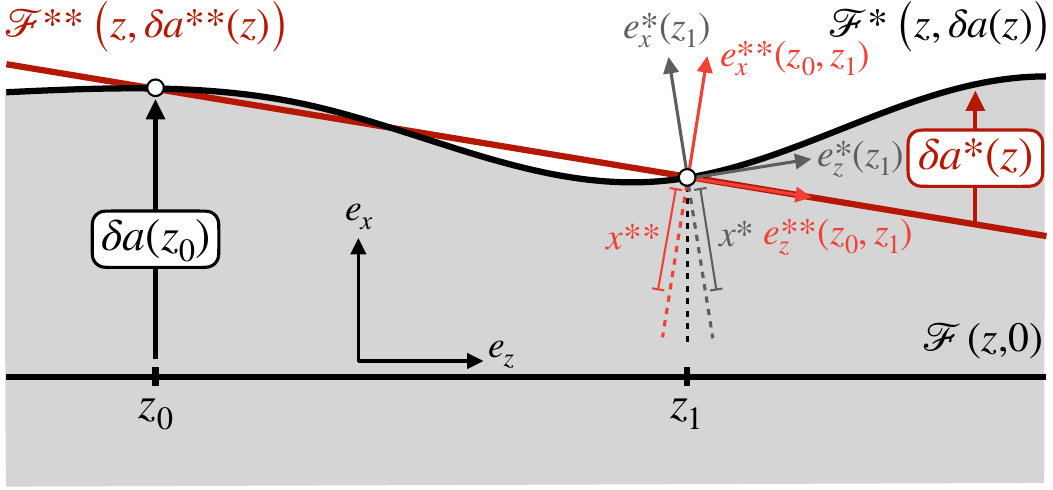}
\caption{The mode I stress intensity factor $k\left(\mathcal{F}^*; z_0; z_1, x^*\right)$ generated at $z=z_0$ by a pair of unitary forces applied along $e_y$ at a distance $x^*$ behind the point $z_1$ of the perturbed crack front $\mathcal{F}^*$ (in black) in the direction of the vector $e_x^*(z_1)$ can be inferred from that generated by a pair of unitary forces applied along $e_y$ at a distance $x_{**}$ behind the point $z_1$ of the auxiliary front $\mathcal{F}^{**}$ (in red) in the direction of the vector $e_x^{**}(z_1)$.}
\label{fig:PerturbedWeightFunction}
\end{figure}

The evolution of the opening displacement and the stress in the vicinity of the perturbed crack front are naturally expressed in the local basis of vectors $\left(e^*_{z}(z), e^*_{x}(z), e_y\right)$, where $e^*_{z}(z)$ is tangent to the crack front $\mathcal{F}^*$ at position $z$, and $e^*_{x}(z)$ is perpendicular to it within the plane $\left(zOx\right)$ and oriented in the direction of propagation (see Fig.~\ref{fig:PerturbedWeightFunction}). Following \cite{favier_coplanar_2006}, we need to define two distinct crack face weight functions:
\begin{itemize}
\item $k^*\left(\mathcal{F}^*; z_0; z_1, x^*\right)$, which corresponds to the stress intensity factor generated at $z=z_0$ by a pair of unitary forces applied along $e_y$ at a distance $x^*$ behind the point $z_1$ of the perturbed crack front $\mathcal{F}^*$ in the direction of the vector $e_x^*(z_1)$;
\item $k\left(\mathcal{F}^*; z_0; z_1, x\right)$, which corresponds to the stress intensity factor generated at $z=z_0$ by a pair of unitary forces applied along $e_y$ at a distance $x$ behind the point $z_1$ of the perturbed crack front $\mathcal{F}^*$ in the direction $e_x$.
\end{itemize}
At first order in the perturbation, \modif{one has:}
\begin{equation}
\label{eq:cfwfEquality}
\modif{k^*\left(\mathcal{F}^*; z_0; z_1, x^*\right) = k\left(\mathcal{F}^*; z_0; z_1, x\right)}
\end{equation}
as the error introduced on the position of the point of application of the forces is of second order in $\delta a$ \citep{favier_coplanar_2006}. Moreover, \cite{rice_weight_1989} showed that provided the crack advance $\delta a(z)$ satisfies the condition:
\begin{equation}
\label{eq:VanishingCrackAdvanceCondition}
\delta a(z_0) = 0 \text{ and } \delta a(z_1) = 0
\end{equation}
$k\left(\mathcal{F}^*; z_0; z_1, x\right)$ can be expressed from the CFWF $k\left(\mathcal{F}; z_0; z_1, x\right)$ of the reference straight front following:
\begin{equation}
\label{eq:PerturbedWeightFunction_Reference}
k\left(\mathcal{F}^*; z_0; z_1, x\right) = k\left(\mathcal{F}; z_0; z_1, x\right) + \dfrac{1}{2\pi} \int_{-\infty}^{+\infty} k\left(\mathcal{F}; z; z_1, x\right) \dfrac{\delta a(z)}{(z-z_0)^2} dz
\end{equation}
where $k\left(\mathcal{F}; z_0; z_1, x\right)$ is known analytically for the semi-infinite coplanar crack with a straight crack front $\Gamma$:
\begin{equation}
\label{eq:StraightWeightFunction}
k\left(\mathcal{F}; z_0; z_1, x\right) = \dfrac{\sqrt{2}}{\pi^{3/2}} \dfrac{\sqrt{x}}{(z_0-z_1)^2 + x^2}
\end{equation}

Yet, the condition \eqref{eq:VanishingCrackAdvanceCondition} is not satisfied for an arbitrary perturbation $\delta a$. To circumvent this difficulty, we follow the ideas of \cite{leblond_second_2012}, and compute the perturbed CFWF associated to $\Gamma^{*}$ from those of a reference crack $\Gamma^{**}$ that results from the combination of translatory motion and a rotation $\delta a^{**}$ making $\delta a^{*}(z) = \delta a(z) - \delta a^{**}(z)$ vanish in $z_0$ and $z_1$:
\begin{equation}
\label{eq:TranslationRotation}
\delta a^{**}(z) = \delta a(z_0) + \dfrac{\delta a(z_1)-\delta a(z_0)}{z_1-z_0}(z-z_0) = \delta a(z_1) + \dfrac{\delta a(z_1)-\delta a(z_0)}{z_1-z_0}(z-z_1)
\end{equation}
In the following, $\left(e^{**}_z(z_0,z_1), e^{**}_x(z_0,z_1), e_y\right)$ denotes the natural basis of vectors associated to the straight crack front $\mathcal{F}^{**}$ of $\Gamma^{**}$, and $\left(z^{**}, x^{**}, y\right)$ the point coordinates in this basis (see Fig.~\ref{fig:PerturbedWeightFunction}). One may then define:
\begin{itemize}
\item $k^{**}\left(\Gamma^*; z_0^{**}, z_1^{**}, x^{**}\right)$, which corresponds to the stress intensity factor generated at $z^{**}=z_0^{**}$ by a pair of unitary forces applied along $e_y$ at a distance $x^{**}$ behind the point $z_1^{**}$ of the perturbed crack front $\mathcal{F}^{**}$ in the direction $e_x^{**}(z_0,z_1)$. Since $\Gamma^{**}$ is a semi-infinite coplanar crack with a straight crack front, $k^{**}\left(\Gamma^*; z_0^{**}, z_1^{**}, x^{**}\right)$ reads:
\begin{equation}
\label{eq:StraightWeightFunction_Rstar}
k^{**}\left(\mathcal{F}; z_0^{**}; z_1^{**}, x^{**}\right) = \dfrac{\sqrt{2}}{\pi^{3/2}} \dfrac{\sqrt{x^{**}}}{\left(z_0^{**}-z_1^{**}\right)^2 + {x^{**}}^2}
\end{equation}
\item $k^{*}\left(\Gamma^*; z_0^{**}, z_1^{**}, x^{**}\right)$, which corresponds to the stress intensity factor generated at $z^{**}=z_0^{**}$ by a pair of unitary forces applied along $e_y$ at a distance $x^{**}$ behind the point $z_1^{**}$ of the perturbed crack front $\mathcal{F}^*$ in the direction $e_x^{**}(z_0,z_1)$. Applying Eq.~\eqref{eq:PerturbedWeightFunction_Reference} to $\Gamma^{**}$ and $\Gamma^*$ yields:
\begin{equation}
\label{eq:PerturbedWeightFunction_Rstar}
\scalemath{0.8}{
\begin{aligned}
k^{**}\left(\Gamma^*; z_0^{**}, z_1^{**}, x^{**}\right) = & \,k^{**}\left(\Gamma^{**}; z_0^{**}, z_1^{**}, x\right) + \frac{1}{2\pi} \mathrm{PV} \int_{-\infty}^{+\infty} k^{**}\left(\mathcal{F}^{**}; z^{**}; z_1^{**}, x^{**}\right) \dfrac{\delta a(z^{**})-\delta a^{**}(z^{**})}{\left(z^{**}-z_0^{**}\right)^2} dz^{**}
\end{aligned}
}
\end{equation}
\end{itemize}
Again, the error on the position $\left(z^{**}, x^{**}\right)$ with respect to $\left(z, x\right)$ is of second order in $\delta a$. In the $\left(e_z, e_x, e_y\right)$ basis, Eq.~\eqref{eq:PerturbedWeightFunction_Rstar} writes as:
\begin{equation}
\label{eq:PerturbedWeightFunction_Raw}
k\left(\Gamma^*; z_0; z_1, x\right) = \dfrac{\sqrt{2}}{\pi^{3/2}} \dfrac{\sqrt{x}}{(z_0-z_1)^2 + x^2} + \dfrac{1}{2\pi} \mathrm{PV} \int_{\infty}^{+\infty} \dfrac{\sqrt{2}}{\pi^{3/2}} \dfrac{\sqrt{x}}{(z-z_1)^2 + x^2} \dfrac{\delta a(z)-\delta a^{**}(z)}{(z-z_0)^2} dz
\end{equation}

Combined with Eq.~\eqref{eq:TranslationRotation}, Eq.~\eqref{eq:PerturbedWeightFunction_Raw} provides a direct way to evaluate Eq.~\eqref{eq:Cohesive_SIF_General} numerically. Yet, as noted by \cite{leblond_second_2012}, the presence of the rational function $1/((z-z_1)^2 + x^2)(z-z_0)^2$ in the integrand makes it unfit for any analytical calculations. One may then decompose it as:
\begin{equation}
\label{eq:PartialFractionDecomposition}
\scalemath{0.78}{
\begin{aligned}
\dfrac{1}{\left((z-z_1)^2 + x^2\right)(z-z_0)^2} = & \dfrac{1}{\left((z_0-z_1)^2 + x^2\right)^2} \left[ \dfrac{(z_0-z_1)^2 + x^2}{(z-z_0)^2} - \dfrac{2(z_0-z_1)}{(z-z_0)} + \dfrac{2(z_0-z_1)(z-z_1)}{(z-z_1)^2 + x^2} + \dfrac{(z_0-z_1)^2 - x^2}{(z-z_1)^2 + x^2} \right]
\end{aligned}
}
\end{equation}
From Eqs.~\eqref{eq:TranslationRotation}, \eqref{eq:PerturbedWeightFunction_Raw}, and \eqref{eq:PartialFractionDecomposition}, one finally gets for the expression of $k\left(\Gamma^*; z_0; z_1, x\right)$:
\begin{equation}
\label{eq:PerturbedWeightFunction}
\scalemath{0.8}{
\begin{aligned}
k\left(\Gamma^*; z_0; z_1, x\right) = & \,k\left(\Gamma; z_0; z_1, x\right) + \delta k\left(\Gamma^*; z_0; z_1, x\right) \\
= &\,\dfrac{\sqrt{2}}{\pi^{3/2}} \dfrac{\sqrt{x}}{(z_0-z_1)^2 + x^2} \left[ 1 + \dfrac{1}{2\pi} \mathrm{PV} \int_{-\infty}^{+\infty} \dfrac{\delta a(z) - \delta a(z_0)}{(z-z_0)^2} dz \right. \\
& + \dfrac{2(z_0-z_1)}{(z_0-z_1)^2 + x^2} \,\dfrac{1}{2\pi} \mathrm{PV} \int_{-\infty}^{+\infty} \left(-\dfrac{1}{z-z_0} + \dfrac{(z-z_1)}{(z-z_1)^2 + x^2} \right)\delta a(z)\, dz \\
& + \left. \dfrac{(z_0-z_1)^2 - x^2}{(z_0-z_1)^2 + x^2} \,\dfrac{1}{2\pi} \mathrm{PV} \int_{-\infty}^{+\infty} \dfrac{\delta a(z) - \delta a(z_1)}{(z-z_1)^2 + x^2} dz + \dfrac{x}{(z_0-z_1)^2 + x^2} \left(\delta a(z_0) - \delta a(z_1)\right) \right]
\end{aligned}
}
\end{equation}
We derived here for the first time the analytical expression for the crack face weight function $k\left(\Gamma^*; z_0; z_1, x\right)$ of a perturbed crack at first-order in the perturbation $\delta a$. \modif{We observe that one retrieves, taking the limit $x \rightarrow 0$ in our Eq.~\eqref{eq:PerturbedWeightFunction}, the Eq.~(10) of \cite{leblond_second_2012} that describes the first-order variations of the fundamental kernel $Z\left(\Gamma^*; z; z'\right) = \sqrt{\pi}/8 \lim\limits_{x \rightarrow 0} k\left(\Gamma^*; z; z', x\right)/\sqrt{x}$.} Our equation provides the fundamental ingredients to extend the model of \cite{rice_first-order_1985} to cohesive materials. Technical details on the derivation of Eq.~\eqref{eq:PerturbedWeightFunction} are given in \ref{app:Compute_CFWF}.

\subsection{First-order variations of the mode I stress intensity factor}
\label{subsec:Model_SIF}

Using Eq.~\eqref{eq:PerturbedWeightFunction}, it is now possible to compute at first order in $\delta a$ the stress intensity factor $K_\mathrm{czm}$ of Eq.~\eqref{eq:Cohesive_SIF_General} generated by the cohesive stress acting in the wake of the perturbed crack front $\mathcal{F}^{*}$. We perform this in the reduced case where the material is \emph{translationally invariant} in the propagation direction $\left(Ox\right)$. In this case, the cohesive stress might be expressed as:
\begin{equation}
\label{eq:CohesiveStress_Evolution}
\sigma\left(z,x\right) = \sigmac(z) \fw\left(x/\omega(z)\right)
\end{equation}
where $\sigmac(z)$ and $\omega(z)$ are the local \emph{strength} and \emph{process zone size} at position $z$, $x$ is the distance to the crack tip located at $\left(z,\delta a(z)\right)$, and $\fw$ is a shape function that relates to the nature of weakening. An example of $\fw$ is given in the inset of Fig.~\ref{fig:PerturbedCrackFront}c for the generic linear traction-separation cohesive law (see \ref{app:Asymptotic_SlipWeakening} for more details).\\

The particular choice of cohesive stress evolution in Eq.~\eqref{eq:CohesiveStress_Evolution}, often referred to as \emph{distance-weakening}, is rather limiting as it does not provide a comprehensive framework to investigate the influence of spatially distributed material heterogeneities on crack propagation. Yet, it provides ways to account for the influence of a finite-size dissipation in the crack wake onto the rupture behavior in a fully analytical manner; see for example \citep{palmer_growth_1973} for an estimate of the process zone size in cohesive materials, and \citep{poliakov_dynamic_2002} for the displacement/strain field ahead of a dynamic cohesive rupture tip.\\

\modif{A more standard formulation of cohesive zone models is to describe material degradation with the local crack opening displacement $\delta$:}
\begin{equation}
\label{eq:CohesiveStress_TractionSeparation}
\modif{\sigma\left(z\right) = \sigmac(z) f_\delta\left(\delta(z)/\deltac(z)\right)}
\end{equation}
\modif{where $\deltac(z)$ is the local critical crack opening, above which cohesive stress are negligible, and $f_\delta$ is a shape function describing material weakening. While we chose here to express  directly $\sigma$ in terms of the variables $(\sigmac, \omega, \fw)$, they usually emerge from the knowledge $(\sigmac, \deltac, f_\delta)$ and the resolution of the structural problem. In particular, the process zone size $\omega$ can be expressed as:}
\begin{equation}
\label{eq:CohesiveStress_ProcessZoneSize}
\modif{\omega = \alpha \dfrac{\mu}{\sigmac}\deltac}
\end{equation}
\modif{where $\mu$ is the shear modulus, and $\alpha$ is a proportionality constant that relates to $f_\delta$. It can be either estimated analytically \citep{barenblatt_processzone_1962} or computed numerically \citep{viesca_numerical_2018}. The formulation in terms of $(\sigmac, \omega, \fw)$ is strictly equivalent to that in $(\sigmac, \deltac, f_\delta)$ for materials that are translationally invariant in the propagation direction $\left(Ox\right)$. The latter formulation would provide a more comprehensive framework to deal with cases where the heterogeneities are random.\\}

In the following, perturbations $\delta a$ of the crack front may arise from the spatial variations of strength $\sigmac$ and process zone size $\omega$ along the front, \modif{the latter being associated with variations of both strength $\sigmac$ and critical crack opening $\deltac$}. We decompose $\sigmac$ and $\omega$ in uniform contributions $\sigmac^0$ and $\omega_0$ associated to a reference homogeneous material, and spatial fluctuations $\delta\sigmac$ and $\delta\omega$:
\begin{equation}
\label{eq:CohesiveStress_Decomposition}
\begin{cases}
\sigmac(z) = \sigmac^0 + \delta \sigmac(z)\\
\omega(z) = \omega_0 + \delta \omega(z)
\end{cases}
\end{equation}
where $\sigmac^0$ and $\omega_0$ correspond to the spatial averages of $\sigmac$ and $\omega$ respectively. We can now insert Eqs.~\eqref{eq:cfwfEquality}, \eqref{eq:PerturbedWeightFunction}, \eqref{eq:CohesiveStress_Evolution}, and \eqref{eq:CohesiveStress_Decomposition} into Eq.~\eqref{eq:Cohesive_SIF_General} that gives the cohesive stress intensity factor $K_\mathrm{czm}(z)$ acting along the perturbed crack front $\mathcal{F}^{*}$. It yields:
\begin{equation}
\label{eq:Cohesive_SIF_Raw}
\scalemath{1}{
\begin{aligned}
K_\mathrm{czm}(z) = & \int_{0}^{+\infty} \int_{-\infty}^{+\infty} \sigma\left(z',x\right) k^*\left(\Gamma^*; z; z', x\right) dz' dx \\
= & \int_{0}^{+\infty} \int_{-\infty}^{+\infty} \sigmac(z')\,\fw\left(x/\omega(z')\right) \left[k\left(\Gamma; z; z', x\right) + \delta k\left(\Gamma^*; z; z', x\right)\right] dz' dx
\end{aligned}
}
\end{equation}
The cohesive SIF $K_\mathrm{czm}(z)$ can be expressed as the sum of a zero-order term $K_\mathrm{czm}^0$, and first-order variations $\delta K_\mathrm{czm}(z)$ that relates to the perturbations $\delta a$, $\delta\sigmac$ and $\delta\omega$. Following \cite{irwin_fracture_1958}'s criterion, $K_\mathrm{czm}^0$ corresponds to the mode I toughness $\KIc^0$ of the reference material when the crack propagates. We show in \ref{app:Compute_SIF} that it writes as:
\begin{equation}
\label{eq:Cohesive_Toughness}
K_\mathrm{czm}^0 = \KIc^0 = \Cw \sqrt{\dfrac{2}{\pi}} \sigmac^0 \omega_0^{1/2}
\end{equation}
where $\Cw = \int_{0}^{+\infty} \fw\left(u\right) u^{-1/2} du$ is a pre-factor that relates to the nature of the weakening. We retrieve here the results of \cite{palmer_growth_1973} that derived the expression of the mode I toughness $\KIc^0$ of a cohesive crack in 2D. This was expected, as three-dimensional crack propagation in a spatially homogeneous reference material can be reduced to a two-dimensional problem.\\

The first-order variations of cohesive stress intensity factor $\delta K_\mathrm{czm}(z)$ can be expressed as:
\begin{align}
\label{eq:Cohesive_SIF_Variations}
\dfrac{\delta K_\mathrm{czm}}{\KIc^0}(z) = & \dfrac{1}{2\pi} \int_{-\infty}^{+\infty} \left[ -\dfrac{\kw_0}{2} + \dfrac{1}{\Cw} \int_{0}^{+\infty} -\dfrac{\dfw\left(u\right)}{u^{1/2}} \left(1-e^{-\kw_0 u}\right) du \right] \dfrac{\widehat{\delta a}\left(k\right)}{\omega_0} e^{ikz} dk \nonumber\\
& + \dfrac{1}{2\pi} \int_{-\infty}^{+\infty} \left[ \dfrac{1}{\Cw} \int_{0}^{+\infty} \dfrac{\fw\left(u\right)}{u^{1/2}} e^{-\kw_0 u} du \right] \dfrac{\widehat{\delta \sigmac}\left(k\right)}{\sigmac^0} e^{ikz} dk \\
& + \dfrac{1}{2\pi} \int_{-\infty}^{+\infty} \left[ \dfrac{1}{\Cw} \int_{0}^{+\infty} -\dfw\left(u\right) u^{1/2} e^{-\kw_0 u} du \right] \dfrac{\widehat{\delta \omega}\left(k\right)}{\omega_0} e^{ikz} dk \nonumber
\end{align}
Details on the derivation of Eq.~\eqref{eq:Cohesive_SIF_Variations} are given in \ref{app:Compute_SIF}. We further observe that Eq.~\eqref{eq:Cohesive_SIF_Variations} takes a much simpler expression in the Fourier space that one can easily build upon to understand the physical implications of a finite-size dissipation on the fracture process. Eq.~\eqref{eq:Cohesive_SIF_Variations} reads in the Fourier space:
\begin{equation}
\label{eq:Cohesive_SIF_Fourier}
\dfrac{\widehat{\delta K_\mathrm{czm}}\left(k\right)}{\KIc^0} = \left(\dfrac{\hat{\mathcal{A}}\left(\kw_0\right)}{\omega_0} - \dfrac{\vert k\vert}{2}\right) \widehat{\delta a}\left(k\right) + \hat{\Sigma}\left(\kw_0\right) \dfrac{\widehat{\delta \sigmac}\left(k\right)}{\sigmac^0} + \hat{\Omega}\left(\kw_0\right) \dfrac{\widehat{\delta \omega}\left(k\right)}{2\omega_0}
\end{equation}
where:
\begin{equation}
\label{eq:Fourier_Cohesive_Prefactors}
\begin{cases}
\hat{\mathcal{A}}\left(\kw_0\right) & = -\dfrac{1}{\Cw} \int_{0}^{+\infty} \dfrac{\dfw\left(u\right)}{u^{1/2}} \left(1-e^{-\kw_0 u}\right) du \\
\hat{\Sigma}\left(\kw_0\right) & = \dfrac{1}{\Cw} \int_{0}^{+\infty} \dfrac{\fw\left(u\right)}{u^{1/2}} e^{-\kw_0 u} du \\
\hat{\Omega}\left(\kw_0\right) & = -\dfrac{2}{\Cw} \int_{0}^{+\infty} \dfw\left(u\right) u^{1/2} e^{-\kw_0 u} du
\end{cases}
\end{equation}
Combining Eqs.~\eqref{eq:LEFM_SIF_Fourier} and \eqref{eq:Cohesive_SIF_Fourier}, one finds the equation ruling crack propagation in heterogeneous cohesive materials:
\begin{equation}
\label{eq:PropagationCriterion}
\scalemath{0.85}{
\KI^0\left[ 1 + \left(\dfrac{1}{\KI^0} \dfrac{\partial \KI^0}{\partial a} -\dfrac{\vert k\vert}{2}\right) \widehat{\delta a}\left(k\right)\right] = \KIc^0\left[1+\left(\dfrac{\hat{\mathcal{A}}\left(\kw_0\right)}{\omega_0} - \dfrac{\vert k\vert}{2}\right) \widehat{\delta a}\left(k\right) + \hat{\Sigma}\left(\kw_0\right) \dfrac{\widehat{\delta \sigmac}\left(k\right)}{\sigmac^0} + \hat{\Omega}\left(\kw_0\right) \dfrac{\widehat{\delta \omega}\left(k\right)}{2\omega_0}\right]
}
\end{equation}

Equation~\eqref{eq:PropagationCriterion} unveils rich physics about the influence of a finite process zone size on the front deformations. \modif{It can be reformulated in terms of variations of strength and critical crack opening, building on Eq.~\eqref{eq:CohesiveStress_ProcessZoneSize} (see Eq.~\eqref{eq:PropagationCriterion_Dc} in \ref{app:Compute_SIF} for more details).} One can also make use of efficient Fast Fourier Transform (FFT) algorithms to solve it efficiently, at a much lower computational expense than more standard simulation methods \citep{geubelle_spectral_1995}. As such, our model shows potential to investigate front deformations induced by heterogeneities at multiple scales. We focus here on two problems: the configurational stability of a crack propagating in a homogeneous yet cohesive material in Section~\ref{sec:StabilityAnalysis}, and the influence of material heterogeneities on the front deformation in Section~\ref{sec:FrontDeformations}. Before doing so, several comments are in order:
\begin{itemize}
\item First, one expects to find back the results of \cite{rice_first-order_1985} and \cite{gao_trapping_1989} in the limit $\omega_0 \rightarrow 0$. The study of the asymptotic behavior of the cohesive pre-factors $\mathcal{A}$, $\Sigma$, and $\Omega$ yields in this limit:
\begin{equation}
\label{eq:Fourier_Cohesive_Asymptotics_0}
\hat{\mathcal{A}}\left(\kw_0\right) \underset{\kw_0 \rightarrow 0}{\sim} \dfrac{\kw_0}{2} 
\text{; }
\hat{\Sigma}\left(\kw_0\right) \underset{\kw_0 \rightarrow 0}{\longrightarrow} 1
\text{ and }
\hat{\Omega}\left(\kw_0\right) \underset{\kw_0 \rightarrow 0}{\longrightarrow} 1
\end{equation}
So that, in the limit of a perfectly brittle material, the contribution of the front deformations $\delta a$ in $K_\mathrm{czm}$ goes to zero. Eq.~\eqref{eq:CohesiveApproach} yields at first-order in $\delta a$, $\delta \sigmac$, and $\delta \omega$:
\begin{equation}
\label{eq:LEFM_limit}
K_\mathrm{lefm}(z) = \KIc^{0} \left[1+\dfrac{\delta \sigmac(z)}{\sigmac^0} + \dfrac{\delta \omega(z)}{2\omega_0} \right] = \KIc(z)
\end{equation}
where we used Eq.~\eqref{eq:Cohesive_Toughness} to link the fluctuations of cohesive properties $\delta \sigmac$ and $\delta \omega$ to the material toughness $\KIc$. We find back \cite{irwin_fracture_1958}'s criterion that describes crack propagation in perfectly brittle materials.
\item Second, the amplitude of the front deformations $\delta$ are multiplied by a cohesive pre-factor $\hat{\mathcal{A}}$ that depends on the product of the wavenumber $k$ and the process zone size $\omega_0$. It means that the crack front accommodates a given perturbation length scale differently depending on its size and that of the process zone.
\item Third, we observe from Eq.~\eqref{eq:Cohesive_SIF_Fourier} that the process zone acts as a high-frequency filter for the variations of strength $\delta \sigmac$ and that of process zone size $\delta \omega$. One may thus expect that the influence of heterogeneities on rupture propagation can be averaged at scales below the average process zone size $\omega_0$, while the influence of asperities at a scale considerably larger than $\omega_0$ can be assessed quantitatively within the perturbative framework of LEFM.
\item Fourth, the cohesive pre-factors $\hat{\mathcal{A}}$, $\hat{\Sigma}$, and $\hat{\Omega}$ relate to the spatial distribution of weakening characterized by $\fw$. As such, distinct behaviors can be expected depending on how the material weakens. The values of $\hat{\mathcal{A}}$, $\hat{\Sigma}$, and $\hat{\Omega}$ and their asymptotic behavior are either given analytically or computed numerically in \ref{app:Asymptotic_Weakenings} for several types of weakening.
\end{itemize}

\section{Stability analysis}
\label{sec:StabilityAnalysis}

As a first application of the newly derived cohesive ``line-tension'' model of Section~\ref{sec:Model}, we revisit the stability analysis of \cite{rice_first-order_1985}. It allows us to focus on the impact of a finite-size dissipation on the front deformations only, considering \emph{homogeneous} yet \emph{cohesive} materials.

\subsection{Crack front stability to sinusoidal perturbations for perfectly brittle materials}
\label{subsec:StabilityAnalysis_Brittle}

\begin{figure}[h]
\centering \includegraphics[width=\textwidth]{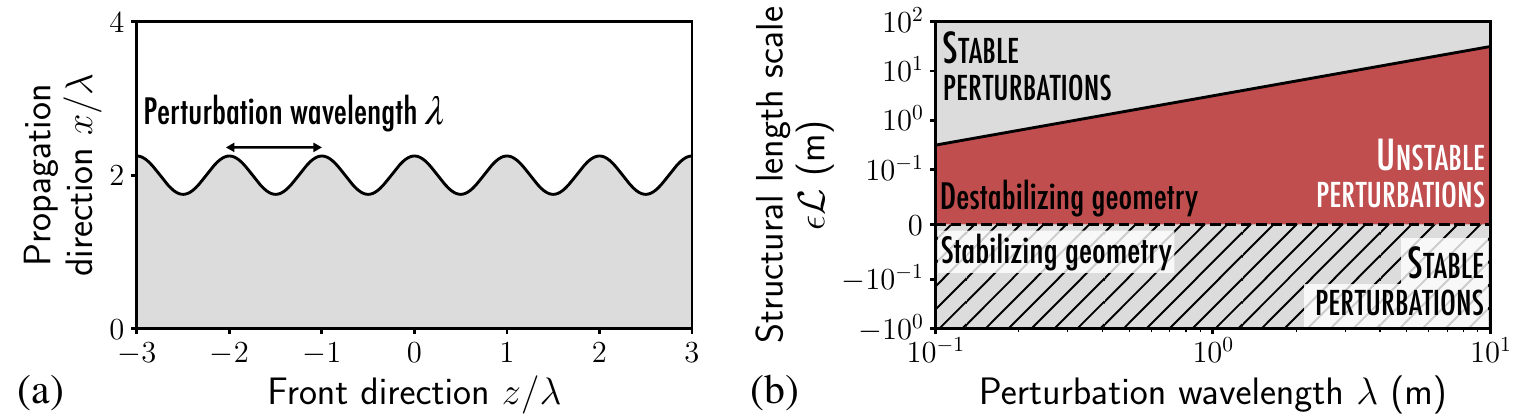}
\caption{\modif{(a) Stability analysis of a sinusoidal crack front (in black solid line) of wavelength $\lambda$ for perfectly brittle materials: (b) the stability of the crack front is controlled by the sign $\epsilon$ and the characteristic distance $\mathcal{L}$ along which the macroscopic $\KI^0$ varies. For stabilizing geometries $\epsilon \leq 0$ (hatched surface), the perturbation is stable (in gray) no matter its wavelength. For destabilizing geometries $\epsilon>0$ (non-hatched surface), the perturbation is unstable (in red) if its wavelength is larger than a critical value $\lambda_\mathrm{c}$, and stable (in gray) otherwise. The critical wavelength $\lambda_\mathrm{c}$ is predicted by Eq.~\eqref{eq:Stability_Brittle_CriticalWavelength}.}}
\label{fig:FrontStability_Brittle}
\end{figure}

The stability of a semi-infinite crack to some front perturbations has been investigated first by \cite{rice_first-order_1985} for perfectly brittle materials. They considered sinusoidal perturbations $\delta a$ that are characterized by their wavenumber $k$, or equivalently by their wavelength $\lambda =2\pi/k$, and their amplitude $A$ (see Fig.~\ref{fig:FrontStability_Brittle}a): 
\begin{equation}
\label{eq:Stability_Brittle_FrontPerturbation}
\delta a(z) = A \cos(kz) 
\end{equation}

Following \cite{rice_first-order_1985}, the stability of the perturbed crack is assessed by looking at the value of mode I SIF at the most advanced points of the crack front $z = n\pi/\lambda$. If this value is found larger than the applied SIF $\KI^0$, crack propagation is considered unstable, and stable otherwise. \cite{rice_first-order_1985} showed that the configurational stability is partially controlled by (i) the sign $\epsilon$ of the variations $\partial \KI^0/\partial a$ of loading with crack advance, and (ii) a structural length scale $\mathcal{L}$ characterizing their intensity. These two quantities are defined by the following equations:
\begin{equation}
\label{eq:Stability_Brittle_StructuralLengthScale}
\mathcal{L} = \left\lvert \KI^0/(\partial \KI^0/\partial a) \right\rvert \text{ and } \epsilon = \mathrm{sgn}(\partial \KI^0/\partial a)
\end{equation}
The expression of the mode I SIF of Eq.~\eqref{eq:LEFM_SIF_Variations} rewrites as:
\begin{equation}
\label{eq:Stability_Brittle_KI}
\KI(z) = \KI^0 \left[1+\left(\frac{1}{\KI^0} \dfrac{\partial \KI^0}{\partial a}-\dfrac{\pi}{\lambda}\right) A \cos(kz) \right] = \KI^0 \left[1+\left(\dfrac{\epsilon}{\mathcal{L}}-\dfrac{\pi}{\lambda}\right) A \cos(kz) \right]
\end{equation}
From Eq.~\eqref{eq:Stability_Brittle_KI}, one observes that the crack is \emph{unconditionally stable} for stabilizing geometries $\epsilon \leq 0$ for which $\KI^0$ decreases with crack advance. When $\KI^0$ increases with crack growth $\epsilon > 0$ (destabilizing geometry), the system is \emph{conditionally stable}: small-wavelength perturbations are stable, while large wavelengths are unstable. This shift in stability occurs for a critical wavelength $\lambda_\mathrm{c}$ that reads:
\begin{equation}
\label{eq:Stability_Brittle_CriticalWavelength}
\lambda_\mathrm{c} = \left[\epsilon \pi\mathcal{L}\right]^+
\end{equation}
where $\left[x\right]^+$ denotes the positive part of the real $x$.

\subsection{Influence of a finite-size dissipation}
\label{subsec:StabilityAnalysis_Cohesive}

One may then wonder how the presence of a finite-size dissipation influences the stability of a crack to sinusoidal perturbations. For homogeneous materials ($\Delta \sigmac = \Delta \omega = 0$), Eq.~\eqref{eq:PropagationCriterion} rewrites as an Irwin-like criterion $\KI^\mathrm{eff} = \KIc^0$, with an effective SIF $\KI^\mathrm{eff}$ that encompasses the influence of finite cohesive stresses behind the crack front. $\KI^\mathrm{eff}$ writes as:
\begin{equation}
\label{eq:Stability_Cohesive_KI_01}
\KI^\mathrm{eff}(z) = \KI^0\left[1+\left(\dfrac{\epsilon}{\mathcal{L}} - \dfrac{\hat{\mathcal{A}}\left(\kw_0\right)}{\omega_0}\right) A \cos(kz)\right]
\end{equation}
\modif{In the remaining of the manuscript, we assume that the material continuously weakens behind the crack front $\dfw \leq 0$. From the definition of $\hat{\mathcal{A}}$ in Eq.~\eqref{eq:Fourier_Cohesive_Prefactors}, one may easily show that $\hat{\mathcal{A}}$ is a (i) \emph{positive} and (ii) \emph{increasing} function of $\kw_0$.} Moreover, in the limit $\abs{k}\omega_0 \rightarrow +\infty$, (iii) $\hat{\mathcal{A}}$ \emph{saturates} to a value $\mathcal{A}_w^\infty$ (see Fig.~\ref{fig:AsymptoticBehavior_LinearDistance}.a) that reads:
\begin{equation}
\label{eq:Fourier_Cohesive_Asymptotics_infinity_A}
\mathcal{A}_w^\infty = \dfrac{1}{\Cw}\int_{0}^{+\infty} -\dfrac{\dfw(u)}{u^{1/2}} du
\end{equation}
One then finds:
\begin{equation}
\label{eq:Stability_Cohesive_KI_02}
\KI^\mathrm{eff}(z) \geq \KI^0\left[1+\left(\dfrac{\epsilon}{\mathcal{L}} - \dfrac{\mathcal{A}_w^\infty}{\omega_0}\right) A \cos(kz)\right]
\end{equation}

\begin{figure}[h]
\centering \includegraphics[width=\textwidth]{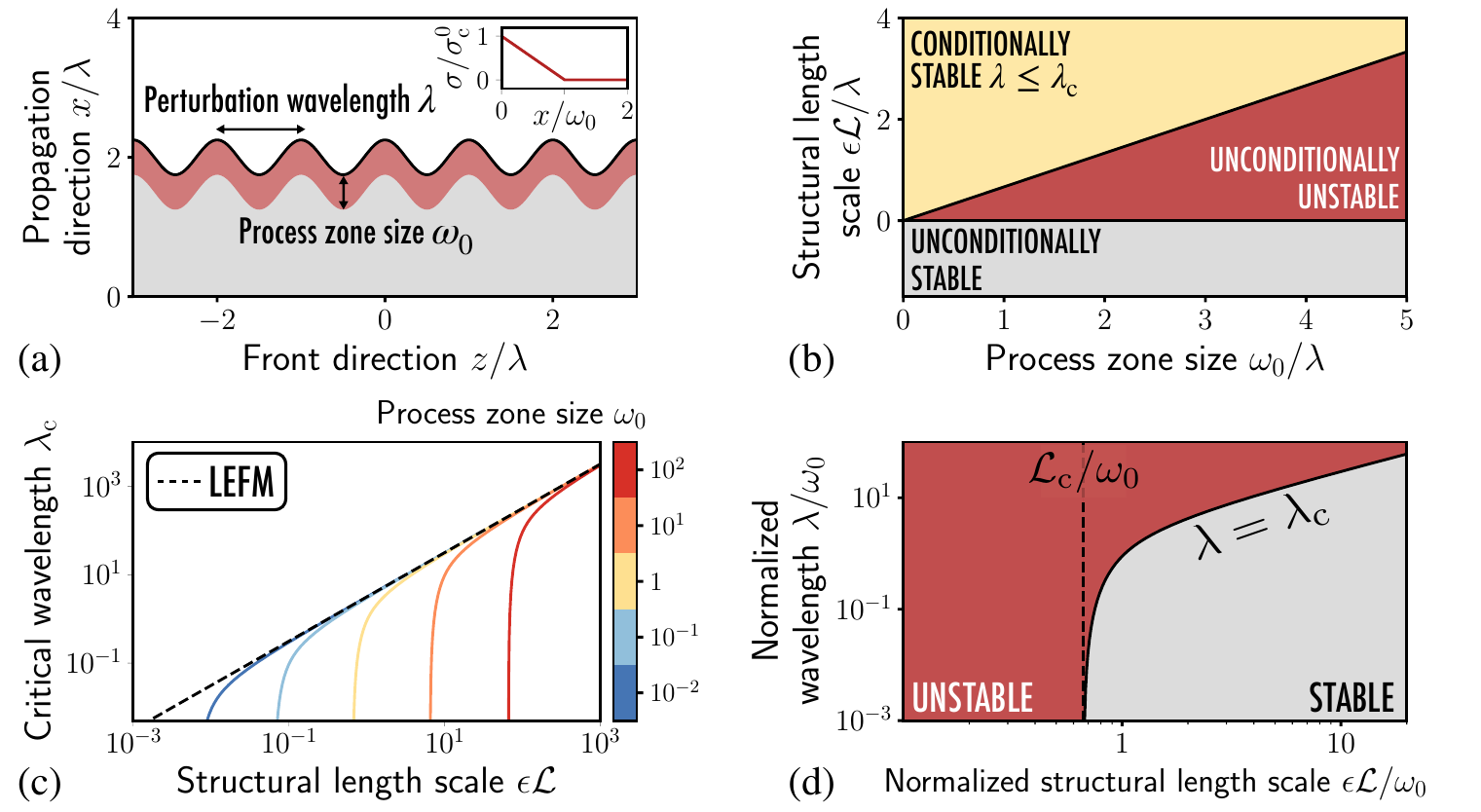}
\caption{\modif{(a) Stability analysis of a sinusoidal crack front (in black solid line) of wavelength $\lambda$ in presence of a finite process zone size $\omega_0$ (in red): (b) the stability of the crack front is controlled by the structural length scale $\mathcal{L}=\KI^0/\vert\partial \KI^0/\partial a\vert$ characteristic of the loading variations and their sign $\epsilon$. For stabilizing geometries $\epsilon \leq 0$, the perturbation is stable no matter its wavelength (in gray; see Regime I in the main text). For destabilizing geometries $\epsilon>0$, the perturbation is stable if its wavelength $\lambda$ is smaller than a critical value $\lambda_\mathrm{c}$ and unstable otherwise (in yellow; see Regime III in the main text). For a large enough process zone size, the critical wavelength $\lambda_\mathrm{c}$ goes to $0$ (in red; see Regime II in the main text), and perturbations of any wavelength are unstable. (c) $\lambda_\mathrm{c}$ depends on both the structural length $\mathcal{L}$ and the process zone size $\omega_0$. (d) When normalized by $\omega_0$ the values of $\lambda_\mathrm{c}$ all collapse to a master curve (in black solid line) predicted by Eq.~\eqref{eq:Stability_Cohesive_CriticalWavelength}. Note that, below a critical value $\mathcal{L}_\mathrm{c}/\omega_0$ of the structural length scale (marked by a vertical dashed black line), perturbations of any wavelength $\lambda$ are unstable (Regime II).}}
\label{fig:FrontStability_Cohesive}
\end{figure}

From Eqs.~\eqref{eq:Stability_Cohesive_KI_01} and \eqref{eq:Stability_Cohesive_KI_02}, one can distinguish three different instability regimes:
\begin{itemize}
\item \textbf{Regime I:} When $\epsilon = -1$ or $0$, the system remains \emph{unconditionally stable} to sinusoidal perturbations due to the positiveness of $\hat{\mathcal{A}}$.
\item \textbf{Regime II:} When $\epsilon = 1$, the system is \emph{unconditionally unstable} to sinusoidal perturbations of any wavelength if the structural length scale $\mathcal{L}$ is smaller than a critical value $\mathcal{L}_\mathrm{c}$, which reads:
\begin{equation}
\label{eq:Stability_Cohesive_CriticalLengthScale}
\mathcal{L}_\mathrm{c} = \dfrac{\omega_0}{\mathcal{A}_w^\infty}
\end{equation}
\item \textbf{Regime III:} When $\epsilon = 1$ and $\mathcal{L} \geq \mathcal{L}_\mathrm{c}$, the system is \emph{conditionally stable}: perturbations of wavelength smaller than a critical value $\lambda_\mathrm{c}$ are stable, while larger ones are unstable. For cohesive materials, $\lambda_\mathrm{c}$ reads:
\begin{equation}
\label{eq:Stability_Cohesive_CriticalWavelength}
\hat{\mathcal{A}}\left(\dfrac{2\pi \omega_0}{\lambda_\mathrm{c}}\right) = \dfrac{\omega_0}{\mathcal{L}}
\end{equation}
In contrast with the perfectly brittle case of Eq.~\eqref{eq:Stability_Brittle_CriticalWavelength} where $\lambda_\mathrm{c}$ only depends on the structural length scale $\mathcal{L}$, it is also influenced here by the process zone size $\omega_0$. We further observe in Fig.~\ref{fig:FrontStability_Cohesive}d that $\lambda_\mathrm{c}$ is smaller than its LEFM value, as $\hat{\mathcal{A}}(\kw_0)/\omega_0 < \abs{k}/2$. As such, the introduction of a finite-size region of dissipation can be associated with a \emph{loss of stiffness} of the crack front. This stiffness loss gets stronger as the wavelength $\lambda$ approaches the process zone size $\omega_0$ (see Fig.~\ref{fig:FrontStability_Cohesive}d and Fig.~\ref{fig:AsymptoticBehavior_LinearDistance} in \ref{app:Asymptotic_Weakenings}).
\end{itemize}
The different regimes and the associated stability diagram are summarized in Fig.~\ref{fig:FrontStability_Cohesive}b-c. As can be seen from the conditions of Eqs.~\eqref{eq:Stability_Cohesive_CriticalLengthScale} and \eqref{eq:Stability_Cohesive_CriticalWavelength}, the stability of a perturbed crack strongly depends on the evolution of $\hat{\mathcal{A}}$ and its asymptotic behavior as $\kw_0 \rightarrow +\infty$. It naturally relates to the spatial distributions of cohesive stress behind the crack front, and so to the function $\fw$. The influence of the precise shape of the cohesive law is discussed in \ref{app:NatureWeakening_Stability}.\\

We conclude this section with some \modif{potential} implications of our results on the stability of perturbed cracks in a \emph{dynamic} setting. We showed that the stability of a perturbed crack front of fixed wavelength $\lambda$ embedded in a destabilizing structure ($\epsilon>0$) is controlled by the size $\omega_0$ of the process zone. A crack in such a configuration could be stable when embedded in a rather brittle material (small $\omega_0/\lambda$) and unstable for a more ductile one (large $\omega_0/\lambda$). \cite{rice_mechanics_1980} showed that the process zone size $\omega_0$ \emph{contracts} as the crack accelerates. One may then imagine situations where a crack oscillates between a \emph{stable} configuration and an \emph{unstable} one: in a first stage, the arrested crack becomes unstable because of some large wavelength perturbations (Regime III). The subsequent decrease in process zone size during the instability could stabilize the system (Regime III to Regime II). As the crack decelerates, the process zone gets larger, and the crack becomes unstable again (Regime II to Regime III). The Lorentz contraction of the process zone with crack velocity may also interact with the dynamic stiffening of the crack front and \modif{crack front waves} observed by \cite{morrissey_perturbative_2000}. \modif{Further investigations are needed to support these ideas. One could build on the spectral method of \cite{geubelle_spectral_1995} to perform efficient numerical simulations of front stability during quasi-static - to - dynamic and dynamic - to - quasi-static transients.}

\section{Deformations of a crack front encountering an obstacle}
\label{sec:FrontDeformations}

We saw that the presence of finite cohesive stress in the crack wake makes the front \emph{more compliant}, especially when the perturbation wavelength is smaller than the process zone size. It is tempting to think that this stiffness loss consequently increases the deformations of a crack front induced by some heterogeneities of fracture energy. Yet, one has to bear in mind that the presence of a finite cohesive zone size also influences the variations of strength and process zone size, from which emerge the fluctuations of fracture energy. To investigate the ultimate impact of these two potentially competing mechanisms, we study the interaction of a crack front with periodic arrays of tough inclusions of increased fracture energy. We recall in Section~\ref{subsec:FrontDeformations_Brittle} the results obtained by \cite{gao_trapping_1989} for perfectly brittle materials. We then investigate the influence of a finite process zone size, first in Section~\ref{subsec:FrontDeformations_Strength} where heterogeneities of fracture energy $\Gc$ are associated with fluctuations of strength $\sigmac$, and second in Section~\ref{subsec:FrontDeformations_ProcessZoneSize} where they emerge from variations of process zone size $\omega$. We discuss in Section~\ref{subsec:FrontDeformations_Mixed} the experimental implications of our findings, and show in Section~\ref{subsec:Chopin2011} that our theory may explain the deformation profiles observed in the peeling experiments of \cite{chopin_crack_2011}, for which the small-scale yielding hypothesis is suspected to break down.

\subsection{Front deformation by a periodic array of obstacles in perfectly brittle materials}
\label{subsec:FrontDeformations_Brittle}

In the general case, the spatial distribution of fracture energy $\Gc$ can be decomposed in the sum of a zero order term $\Gc^0$ corresponding to its spatial average, and a first-order term of fluctuations $\delta \Gc$:
\begin{equation}
\label{eq:FrontDeformation_Brittle_FractureEnergy}
\Gc(z) = \Gc^0 + \delta \Gc(z)
\end{equation}
As the crack interacts with the inclusions, front deformations $\delta a$ arise from the spatial variations $\delta \Gc$ of fracture energy. Building on \cite{griffith_phenomena_1921}'s criterion and \cite{irwin_fracture_1958}'s formula, Eq.~\eqref{eq:LEFM_SIF_Fourier} yields at zero order in the perturbation $\delta a$:
\begin{equation}
\label{eq:FrontDeformation_Brittle_EffectiveToughness}
G^0 = \dfrac{1-\nu^2}{E} {\KI^0}^2 = \Gc^0
\end{equation}
We find back here the results of \cite{roux_effective_2003}, who showed that the effective fracture energy $G^0$ of a heterogeneous brittle material corresponds to its spatial average $\Gc^0$ in the \emph{weak pinning} limit of infinitely long defects. At first order, one finds:
\begin{equation}
\label{eq:FrontDeformation_Brittle_Deformations}
\widehat{\delta a}(k) = - \dfrac{1}{\abs{k}} \dfrac{\widehat{\delta \Gc}(k)}{\Gc^0}
\end{equation}

\begin{figure}[h]
\centering \includegraphics[width=\textwidth]{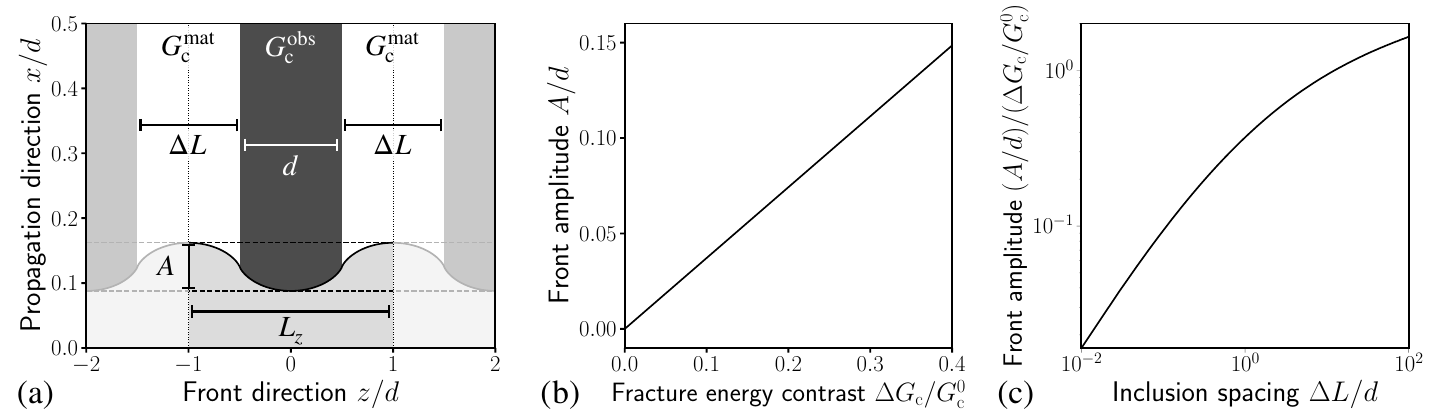}
\caption{(a) The crack front (in solid black line) is deformed by a periodic array of tough obstacles (in gray) of width $d$, separated by a distance $\Delta L$. The obstacles have a fracture energy $G_\mathrm{c}^\mathrm{obs}$ slightly larger than that of the matrix $G_\mathrm{c}^\mathrm{mat}$. The front deformations are characterized by their amplitude $A$, defined as the distance between the most advanced point of the front and the less advanced one. (b) For a \emph{perfectly brittle} material, $A$ increases \emph{linearly} with the fracture energy contrast $\Delta G_\mathrm{c}/G_\mathrm{c}^0$. (c) $A$ depends non-linearly on the ratio of the obstacle spacing $\Delta L$ to their size $d$.}
\label{fig:FrontDeformation_Brittle}
\end{figure}

Eq.~\eqref{eq:FrontDeformation_Brittle_Deformations} links the stationary shape $\delta a$ of the front deformations to the local fluctuations of fracture energy $\delta \Gc$. \cite{gao_trapping_1989} investigated the deformation of the crack front resulting from its interaction with a periodic array of tough obstacles of width $d$ separated by a distance $\Delta L$ (see Fig.~\ref{fig:FrontDeformation_Brittle}). The spatial distribution of fracture energy writes as:
\begin{equation}
\label{eq:FrontDeformation_Brittle_PeriodicArray}
\Gc(z) = 
\begin{cases}
\Gc^\mathrm{obs} & \text{ if } x \in\left[nL_z-\frac{d}{2},(n+1)L_z+\frac{d}{2}\right) \\
\Gc^\mathrm{mat} & \text{ otherwise}
\end{cases}
\end{equation}
where the fracture energy $\Gc^\mathrm{obs}$ of the obstacles is larger than that $\Gc^\mathrm{mat}$ of the embedding matrix, and $L_z = \Delta L + d$ corresponds to the spatial period of the distribution. One may additionally characterize the fracture energy field through the two quantities:
\begin{equation}
\label{eq:FrontDeformation_Brittle_Properties}
\begin{cases}
\Gc^0  = (\Delta L/L_z)\,\Gc^\mathrm{mat} + (d/L_z)\,\Gc^\mathrm{obs}\\
\Delta \Gc = \Gc^\mathrm{obs}-\Gc^\mathrm{mat}
\end{cases}
\end{equation}
where $\Delta \Gc$, the fracture energy contrast, is a parameter that is assumed small in the following.

A quantity of interest is the amplitude $A$ of the front deformations, defined as the distance between the most advanced point of the front and the less advanced one (see Fig.~\ref{fig:FrontDeformation_Brittle}a). From Eq.~\eqref{eq:FrontDeformation_Brittle_Deformations}, one may notice that $A$ increases linearly with $\Delta \Gc/\Gc^0$ (see Fig.~\ref{fig:FrontDeformation_Brittle}b). The dependence of $A$ with the inclusion spacing $\Delta L$ is non-linear, and can be computed numerically from Eq.~\eqref{eq:FrontDeformation_Brittle_Deformations} for spatial distributions of $\Gc$ described by Eq.~\eqref{eq:FrontDeformation_Brittle_PeriodicArray} (see Fig.~\ref{fig:FrontDeformation_Brittle}c).

One may then wonder how these observations fare in the framework of cohesive materials, where the fracture properties are not solely described by the fracture energy $\Gc$, but rather by the strength $\sigmac$ and the process zone size $\omega$. Using \cite{irwin_fracture_1958}'s formula combined with Eq.~\eqref{eq:Cohesive_Toughness}, one finds:
\begin{equation}
\label{eq:FrontDeformation_Cohesive_FractureEnergy}
\Gc^0 = \dfrac{1-\nu^2}{E}{\KIc^0}^2 = \dfrac{1-\nu^2}{E} \dfrac{2}{\pi}\Cw^2 {\sigmac^0}^2 \omega_0
\end{equation}
As such, the variations of fracture energy $\delta \Gc$ in cohesive materials can either emerge from the variations of strength $\delta \sigmac$, or from that of process zone size $\delta \omega$. At first order in those two quantities, one has:
\begin{equation}
\label{eq:FrontDeformation_Cohesive_FractureEnergyVariations}
\dfrac{\delta \Gc(z)}{\Gc^0} = 2\dfrac{\delta \sigmac(z)}{\sigmac^0} + \dfrac{\delta \omega(z)}{\omega_0}
\end{equation}
Next, we explore two limit cases: that of Section~\ref{subsec:FrontDeformations_Strength} where heterogeneities of fracture energy solely emerge from local variations of strength, and that of Section~\ref{subsec:FrontDeformations_ProcessZoneSize} where they are associated with fluctuations of process zone size only. 

\subsection{Heterogeneities of strength}
\label{subsec:FrontDeformations_Strength}

We first consider the case where the heterogeneities of fracture energy emerge from variations of strength only. The strength of the obstacles $\sigmac^\mathrm{obs}$ is larger than that of the matrix $\sigmac^\mathrm{mat}$, while they are equally brittle $\omega^\mathrm{obs} = \omega^\mathrm{mat} = \omega_0$. \modif{Note that in the standard formulation of cohesive-zone models that deal with the pair $(\sigma_\mathrm{c}, \deltac)$ instead of $(\sigma_\mathrm{c}, \omega)$, this corresponds to a simultaneous proportional increase of  strength $\sigma_\mathrm{c}$ and critical opening $\deltac$.}

We define the average strength $\sigmac^0$ and the strength contrast $\Delta \sigmac$ as:
\begin{equation}
\label{eq:FrontDeformation_Strength_Properties}
\begin{cases}
\sigmac^0 = (\Delta L/L_z)\,\sigmac^\mathrm{mat} + (d/L_z)\,\sigmac^\mathrm{obs}\\
\Delta \sigmac = \sigmac^\mathrm{obs}-\sigmac^\mathrm{mat}
\end{cases}
\end{equation}
The resulting fluctuations of fracture energy are linked to the variations of strength following:
\begin{equation}
\Delta \Gc/\Gc^0 = 2\Delta \sigmac/\sigmac^0
\end{equation}
The propagation criterion of \eqref{eq:PropagationCriterion} yields at first order in the perturbations $\delta a$ and $\delta \sigmac$:
\begin{equation}
\label{eq:FrontDeformation_Cohesive_Strength}
-\hat{\mathcal{A}}(\kw_0) \dfrac{\widehat{\delta a}(k)}{\omega_0} = \hat{\Sigma}(\kw_0) \dfrac{\widehat{\delta \sigmac}(k)}{\sigmac^0} = \dfrac{\delta \Gc^\mathrm{eff}(k)}{2\Gc^0}
\end{equation}
meaning that the front deformation increases linearly with the normalized strength contrast $\Delta \sigmac/\sigmac^0$, and so with the normalized fracture energy contrast $\Delta \Gc/\Gc^0$.\\

\begin{figure}[!h]
\centering \includegraphics[width=\textwidth]{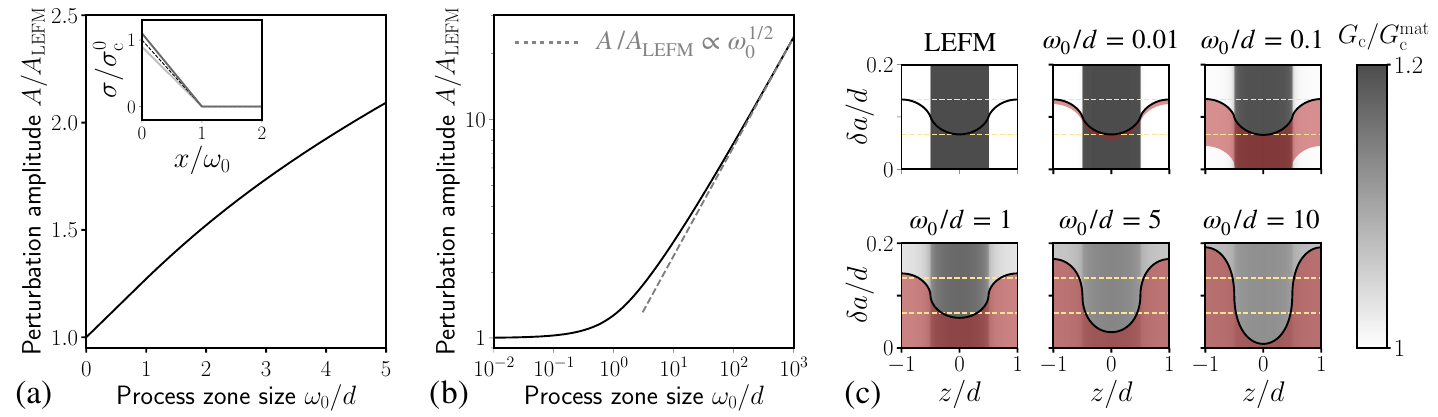}
\caption{\modif{(a) Influence of the average process zone size $\omega_0$ on the amplitude $A$ of the front deformations for a periodic array of obstacles of width $d$; Inset: the obstacles (in solid dark gray line) are stronger than the matrix $\sigma_\mathrm{c}^\mathrm{obs} = 1.1 \sigma_\mathrm{c}^\mathrm{mat}$ (in solid light gray line), but they are equally brittle $\omega^\mathrm{obs} = \omega^\mathrm{mat}$, so that $G_\mathrm{c}^\mathrm{obs} \simeq 1.2 G_\mathrm{c}^\mathrm{mat}$. For moderate values of $\omega_0/d$, the amplitude $A$ increases almost linearly with the process zone size $\omega_0$ from its reference LEFM value. (b) For larger values of $\omega_0/d$, it scales as $\propto (\omega_0/d)^{1/2}$. (c) The presence of a finite process zone size smooths out the small-scale variations of strength, so that the amplitude of the apparent fracture energy field (in gray scale) is decreased from its material value (LEFM case). Yet, the decrease in apparent fracture energy amplitude for increasing process zone size (in red) is weaker than the associated decrease in front stiffness, so that the deformation amplitude (extremal points of the crack front in solid black line) is ultimately larger than in the LEFM case (in yellow dashed line).}}
\label{fig:FrontDeformation_Strength_ProcessZone}
\end{figure}

The presence of a cohesive stress in the crack wake influences the amplitude of the front deformations through two \emph{competing} mechanisms. We saw in Section~\ref{sec:StabilityAnalysis} that (i) it \emph{decreases the front stiffness} through the cohesive pre-factor $\hat{\mathcal{A}}$. We now observe from Eq.~\ref{eq:FrontDeformation_Cohesive_Strength} that (ii) it also \emph{smooths out the fluctuations of strength} $\delta \sigmac$ ($\hat{\Sigma}(\kw_0) \leq 1$), leaving its average value $\sigmac^0$ unchanged. The overall decrease in magnitude of the strength fluctuations is mostly prevalent at scales lower than the average process zone size $\omega_0$ (see the evolution of $\hat{\Sigma}$ in Fig.~\ref{fig:AsymptoticBehavior_LinearDistance}b). While the first effect leads to an increase of the front deformations with respect to the perfectly brittle case, the second effect should decrease it as the crack experiences an effectively smoother distribution $\delta \Gc^\mathrm{eff}$ of fracture energy. Examples of the effective fluctuations of fracture energy $\delta \Gc^\mathrm{eff}$ are shown in Fig.~\ref{fig:FrontDeformation_Strength_ProcessZone}c for several values of process zone size to obstacle width ratio $\omega_0/d$.\\

We show in Fig.~\ref{fig:FrontDeformation_Strength_ProcessZone}a the evolution of the front deformation amplitude $A$ with the process zone size $\omega_0$. The competition between the two opposite effects mentioned above results in an overall \emph{increase} of the amplitude of the front deformations with the process zone size. \modif{This increase in front amplitude is observed no matter the inclusion spacing (see Fig.~\ref{fig:FrontDeformation_Strength_InclusionSpacing} and \ref{app:InclusionSpacing_Heterogeneities} for more details).}

In accordance with the asymptotic behaviors derived in Eq.~\eqref{eq:Fourier_Cohesive_Asymptotics_0}, $A$ converges to its LEFM value $A_\mathrm{LEFM}$ as the ratio between the process zone size $\omega_0$ and the obstacle width $d$ goes to zero (see Fig.~\ref{fig:FrontDeformation_Strength_ProcessZone}a). We further notice that the amplitude of the deformations grows as $(\omega_0/d)^{1/2}$ for process zone sizes much larger than the obstacle width  (see Fig.~\ref{fig:FrontDeformation_Strength_ProcessZone}b). This was expected from the asymptotic behavior of $\hat{\Sigma}$ when $\kw_0 \rightarrow +\infty$:
\begin{equation}
\label{eq:Fourier_Cohesive_Asymptotics_infinity_Sigma}
\hat{\Sigma}\left(\kw_0\right) \underset{\kw_0 \rightarrow +\infty}{\sim} \dfrac{\Sigma_w^\infty}{(\kw_0)^{1/2}} \text{, with } \Sigma_w^\infty = \dfrac{\sqrt{\pi}\fw(0)}{\Cw}
\end{equation}
The front deformations are thus amplified by the existence of cohesive stresses, as the gain in front compliance overcomes the smoothing of heterogeneities by the process zone. This behavior and the associated scaling are retrieved independently of how the material weakens behind the crack front (see Fig.~\ref{fig:NatureWeakening_Strength} and \ref{app:NatureWeakening_FrontDeformations} for more details).

\modif{A large process zone size lead to a localization of the deformation at the edges of the defect (see Fig.~\ref{fig:FrontDeformation_Strength_ProcessZone}c, and Eq.~\eqref{eq:FrontDeformation_Strength_SingleDefect} for an analytical expression of the front deformations in the limit case of a single defect embedded in an infinite matrix).} Similar deformation patterns have been previously observed in peeling experiments of a silicon elastomer block from a heterogeneous glass substrate \citep{chopin_crack_2011}. They are not ubiquitous, as other peeling experiments lead to deformations closer to the LEFM predictions \citep{patinet_pinning_2013, vasoya_experimental_2016}. Our theory may then provide a comprehensive framework to bridge various experimental observations where the small-scale yielding condition of LEFM may not always be met. We explore further this avenue in the final Section~\ref{subsec:Chopin2011}.

\subsection{Heterogeneities of process zone size}
\label{subsec:FrontDeformations_ProcessZoneSize}

We explore next the situation where the heterogeneities of fracture energy solely emerge from variations of process zone size. The process zone size of the obstacles $\omega^\mathrm{obs}$ is now larger than that of the matrix $\omega^\mathrm{mat}$, but both materials are equally strong $\sigmac^\mathrm{obs} = \sigmac^\mathrm{mat} = \sigmac^0$. \modif{In the standard formulation of cohesive-zone models that deal with the pair $(\sigma_\mathrm{c}, \deltac)$ instead of $(\sigma_\mathrm{c}, \omega)$, this corresponds to an increase of critical opening $\deltac$ only.}

We define the average process zone size $\omega^0$ and the size contrast $\Delta \omega$ as:
\begin{equation}
\label{eq:FrontDeformation_ProcessZoneSize_Properties}
\begin{cases}
\omega_0 = (\Delta L/L_z)\,\omega^\mathrm{mat} + (d/L_z)\,\omega^\mathrm{obs}\\
\Delta \omega = \omega^\mathrm{obs}-\omega^\mathrm{mat}
\end{cases}
\end{equation}
The resulting fluctuations of fracture energy are linked to the variations of process zone size following:
\begin{equation}
\Delta \Gc/\Gc^0 = \Delta \omega/\omega_0
\end{equation}
The propagation criterion of \eqref{eq:PropagationCriterion} yields at first order in the perturbations $\delta a$ and $\delta \omega$:
\begin{equation}
\label{eq:FrontDeformation_Cohesive_ProcessZoneSize}
-\hat{\mathcal{A}}(\kw_0) \dfrac{\widehat{\delta a}(k)}{\omega_0} = \hat{\Omega}(\kw_0) \dfrac{\widehat{\delta \omega}(k)}{2\omega_0} = \dfrac{\delta \Gc^\mathrm{eff}(k)}{2 \Gc^0}
\end{equation}
meaning that the front deformations increase linearly with the normalized process zone size contrast $\Delta \omega/\omega_0$, and so with the normalized fracture energy contrast $\Delta \Gc/\Gc^0$.\\

\begin{figure}[!h]
\centering \includegraphics[width=\textwidth]{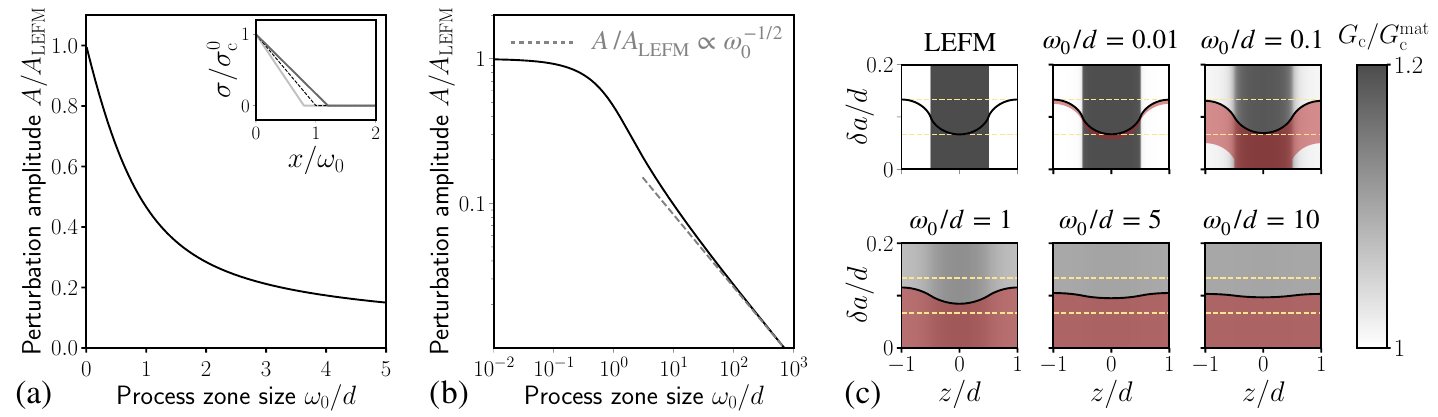}
\caption{\modif{(a) Influence of the average process zone size $\omega_0$ on the amplitude $A$ of the front deformations for a periodic array of obstacles of width $d$; Inset: the obstacles (in solid dark gray line) are more ductile than the matrix $\omega^\mathrm{obs} = 1.2 \omega^\mathrm{mat}$ (in solid light gray line), but they are equally strong $\sigmac^\mathrm{obs} = \sigmac^\mathrm{mat}$, so that $G_\mathrm{c}^\mathrm{obs} \simeq 1.2 G_\mathrm{c}^\mathrm{mat}$. For moderate values of $\omega_0/d$, the deformation amplitude $A$ decreases almost linearly with the process zone size $\omega_0$ from its reference LEFM value. (b) For larger values of $\omega_0/d$, it scales as $\propto (\omega_0/d)^{-1/2}$. (c) The presence of a finite process zone size smooths out the small-scale variations of process zone size, so that the apparent amplitude of the fracture energy field (in gray scale) is decreased from its material value (LEFM case). Furthermore, the decrease in apparent fracture energy amplitude for increasing process zone size (in red) is stronger than the associated decrease in front stiffness, so that the deformation amplitude (extremal points of the crack front in solid black line) is ultimately smaller than in the LEFM case (in yellow dashed line).}}
\label{fig:FrontDeformation_ProcessZoneSize_ProcessZone}
\end{figure}

Again, the amplitude $A$ of the front deformations and its evolution with the average process zone size $\omega_0$ result from the competition between an increased compliance of the crack front, which promotes front deformations, and the decrease in magnitude of the effective fracture energy fluctuations $\delta \Gc^\mathrm{eff}$, which smooths out the front. Surprisingly, we observe in Fig.~\ref{fig:FrontDeformation_ProcessZoneSize_ProcessZone}a that the amplitude $A$ \emph{decreases} with the average process zone size $\omega_0$ for heterogeneities of process zone size. This is explained by the asymptotic behavior of the cohesive pre-factor $\hat{\Omega}$ when $\kw_0 \rightarrow +\infty$:
\begin{equation}
\label{eq:Fourier_Cohesive_Asymptotics_infinity_Omega}
\hat{\Omega}\left(\kw_0\right) \underset{\kw_0 \rightarrow +\infty}{\sim} \dfrac{\Omega_w^\infty}{(\kw_0)^{3/2}} \text{, with } \Omega_w^\infty = -\dfrac{\sqrt{\pi}\dfw(0)}{\Cw}
\end{equation}
In that case, the increase in front compliance with the average process zone size cannot compensate for the sharp decrease in magnitude of the process zone size fluctuations. It results in front deformations vanishing as $(\omega_0/d)^{-1/2}$ for average process zone sizes larger than the obstacle width (see Fig.~\ref{fig:FrontDeformation_ProcessZoneSize_ProcessZone}b). \modif{This decrease in front amplitude is observed no matter the inclusion spacing (see Fig.~\ref{fig:FrontDeformation_ProcessZoneSize_InclusionSpacing} and \ref{app:InclusionSpacing_Heterogeneities} for more details). However, the associated scaling strongly depends on how the material weakens behind the crack front (see Fig.~\ref{fig:NatureWeakening_ProcessZoneSize}).} This is explored in more details in \ref{app:NatureWeakening_FrontDeformations}.\\

\modif{The deformation patterns showed in Fig.~\ref{fig:FrontDeformation_ProcessZoneSize_ProcessZone} do not correspond to front configurations observed experimentally. This is expected as real materials usually display variations of both strength and process zone size, and the influence of \emph{amplified} spatial variations of strength should often prevail over that \emph{vanishing} of process zone size in the limit $\omega_0/d \ll 1$ of large process zone size. } 

Overall, our model shed light on the strong influence of the nature of heterogeneities on the front deformations. \modif{When extended to a dynamic setting, it helps to rationalize the front deformations measured in numerical simulations of dynamic rupture where a crack interacts with heterogeneities of cohesive properties \citep{roch_dynamic_2022}.} It may also explain changes in front roughness observed in quasi-static simulations of coplanar crack propagation based on cohesive zone models \citep{sevillano_roughness_2007}. In accordance with our findings, the authors show that crack fronts interacting with heterogeneities of strength get much rougher than those interacting with obstacles of larger process zone size.

\subsection{Implications for the measurement of fracture energy variations from front deformations}
\label{subsec:FrontDeformations_Mixed}

We saw that the front deformations are \emph{amplified} by the presence of a finite process zone for heterogeneities of strength (see Section \ref{subsec:FrontDeformations_Strength}), while it is found \emph{vanishing} for heterogeneities of process zone size (see Section \ref{subsec:FrontDeformations_ProcessZoneSize}). The nature of the heterogeneities thus strongly influences the way cracks distort when interacting with material heterogeneities. Yet, both types of heterogeneities impact similarly the overall value of fracture energy (see Eq.~\eqref{eq:FrontDeformation_Cohesive_FractureEnergyVariations}). As such, one may wonder if the amplitude of the front deformations constitutes a robust measure of the fracture energy contrast.\\

To tackle this issue, we study the general case where heterogeneities of fracture energy emerge from fluctuations of both strength and process zone size. \modif{Together with Eqs.~\eqref{eq:LEFM_SIF_Variations} and \eqref{eq:Cohesive_Toughness}, the condition~\eqref{eq:CohesiveApproach} of finiteness of the stress at the crack front yields at zero order in the perturbations $\delta a$, $\delta \sigmac$ and $\delta \omega$}:
\begin{equation}
\label{eq:FrontDeformation_Cohesive_EffectiveToughness}
G^0 = \Gc^0
\end{equation}
In other words, the presence of a cohesive zone does not influence the effective toughness of a heterogeneous material in the \emph{weak pinning} regime. 

\begin{figure}[!h]
\centering \includegraphics[width=\textwidth]{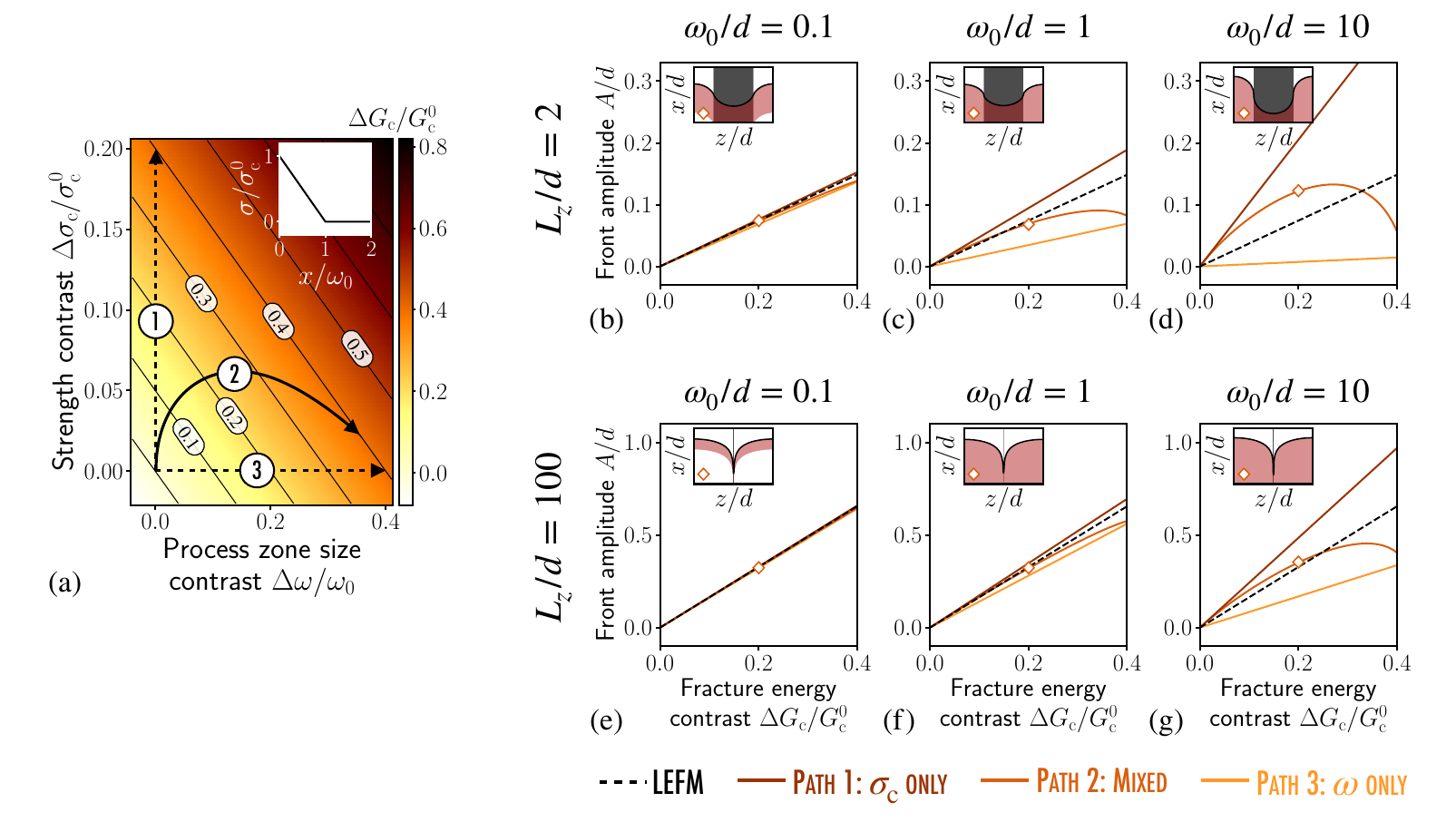}
\caption{(a) An increase of material fracture energy can be achieved either by increasing its strength at constant brittleness (see dashed vertical line of path 1), by increasing its ductility at constant strength (see dashed horizontal line of path 3), or by any mixed path combining an increase of strength with that of ductility (see curved black line of path 2). (b-g) While the front amplitude increases linearly with the ratio of the fracture energy of the obstacles $\Gc^\mathrm{obs}$ to that of the matrix $\Gc^\mathrm{mat}$ in the limit cases of LEFM (dashed black line), and path 1 (solid brown line) \& 3 (solid light orange line), it usually evolves non-linearly with the fracture energy contrast $\Delta G_\mathrm{c}/G_\mathrm{c}^0$ in presence of a finite-size dissipation (see path 2 in solid dark orange line). The amplitude of the front deformations may vary between the two curves associated with paths 1 \& 3 (light yellow area). These variations increase with the process zone size $\omega$ (see b \& e, c \& f, d \& g), but decrease with the obstacle spacing $L_z$ (see b-d \& e-g). Insets: front deformation profiles for $\Delta \Gc/\Gc^0 = 1.2$.}
\label{fig:FrontDeformation_FractureEnergy}
\end{figure}

At first order, one finds:
\begin{equation}
\label{eq:FrontDeformation_Cohesive_StrenghAndProcessZoneSize}
\widehat{\delta a}(k) = - \omega_0 \dfrac{\hat{\Sigma}(\kw_0)}{\hat{\mathcal{A}}(\kw_0)} \dfrac{\widehat{\delta \sigmac}(k)}{\sigmac^0} - \omega_0 \dfrac{\hat{\Omega}(\kw_0)}{\hat{\mathcal{A}}(\kw_0)} \dfrac{\widehat{\delta \omega}(k)}{2\omega_0}
\end{equation}
One may then define multiple paths of increasing fracture energy contrast $\Delta \Gc^0/\Gc^0$ from a path of increasing strength (Path 1 of Fig.~\ref{fig:FrontDeformation_FractureEnergy}a) to one of increasing process zone size (Path 3 of Fig.~\ref{fig:FrontDeformation_FractureEnergy}a). We observe in Fig.~\ref{fig:FrontDeformation_FractureEnergy}b-d that the former constitutes an upper bound of the front deformation amplitude $A$, while the latter represents a lower bound. Mixed paths (Path 2 of Fig.~\ref{fig:FrontDeformation_FractureEnergy}a) are found in between. When the average process zone size $\omega_0$ is way smaller than the obstacle width $d$ (see Fig.~\ref{fig:FrontDeformation_FractureEnergy}b), the two bounds converge to the LEFM value and the amplitude of the front deformation constitutes a good proxy for the contrast $\Delta \Gc/\Gc^0$ of the fracture energy \citep{patinet_pinning_2013}. For larger process zone sizes, the two bounds depart from one another (see yellow area on Fig.~\ref{fig:FrontDeformation_FractureEnergy}b-d). Yet, a lower dispersion on $\Delta \Gc/\Gc^0$  can be observed when the obstacle spacing $\Delta L$ is larger than the process zone size $\omega_0$ (see Fig.~\ref{fig:FrontDeformation_FractureEnergy}e-g). This is explained by the fact that the spectrum of the fracture energy fluctuations is then mostly carried by large-wavelengths. As such, one should rather measure a contrast in fracture energy from experiments of a crack interacting with a single defect \citep{patinet_pinning_2013} rather than arrays of close obstacles \citep{dalmas_pinning_2009}.\\

It is also worth noticing that a non-linear dependence of the deformation amplitude $A$ with the contrast of fracture energy can be observed for cohesive materials (see Path 2 of Fig.~\ref{fig:FrontDeformation_FractureEnergy}b-g), even within the linear theory. This is solely due to the differential impact of heterogeneities of strength or process zone size on the front deformations.

\subsection{Comparison with peeling experiments along heterogeneous interfaces}
\label{subsec:Chopin2011}

We conclude this section by comparing the output of our model to deformation profiles obtained in experiments. The first-order theory of \cite{rice_first-order_1985} has been shown to reproduce a wide variety of experimental observations of crack front deformations in experiments of peeling or fracture of heterogeneous interfaces \citep{dalmas_pinning_2009}, even if more refined models taking into account the finite thickness of the fracture specimen \citep{legrand_coplanar_2011} or higher-order terms \citep{vasoya_second_2013} yield more accurate results (see \citep{patinet_pinning_2013} and \citep{vasoya_experimental_2016} respectively). \cite{chopin_crack_2011} noted nonetheless some discrepancies between the LEFM predictions and the deformation profile of the crack front in their peeling experiments of a silicon elastomer block from a patterned glass substrate, in particular close to the obstacle. We show here that they may emerge from a non-negligible process zone size, and that our model of Eq.~\eqref{eq:PerturbedWeightFunction} provides a better fit for the experimental profile of crack front deformations in cases where the small-scale yielding assumption breaks down.

\begin{figure}[!h]
\centering \includegraphics[width=\textwidth]{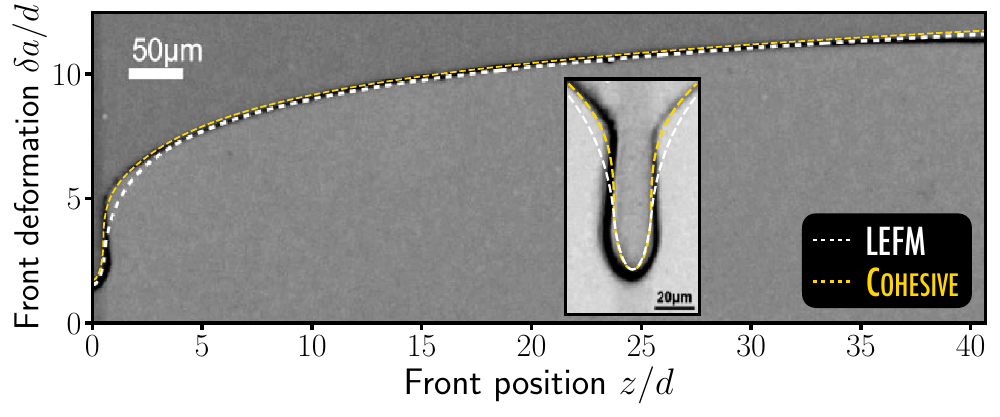}
\caption{Prediction of the crack front deformations in peeling experiments of a silicon elastomer block from a patterned glass substrate (adapted from \citep{chopin_crack_2011}). A $20 \mu m$-large tough obstacle is placed at the center of the $22 mm$-large block, and deforms the crack front. \cite{chopin_crack_2011}'s LEFM fit of the front deformations  is plotted in white dashed line for $\Delta \KIc/\KIc^0 \simeq 3$. Our cohesive fit of Eq.~\ref{eq:PropagationCriterion} for $\Delta \KIc/\KIc^0 = 2.9$ ($\Delta \sigmac/\sigmac^0 = 1.65$ and $\Delta \omega/\omega_0 = 2.5$) and $\omega_0/d = 2.7$ is plotted in yellow dashed line; Inset: Zoom on the front deformations near the obstacle. The large deformations observed at the edges of the defect are a signature of the cohesive nature of the interface.}
\label{fig:Chopin2011_CohesiveFit}
\end{figure}

\cite{chopin_crack_2011} performed peeling experiments of a silicon elastomer block from a patterned glass substrate. The $22$ or $72\mathrm{mm}$-wide and \modif{$10\mathrm{mm}$-thick} elastomer block is made of black cross-linked PDMS (Sylgard170, Dow Corning) of Young's modulus $E \simeq 2\mathrm{MPa}$. It is peeled from a rigid substrate along a heterogeneous interface consisting of a $20\mathrm{\mu m}$-wide PDMS-glass obstacle embedded in a weak PDMS-chromium layer. The toughness contrast between the two types of interfaces has been estimated to $\Delta \KIc/\KIc^0 = 5-9$ from independent peeling tests along homogeneous interfaces. During the experiments, the crack front is pinned by the tough obstacle, and front deformations emerge from the interaction between the crack and the heterogeneous toughness field. \cite{chopin_crack_2011} showed that the deformation profile predicted by LEFM for a single tough defect (see Eq.~\eqref{eq:FrontDeformation_Brittle_SingleDefect}) corresponds to that observed in the experiments, when looked at far away from the defect (see dashed white line in Fig.~\ref{fig:Chopin2011_CohesiveFit}). Yet, two major discrepancies were observed: (i) the LEFM best-fit of the deformation profile corresponds to a toughness contrast $\Delta \KIc/\KIc^0 \simeq 3$, which is lower than the one estimated from the interfacial properties $\Delta \KIc/\KIc^0 = 5-9$; (ii) the experimental crack front displays much sharper variations at the edges of the obstacle that those predicted by LEFM. \modif{The first discrepancy can be attributed to finite-thickness and second-order effects, as they have been shown to reduce crack front deformations \citep{vasoya_experimental_2016}.} We argue here that the second discrepancy relates to a non-negligible process zone size in their experiments.\\

Indeed, we saw in Section~\ref{subsec:FrontDeformations_Strength} that a finite process zone size leads to an overall increase of the amplitude of the front deformations. In that case, the front deformations are concentrated at the obstacles edges (see Fig~\ref{fig:FrontDeformation_Strength_ProcessZone}), as observed in the experiments of \cite{chopin_crack_2011}. \modif{While the localization of the front  deformation is quantitatively grasped by our cohesive framework (see Eq.~\eqref{eq:FrontDeformation_Strength_SingleDefect} describing the front deformations for the single defect and large process zone size), this feature cannot be predicted either by the finite-size effects emerging from the finite thickness of the PDMS block with respect to the defect size \citep{legrand_coplanar_2011} or from higher-order effects \citep{vasoya_experimental_2016}}.

The experimental front profile is best fitted by our model (see dashed yellow line of Fig~\ref{fig:FrontDeformation_Strength_ProcessZone}) for a process zone size $\omega_0 \simeq 2.7 d = 54\mathrm{\mu m}$, a strength contrast $\Delta \sigmac/\sigmac^0 = 1.65$ and a process zone size contrast $\Delta \omega/\omega_0 = 2.5$, which are equivalent to a toughness contrast $\Delta \KIc/\KIc^0 = 2.9$. The size of the process zone $\omega_0 \simeq  54\mathrm{\mu m}$ is compatible with values of the adhesive length of soft materials \citep{creton_fracture_2016}. Moreover, our measurements yield a value of toughness contrast similar to that of \cite{chopin_crack_2011} based on LEFM. This was somehow expected as, in the case of the single obstacle, the amplitude of the front deformations is mostly controlled by the toughness contrast $\Delta \KIc/\KIc^0$ when $\omega_0 \simeq 54 \mathrm{\mu m} \ll \Delta L \simeq 22 \mathrm{mm}$ (see Section~\ref{subsec:FrontDeformations_Mixed}). \modif{Promising directions for a better match of the experimental data would consist in accounting for the influence of the finite thickness of the PDMS block, and second-order terms. Following the theory of \cite{legrand_coplanar_2011}, the former effect is expected to be rather small in the experiments of \citep{chopin_crack_2011}, as the thickness of the PDMS block ($h \simeq 10\mathrm{mm}$) is much larger than the defect width ($d \simeq 20\mathrm{\mu m}$). On the contrary, the influence of second-order terms could be meaningful as the toughness contrast between the matrix and the obstacle is rather large ($\Delta \KIc/\KIc^0 = 5-9$), and they have been shown to reduce significantly crack front deformations even for contrasts as small as $1$ \citep{vasoya_experimental_2016}. Note that mixed mode effects due to a friction-induced shear loading can be observed in the experimental setup of \cite{chopin_crack_2011}, and may also explain some discrepancies between the experimental observations and the theoretical predictions.}

\section{Conclusion}

This study provides a theoretical framework based on a perturbative approach of LEFM that describes the influence of a finite process zone on the deformations of a crack front by heterogeneities of fracture properties. Namely, we extended in equation~\eqref{eq:PropagationCriterion} the first-order theory of \cite{rice_first-order_1985} designed for perfectly brittle materials to the broader case of cohesive materials, where crack advance is resisted by cohesive stress in its wake. %This was performed by deriving in Eq.~\eqref{eq:PerturbedWeightFunction} the analitical expression for the crack face weight function of a semi-infinite crack perturbed with its plane, at first order in the perturbation. 
Our model allowed us to revisit \cite{rice_first-order_1985}'s problem of crack front stability to sinusoidal perturbations and \cite{gao_trapping_1989}'s of crack pinning by an array of tough obstacles, and to stress out the influence of a finite process zone on these matters.\\

We first showed that cohesive cracks accommodate a perturbation differently depending on the size of its wavelength with respect to that of the process zone. In particular, perturbations of wavelength smaller than the process zone size are amplified with respect to the LEFM case, while perturbations of wavelength larger than the process zone size are left unchanged. This was interpreted as a \emph{stiffness loss of the crack front} due to the presence of cohesive stresses in the crack wake. As a result, cracks may be unstable to perturbations of any wavelength if embedded in a \emph{destabilizing} structure, where the mode I SIF strongly increases with crack advance. 

The loss in front stiffness may then lead to amplified front deformations of a crack interacting with periodic arrays of tough obstacles. We showed that it was not always the case, as the process zone also \emph{smooths out the local fluctuations of strength and process zone size} that occur at a scale lower than the process zone size. Interestingly, we showed that the gain in front compliance overcomes the smoothing of perturbations in the case of heterogeneities of strength, while the decrease in amplitude of the fluctuations is found sharper than the stiffness loss in the case of heterogeneities of process zone size. As a result, the front deformations are \emph{amplified} by the presence of a process zone when the increased fracture energy is associated with a higher strength of the obstacle, and \emph{reduced} when it emerges from a larger process zone size. 

Overall, our theory reconciles the wide variety of front profiles observed in experiments, when the small-scale yielding hypothesis of linear elastic fracture mechanics breaks down, as in \citep{chopin_crack_2011}.\\

Future works may focus on the extension of our model to disordered microstructures that are no more invariant in the front direction. A particular attention should then be devoted to the interaction between the finite process zone size, characteristics of the dissipation, and the Larkin length that emerges from the disorder \citep{larkin_pinning_1979}. This may ultimately provide insights on the influence of a finite process zone size on the effective fracture properties of quasi-brittle materials. \modif{Another promising direction is to investigate crack front deformations in a dynamic setting, where the contraction of the process zone with crack velocity \citep{rice_mechanics_1980} interacts with the dynamic stiffening of the crack front observed by \cite{morrissey_perturbative_2000}. This is done in a follow-up study, in which our framework is extended to dynamic fracture, and theoretical predictions are compared to the results of numerical simulations \citep{roch_dynamic_2022}.}

\section*{Acknowledgments}
The Python script used to generate the figures and the associated data are available at:\\ https://doi.org/10.5281/zenodo.6669527. The authors gratefully thank Jean-Baptiste Leblond for stimulating and fruitful discussions, and Julien Chopin for permission to adapt Fig.~\ref{fig:Chopin2011_CohesiveFit} from their work \citep{chopin_crack_2011}.

\section*{CRediT authorship contribution statement}
\noindent
\textbf{Mathias Lebihain:} Conceptualization, Formal analysis, Investigation, Visualization, Software, Writing - Original Draft;
\textbf{Thibault Roch:} Conceptualization, Investigation, Writing - Review \& Editing;
\textbf{Jean-Fra\c{c}ois Molinari:} Conceptualization, Funding acquisition, Writing - Review \& Editing

\appendix
\section{Calculation of $k\left(\mathcal{F}^*; z_0; z_1, x\right)$}
\label{app:Compute_CFWF}
This appendix aims at establishing Eq.~\eqref{eq:PerturbedWeightFunction} from Eqs.~\eqref{eq:TranslationRotation}, \eqref{eq:PerturbedWeightFunction_Raw}, and \eqref{eq:PartialFractionDecomposition}. In particular, we want to find a more suitable expression for:
\begin{equation}
\label{eq:CFWF_Raw}
I\left(z_0,z_1\right) = \dfrac{1}{2\pi} \mathrm{PV} \int_{-\infty}^{+\infty} \dfrac{\delta a(z)-\delta a^{**}(z)}{\left((z-z_1)^2 + x^2\right) (z-z_0)^2} dz
\end{equation}
As noted by \cite{leblond_second_2012}, each of the integrals that comprise $I\left(z_0,z_1\right)$ individually diverges at infinity, so that it is necessary to integrate first between finite bounds $-L_{-}$ and $L_{+}$ before taking the limit $L_{-}, L_{+} \rightarrow +\infty$ in the final combination of integrals. 

\noindent \textbf{Terms in $1/(z-z_0)^2$} -- They read with a pre-factor $1/((z_0-z_1)^2 + x^2)$:
\begin{equation}
\label{eq:CFWF_Bounded_1}
\scalemath{1}{
\mathrm{PV} \int_{-L_{-}}^{L_{+}} \dfrac{\delta a(z)-\delta a^{**}(z)}{(z-z_0)^2} dz = \mathrm{PV} \int_{-L_{-}}^{L_{+}} \dfrac{\delta a(z)-\delta a(z_0)}{(z-z_0)^2} dz - \mathrm{ln}\left[\dfrac{L_{+}-z_0}{L_{-}+z_0}\right] \dfrac{\delta a(z_0)-\delta a(z_1)}{z_0-z_1}
}
\end{equation}
\textbf{Terms in $1/(z-z_0)$} -- They read with a pre-factor $-2(z_0-z_1)/((z_0-z_1)^2 + x^2)^2$:
\begin{equation}
\label{eq:CFWF_Bounded_2}
\scalemath{0.9}{
\int_{-L_{-}}^{L_{+}} \dfrac{\delta a(z)-\delta a^{**}(z)}{(z-z_0)} dz = \mathrm{PV} \int_{-L_{-}}^{L_{+}} \dfrac{\delta a(z)}{(z-z_0)} dz - \mathrm{ln}\left[\dfrac{L_{+}-z_0}{L_{-}+z_0}\right] \delta a(z_0) - \left(L_{+}+L_{-}\right) \dfrac{\delta a(z_0)-\delta a(z_1)}{z_0-z_1}
}
\end{equation}
\textbf{Terms in $1/((z-z_1)^2 + x^2)^2$} -- They read with a pre-factor $((z_0-z_1)^2 - x^2)/((z_0-z_1)^2 + x^2)^2$:
\begin{equation}
\label{eq:CFWF_Bounded_3}
\scalemath{0.9}{
%Prefactor \dfrac{\left((z_0-z_1)^2 - x^2\right)}{\left((z_0-z_1)^2 + x^2\right)^2} 
\mathrm{PV} \int_{-L_{-}}^{L_{+}} \dfrac{\delta a(z)-\delta a^{**}(z)}{\left((z-z_1)^2 + x^2\right)} = \mathrm{PV} \int_{-L_{-}}^{L_{+}} \dfrac{\delta a(z)-\delta a(z_1)}{\left((z-z_1)^2 + x^2\right)} dz - \mathrm{ln}\left[\sqrt{\dfrac{(L_{+}-z_1)^2+x^2}{(L_{-}+z_1)^2+x^2}}\right] \dfrac{\delta a(z_0)-\delta a(z_1)}{z_0-z_1}
}
\end{equation} 
\textbf{Terms in $(z-z_1)/((z-z_1)^2 + x^2)^2$} -- They read 
with a pre-factor $2(z_0-z_1)/((z_0-z_1)^2 + x^2)^2$:
\begin{equation}
\label{eq:CFWF_Bounded_4}
\scalemath{0.8}{
\begin{aligned}
%Prefactor \dfrac{2(z_0-z_1)}{\left((z_0-z_1)^2 + x^2\right)^2} 
\int_{-L_{-}}^{L_{+}} \dfrac{(z-z_1)}{\left((z-z_1)^2 + x^2\right)}\left[\delta a(z)-\delta a^{**}(z)\right] dz = \mathrm{PV} \int_{-L_{-}}^{L_{+}} \dfrac{(z-z_1)}{\left((z-z_1)^2 + x^2\right)} \delta a(z) dz - \mathrm{ln}\left[\sqrt{\dfrac{(L_{+}-z_1)^2+x^2}{(L_{-}+z_1)^2+x^2}}\right] \delta a(z_1) \\
- \left(L_{+}+L_{-}\right) \dfrac{\delta a(z_0)-\delta a(z_1)}{z_0-z_1} + x\left[\mathrm{arctan}\left(\dfrac{L_{+}-z_1}{x}\right)+\mathrm{arctan}\left(\dfrac{L_{-}+z_1}{x}\right)\right] \dfrac{\delta a(z_0)-\delta a(z_1)}{z_0-z_1}
\end{aligned}
}
\end{equation}
We observe that the terms proportional to $(L_{+}+L_{-})$ of Eqs~\eqref{eq:CFWF_Bounded_2} and \eqref{eq:CFWF_Bounded_4} cancel out, we get:
\begin{equation}
\label{eq:CFWF_Bounded_Regrouped}
\scalemath{0.7}{
\begin{aligned}
\mathrm{PV} \int_{-L_{-}}^{L_{+}} \dfrac{\delta a(z)-\delta a^{**}(z)}{\left((z-z_1)^2 + x^2\right) (z-z_0)^2} dz = \left[\frac{1}{(z_0-z_1)^2 + x^2} \mathrm{PV} \int_{-L_{-}}^{L_{+}} \dfrac{\delta a(z) - \delta a(z_0)}{(z-z_0)^2} dz - \frac{2(z_0-z_1)}{\left((z_0-z_1)^2 + x^2\right)^2} \mathrm{PV} \int_{-L_{-}}^{L_{+}} \dfrac{\delta a(z)}{z-z_0} dz \right.\\
+ \frac{(z_0-z_1)^2 - x^2}{\left((z_0-z_1)^2 + x^2\right)^2} \mathrm{PV} \int_{-L_{-}}^{L_{+}} \dfrac{\delta a(z) - \delta a(z_1)}{(z-z_1)^2 + x^2} dz + \frac{2(z_0-z_1)}{\left((z_0-z_1)^2 + x^2\right)^2} \mathrm{PV} \int_{-L_{-}}^{L_{+}} \frac{(z-z_1)}{(z-z_1)^2 + x^2} \delta a(z) dz \\
+ \frac{(z_0-z_1)^2 - x^2}{\left((z_0-z_1)^2 + x^2\right)^2} \left[\mathrm{ln}\left(\dfrac{L_{+}-z_0}{L_{-}+z_0}\right) - \mathrm{ln}\left(\sqrt{\dfrac{(L_{+}-z_1)^2+x^2}{(L_{-}+z_1)^2+x^2}}\right) \right] \dfrac{\delta a \left(z_0\right)}{(z_0-z_1)} \\
+ \frac{1}{(z_0-z_1)^2 + x^2} \left[\mathrm{ln}\left(\sqrt{\dfrac{(L_{+}-z_1)^2+x^2}{(L_{-}+z_1)^2+x^2}}\right) - \mathrm{ln}\left(\dfrac{L_{+}-z_0}{L_{-}+z_0}\right) \right] \dfrac{\delta a \left(z_1\right)}{(z_0-z_1)} \\
\left. + \frac{2 x}{\left((z_0-z_1)^2 + x^2\right)^2} \left[\mathrm{arctan}\left(\dfrac{L_{+}-z_1}{x}\right)+\mathrm{arctan}\left(\dfrac{L_{-}+z_1}{x}\right)\right] \left(\delta a(z_0) - \delta a(z_1)\right)\right] 
\end{aligned}
}
\end{equation}
We then notice that the logarithmic terms go to zero when $L_{-}, L_{+} \rightarrow +\infty$. In this limit, one finds:
\begin{equation}
\label{eq:CFWF_Limit_Regrouped}
\scalemath{0.88}{
\begin{aligned}
\mathrm{PV} \int_{-\infty}^{+\infty} & \dfrac{\delta a(z)-\delta a^{**}(z)}{\left((z-z_1)^2 + x^2\right) (z-z_0)^2} dz = \left[ \frac{1}{(z_0-z_1)^2 + x^2} \mathrm{PV} \int_{-\infty}^{+\infty} \dfrac{\delta a(z) - \delta a(z_0)}{(z-z_0)^2} dz \right. \\
& - \frac{2(z_0-z_1)}{\left((z_0-z_1)^2 + x^2\right)^2} \mathrm{PV} \int_{-\infty}^{+\infty} \dfrac{\delta a(z)}{z-z_0} dz + \frac{(z_0-z_1)^2 - x^2}{\left((z_0-z_1)^2 + x^2\right)^2} \mathrm{PV} \int_{-\infty}^{+\infty} \dfrac{\delta a(z) - \delta a(z_1)}{(z-z_1)^2 + x^2} dz \\
& + \frac{2(z_0-z_1)}{\left((z_0-z_1)^2 + x^2\right)^2} \mathrm{PV} \int_{-\infty}^{+\infty} \frac{(z-z_1)}{(z-z_1)^2 + x^2} \delta a(z) dz + \left. \frac{2\pi x}{\left((z_0-z_1)^2 + x^2\right)^2} \left(\delta a(z_0) - \delta a(z_1)\right) \right]
\end{aligned}
}
\end{equation}
One finally gets Eq.~\eqref{eq:PerturbedWeightFunction} from Eqs.~\eqref{eq:PerturbedWeightFunction_Raw} and \eqref{eq:CFWF_Limit_Regrouped}.

\section{Calculation of $\delta K_\mathrm{czm}(z)$ and its Fourier transform}
\label{app:Compute_SIF}

In this appendix, we derive Eqs.~\eqref{eq:Cohesive_Toughness} and \eqref{eq:Cohesive_SIF_Variations} from Eqs.~\eqref{eq:PerturbedWeightFunction} and \eqref{eq:Cohesive_SIF_Raw}. To do so, we first express $k\left(\Gamma^*; z_0; z_1, x\right)$ as:
\begin{equation}
\label{eq:Derivation_SIF_01}
k\left(\Gamma^*; z_0; z_1, x\right) = k_0\left(z_0, z_1, x\right) \left[ 1 + \delta\mathcal{K}\left(z_0; z_1, x\right)\right]
\end{equation}
where $k_0$ corresponds to the CFWF associated to the reference straight front $\Gamma$:
\begin{equation}
\label{eq:Derivation_SIF_02}
k_0\left(z_0, z_1, x\right) = \dfrac{\sqrt{2}}{\pi^{3/2}} \dfrac{\sqrt{x}}{(z_0-z_1)^2 + x^2} 
\end{equation}
and $\delta\mathcal{K}$ can be decomposed into five different first-order terms following Eq.~\eqref{eq:PerturbedWeightFunction}:
\begin{equation}
\label{eq:Derivation_SIF_03}
\scalemath{0.8}{
\begin{aligned}
\delta\mathcal{K}\left(z_0, z_1, x\right) = & \,\delta\mathcal{K}_1\left(z_0, z_1, x\right) + \delta\mathcal{K}_2\left(z_0, z_1, x\right) + \delta\mathcal{K}_3\left(z_0, z_1, x\right) + \delta\mathcal{K}_4\left(z_0, z_1, x\right) + \delta\mathcal{K}_5\left(z_0, z_1, x\right) \\
= & \frac{1}{2\pi} \mathrm{PV} \int_{z} \dfrac{\delta a(z) - \delta a(z_0)}{(z-z_0)^2} dz - \frac{2(z_0-z_1)}{(z_0-z_1)^2 + x^2} \,\dfrac{1}{2\pi} \mathrm{PV} \int_{z} \dfrac{\delta a(z)}{z-z_0} dz \\
& + \frac{2(z_0-z_1)}{(z_0-z_1)^2 + x^2} \,\dfrac{1}{2\pi} \mathrm{PV} \int_{z} \dfrac{(z-z_1)}{(z-z_1)^2 + x^2} \,\delta a(z) dz + \frac{(z_0-z_1)^2 - x^2}{(z_0-z_1)^2 + x^2} \,\dfrac{1}{2\pi} \mathrm{PV} \int_{z} \dfrac{\delta a(z) - \delta a(z_1)}{(z-z_1)^2 + x^2} dz \\
& + \frac{x}{(z_0-z_1)^2 + x^2} \left(\delta a(z_0) - \delta a(z_1)\right) 
\end{aligned}
}
\end{equation}

\noindent \textbf{Decomposition of the cohesive SIF} -- Eq.~\eqref{eq:Cohesive_SIF_Raw} rewrites as:
\begin{equation}
\label{eq:Derivation_SIF_04}
K_\mathrm{czm}(z_0) = \int_{v} \int_{z_1} \sigmac(z_1)\,\fw\left(u\right)\,\omega(z_1) k_0\left(z_0, z_1, \omega(z_1)u\right) \left[ 1 + \delta\mathcal{K}\left(z_0, z_1, \omega(z_1)u\right)\right] dz_1 du
\end{equation}
where $\omega(z_1) k_0\left(z_0, z_1, \omega(z_1)u\right)$ can be rewritten as the sum of a zero-order term and a first-order term in $\delta \omega$:
\begin{equation}
\label{eq:Derivation_SIF_05}
\scalemath{1}{
\begin{aligned}
\omega(z_1) k_0\left(z_0, z_1, \omega(z_1)u\right) = & \dfrac{\sqrt{2}}{\pi^{3/2}}\dfrac{\omega(z_1)^{3/2} u^{1/2}}{(z_0-z_1)^2+(\omega(z_1)u)^2} \\
= & \,\omega_0\, k_0\left(z_0, z_1, \omega_0 u\right) \left[ 1 + \left(\dfrac{3}{2} - \dfrac{2\omega_0^2 u^2}{(z_0-z_1)^2+(\omega_0u)^2}\right)\dfrac{\delta\omega(z_1)}{\omega_0}\right] \end{aligned}
}
\end{equation}
and $\delta\mathcal{K}\left(z_0; z_1, \omega(z_1)u\right)=\delta\mathcal{K}\left(z_0; z_1, \omega_0u\right)$ since it is already composed of first-order terms in $\delta a$.\\

\noindent \textbf{Zero-order terms} -- In the end, one finds back from Eq.~\eqref{eq:Derivation_SIF_04} at order 0:
\begin{equation}
\label{eq:Derivation_SIF_06}
\scalemath{1}{
\begin{aligned}
K_\mathrm{czm}^0 = & \int_{u} \int_{z_1} \sigmac^0\,\fw\left(u\right)\,\omega_0 k_0\left(z_0, z_1, \omega_0 u\right) dz_1 du \\
= & \left[\int_{u} \dfrac{\fw\left(u\right)}{u^{1/2}} du\right] \sqrt{\dfrac{2}{\pi}} \sigmac^0 \omega_0^{1/2} = \Cw \sqrt{\dfrac{2}{\pi}} \sigmac^0 \omega_0^{1/2} = \KIc^0
\end{aligned}
}
\end{equation}
which corresponds to Eq.~\eqref{eq:Cohesive_Toughness}. We find back the results of \cite{palmer_growth_1973} derived in the 2D case. This was expected since the zero-order terms relate to the reference semi-infinite crack $\Gamma$ with a straight front $\mathcal{F}$ embedded in a homogeneous medium, which is translationally invariant in the $\left(Oz\right)$ direction.\\

\noindent \textbf{First-order terms} -- At order 1 in $\delta \sigmac$, $\delta \omega$ and $\delta a$, one gets:
\begin{equation}
\label{eq:Derivation_SIF_07}
\scalemath{0.9}{
\begin{aligned}
\delta K_\mathrm{czm}(z_0) = & \int_{u} \int_{z_1}
\dfrac{\sqrt{2}}{\pi^{3/2}}\dfrac{\sigmac^0\omega_0^{3/2} u^{1/2}}{(z_0-z_1)^2+(\omega_0 u)^2} \fw(u) \dfrac{\delta \sigmac(z_1)}{\sigmac^0} \,dz_1 du \\
+ & \int_{u} \int_{z_1}
\dfrac{\sqrt{2}}{\pi^{3/2}}\dfrac{\sigmac^0\omega_0^{3/2} u^{1/2}}{(z_0-z_1)^2+(\omega_0 u)^2} \fw(u) \left[\dfrac{3}{2} - \dfrac{2\omega_0^2 u^2}{(z_0-z_1)^2+(\omega_0 u)^2}\right]\dfrac{\delta\omega(z_1)}{\omega_0} \,dz_1 du \\
+ & \int_{u} \int_{z_1}
\dfrac{\sqrt{2}}{\pi^{3/2}}\dfrac{\sigmac^0\omega_0^{3/2} u^{1/2}}{(z_0-z_1)^2+(\omega_0 u)^2} \fw(u) \left[\sum_i \delta\mathcal{K}_i\left(z_0, z_1, \omega_0 u\right)\right] \,dz_1 du
\end{aligned}
}
\end{equation}
The next step is to calculate each of the 7 integrals of Eq.~\eqref{eq:Derivation_SIF_07}. The calculations build upon the Fourier representation $\widehat{\delta \sigmac}$, $\widehat{\delta \omega}$, and $\widehat{\delta a}$ of $\delta \sigmac$, $\delta \omega$, and $\delta a$, and the expression of the following integrals (see Eqs.~(3.723.2), (3.723.3), (3.729.1), (3.729.2) and (3.729.3) of \cite{gradshteyn_integrals_2014}):
\begin{equation}
\label{eq:UsefulIntegrals}
\scalemath{0.9}{
\begin{aligned}
& \int_{-\infty}^{+\infty} \dfrac{e^{ikv}}{v} dv = i\pi \mathrm{sgn}(k) & \text{ and } & \int_{-\infty}^{+\infty} \dfrac{e^{ikv}}{v^2+(\omega_0u)^2} dv = \dfrac{\pi}{\omega_0u} e^{-\kw_0 u} \\
& \int_{-\infty}^{+\infty} \dfrac{v e^{ikv}}{v^2+(\omega_0u)^2} dv = i\pi e^{-\kw_0 u} & \text{ and } & \int_{-\infty}^{+\infty} \dfrac{e^{ikv}}{\left[v^2+(\omega_0u)^2\right]^2} dv = \dfrac{\pi}{2}\dfrac{1+\kw_0 u}{(\omega_0u)^3} e^{-\kw_0 u} \\
& \int_{-\infty}^{+\infty} \dfrac{v e^{ikv}}{\left[v^2+(\omega_0u)^2\right]^2} dv = i\dfrac{\pi}{2}\dfrac{\vert k \vert}{\omega_0u} e^{-\kw_0 u} & \text{ and } & \int_{-\infty}^{+\infty} \dfrac{v^2-(\omega_0u)^2}{\left[v^2+(\omega_0u)^2\right]^2} e^{ikv} dv = -\pi \vert k \vert e^{-\kw_0 u}
\end{aligned}
}
\end{equation}
where $\mathrm{sgn}(k)$ denotes the sign of the real number $k$.\\

\noindent \textit{Terms in $\delta \sigmac$\,} -- We denote $\delta K_\mathrm{czm}^{\sigma}$ the contributions of $\delta \sigmac$ to the variations $\delta K_\mathrm{czm}$ of the cohesive SIF. From Eq.~\eqref{eq:Derivation_SIF_07}, it reads:
\begin{equation}
\label{eq:Derivation_SIF_08}
\scalemath{0.95}{
\begin{aligned}
\delta K_\mathrm{czm}^\sigma(z_0) = & \int_{u} \int_{z_1}
\dfrac{\sqrt{2}}{\pi^{3/2}}\dfrac{\sigmac^0\omega_0^{3/2} u^{1/2}}{(z_0-z_1)^2+(\omega_0 u)^2} \fw(u) \dfrac{\delta \sigmac(z_1)}{\sigmac^0} \,dz_1 du \\
= & \int_{u} \left[du \dfrac{\sqrt{2}}{\pi^{3/2}} \sigmac^0\omega_0^{3/2} \,u^{1/2}\fw(u) \,\frac{1}{2\pi}\int_{k} dk \dfrac{\widehat{\delta \sigmac}(k)}{\sigmac^0} e^{ikz_0} \left(\int_{z_1} \dfrac{e^{ik(z_1-z_0)}}{(z_1-z_0)^2 + (\omega_0 u)^2} dz_1\right)\right] \\
= & \sqrt{\dfrac{2}{\pi}} \sigmac^0\omega_0^{1/2} \frac{1}{2\pi}\int_{k} \left[ \int_{u} \dfrac{\fw\left(u\right)}{u^{1/2}} e^{-\kw_0 u} du \right] \dfrac{\widehat{\delta \sigmac}(k)}{\sigmac^0} e^{ikz_0} dk \\
= & \,\KIc^{0} \cdot \frac{1}{2\pi}\int_{k} \left[ \dfrac{1}{\Cw} \int_{u} \dfrac{\fw\left(u\right)}{u^{1/2}} e^{-\kw_0 u} du \right] \dfrac{\widehat{\delta \sigmac}(k)}{\sigmac^0} e^{ikz_0} dk
\end{aligned}
}
\end{equation}

\noindent \textit{Terms in $\delta \omega$\,} -- We denote $\delta K_\mathrm{czm}^{\omega}$ the contributions of $\delta \omega$ to the variations $\delta K_\mathrm{czm}$ of the cohesive SIF. From Eq.~\eqref{eq:Derivation_SIF_07}, it reads:
\begin{equation}
\label{eq:Derivation_SIF_09}
\scalemath{0.95}{
\begin{aligned}
\delta K_\mathrm{czm}^\omega (z_0) = & \int_{u} \int_{z_1}
\dfrac{\sqrt{2}}{\pi^{3/2}}\dfrac{\sigmac^0\omega_0^{3/2} u^{1/2}}{(z_0-z_1)^2+(\omega_0 u)^2} \fw(u) \left[\dfrac{3}{2} - \dfrac{2\omega_0^2 u^2}{(z_0-z_1)^2+(\omega_0 u)^2}\right]\dfrac{\delta\omega(z_1)}{\omega_0} \,dz_1 du \\
= & \int_{u} \left[du \dfrac{\sqrt{2}}{\pi^{3/2}} \sigmac^0\omega_0^{3/2} \,u^{1/2}\fw(u) \,\frac{1}{2\pi}\int_{k} dk \dfrac{3}{2}\dfrac{\widehat{\delta \omega}(k)}{\omega_0} e^{ikz_0} \left(\int_{z_1} \dfrac{e^{ik(z_1-z_0)}}{(z_1-z_0)^2 + (\omega_0 u)^2} dz_1\right)\right] \\
- & \int_{u} \left[du \dfrac{\sqrt{2}}{\pi^{3/2}} \sigmac^0\omega_0^{3/2} \,u^{1/2}\fw(u) \,\frac{1}{2\pi}\int_{k} dk \dfrac{\widehat{\delta \omega}(k)}{\omega_0} e^{ikz_0} \left(\int_{z_1} \dfrac{2(\omega_0 u)^2 e^{ik(z_1-z_0)}}{\left[(z_1-z_0)^2 + (\omega_0 u)^2\right]^2} dz_1\right)\right] \\
= & \,\KIc^{0} \cdot \frac{1}{2\pi}\int_{k} \left[ \dfrac{1}{\Cw} \int_{u} \dfrac{\fw\left(u\right)}{u^{1/2}} \left(\frac{1}{2}-\kw_0 u\right) e^{-\kw_0 u} du \right] \dfrac{\widehat{\delta \omega}(k)}{\omega_0} e^{ikz_0} dk \\
\underset{\mathrm{I.B.P/} u}{=} & \,\KIc^{0} \cdot \frac{1}{2\pi}\int_{k} \left[ \dfrac{1}{\Cw} \int_{u} -\dfw\left(u\right) u^{1/2} e^{-\kw_0 u} du \right] \dfrac{\widehat{\delta \omega}(k)}{\omega_0} e^{ikz_0} dk
\end{aligned}
}
\end{equation}
where ``$\mathrm{I.B.P/} u$" denotes the integration by parts with respect to the variable $u$.\\

\noindent \textit{Terms in $\delta a$\,} -- We denote $\delta K_\mathrm{czm}^{a,i}$ the contributions of $\delta \mathcal{K}_i$ to the variations $\delta K_\mathrm{czm}$ of the cohesive SIF. From Eqs.~\eqref{eq:Derivation_SIF_03} and \eqref{eq:Derivation_SIF_07}, $\delta K_\mathrm{czm}^{a,1}$ reads:
\begin{equation}
\label{eq:Derivation_SIF_10}
\scalemath{0.82}{
\begin{aligned}
\delta K_\mathrm{czm}^{a,1} (z_0) = & \int_{u} \int_{z_1}
\dfrac{\sqrt{2}}{\pi^{3/2}}\dfrac{\sigmac^0\omega_0^{3/2} u^{1/2}}{(z_0-z_1)^2+(\omega_0 u)^2} \fw(u) \left[\dfrac{1}{2\pi} \mathrm{PV} \int_{z} \dfrac{\delta a(z) - \delta a(z_0)}{(z-z_0)^2} dz\right] \,dz_1 du \\
= & \int_{u} \int_{z_1}
\dfrac{\sqrt{2}}{\pi^{3/2}}\dfrac{\sigmac^0\omega_0^{3/2} u^{1/2}}{(z_0-z_1)^2+(\omega_0 u)^2} \fw(u) \left[\dfrac{1}{2\pi} \mathrm{PV} \int_{z} \dfrac{\delta a'(z)}{z-z_0} dz\right] \,dz_1 du \\
= & \int_{u} \left[du \dfrac{\sqrt{2}}{\pi^{3/2}} \sigmac^0\omega_0^{3/2} \,u^{1/2}\fw(u) \,\frac{1}{(2\pi)^2} \int_{k} dk \,ik\,\widehat{\delta a}(k) e^{ikz_0} \left(\int_{z_1} \dfrac{dz_1}{(z_0-z_1)^2+(\omega_0 u)^2}\right) \left(\int_{z} \dfrac{e^{ik(z-z_0)}}{z-z_0} dz\right)\right] \\
= & \,\KIc^{0} \cdot \frac{1}{2\pi}\int_{k} -\dfrac{\vert k \vert}{2} \widehat{\delta a}(k) e^{ikz_0} dk
\end{aligned}
}
\end{equation}
One may then show that:
\begin{equation}
\label{eq:Derivation_SIF_11}
\delta K_\mathrm{czm}^{a,2}(z)=0
\end{equation}
because the pre-factor in front of the integral over $z$ of $\delta \mathcal{K}_2$ is an even function of $(z_0-z_1)$, so that the integral over $z_1$ in $\delta K_\mathrm{czm}^{a,2}$ equates to zero. It is not the case for $\delta K_\mathrm{czm}^{a,3}$, which reads:
\begin{equation}
\label{eq:Derivation_SIF_12}
\scalemath{0.7}{
\begin{aligned}
\delta K_\mathrm{czm}^{a,3} (z_0) = & \int_{u} \int_{z_1} \dfrac{\sqrt{2}}{\pi^{3/2}}\sigmac^0\omega_0^{3/2} u^{1/2} \fw(u) \frac{2(z_0-z_1)}{\left[(z_0-z_1)^2+(\omega_0 u)^2\right]^2} \left[ \dfrac{1}{2\pi} \mathrm{PV} \int_{z} \frac{(z-z_1)}{(z-z_1)^2 + (\omega_0 u)^2} \,\delta a(z) dz \right] \,dz_1 du \\
= & \int_{u} \int_{z_1} \dfrac{\sqrt{2}}{\pi^{3/2}}\sigmac^0\omega_0^{3/2} u^{1/2} \fw(u) \frac{2(z_0-z_1)}{\left[(z_0-z_1)^2+(\omega_0 u)^2\right]^2} \dfrac{1}{(2\pi)^2} \int_{k} dk \,\widehat{\delta a}(k) e^{ikz_1} \left[\int_{z} \frac{(z-z_1)}{(z-z_1)^2 + (\omega_0 u)^2} e^{ik(z-z_1)} dz \right] \,dz_1 du \\
= & \int_{u} \dfrac{\sqrt{2}}{\pi^{3/2}}\sigmac^0\omega_0^{3/2} u^{1/2} \fw(u) \dfrac{1}{2\pi} \int_{k} dk \,e^{-\kw_0 u} \widehat{\delta a}(k) e^{ikz_0} i \left[ \int_{z_1} \frac{(z_0-z_1)}{\left[(z_0-z_1)^2 + (\omega_0 u)^2\right]^2} e^{ik(z_1-z_0)} dz_1 \right] du \\
= & \,\KIc^{0} \cdot \frac{1}{2\pi}\int_{k} \left[ \dfrac{1}{\Cw} \int_{u} \dfrac{\fw\left(u\right)}{u^{1/2}} e^{-2\kw_0 u} du \right] \dfrac{\vert k \vert}{2}\widehat{\delta a}(k) e^{ikz_0} dk
\end{aligned}
}
\end{equation}
The process for $\delta K_\mathrm{czm}^{a,4}$ is similar. One finds:
\begin{equation}
\label{eq:Derivation_SIF_13}
\scalemath{0.75}{
\begin{aligned}
\delta K_\mathrm{czm}^{a,4} (z_0) = & \int_{u} \int_{z_1} \dfrac{\sqrt{2}}{\pi^{3/2}}\sigmac^0\omega_0^{3/2} u^{1/2} \fw(u) \frac{(z_0-z_1)^2-(\omega_0 u)^2}{\left[(z_0-z_1)^2+(\omega_0 u)^2\right]^2} \left[ \dfrac{1}{2\pi} \mathrm{PV} \int_{z} \frac{\delta a(z)-\delta a(z_1)}{(z-z_1)^2 + (\omega_0 u)^2} \,dz \right] \,dz_1 du \\
= & \int_{u} \int_{z_1} \dfrac{\sqrt{2}}{\pi^{3/2}}\sigmac^0\omega_0^{3/2} u^{1/2} \fw(u) \frac{(z_0-z_1)^2-(\omega_0 u)^2}{\left[(z_0-z_1)^2+(\omega_0 u)^2\right]^2} \dfrac{1}{(2\pi)^2} \int_{k} dk \,\widehat{\delta a}(k) e^{ikz_1} \left[\int_{z} \frac{e^{ik(z-z_1)}-1}{(z-z_1)^2 + (\omega_0 u)^2} dz \right] \,dz_1 du \\
= & \int_{u} \dfrac{\sqrt{2}}{\pi^{3/2}}\sigmac^0\omega_0^{1/2} \dfrac{\fw(u)}{u^{1/2}} \dfrac{1}{2\pi} \int_{k} dk \,(e^{-\kw_0 u}-1) \frac{1}{2}\widehat{\delta a}(k) e^{ikz_0} \left[ \int_{z_1} \frac{(z_0-z_1)^2-(\omega_0 u)^2}{\left[(z_0-z_1)^2 + (\omega_0 u)^2\right]^2} e^{ik(z_1-z_0)} dz_1 \right] du \\
= & \,\KIc^{0} \cdot \frac{1}{2\pi}\int_{k} \left[ \dfrac{1}{\Cw} \int_{u} \dfrac{\fw\left(u\right)}{u^{1/2}} \left(1-e^{-\kw_0 u}\right) e^{-\kw_0 u} du \right] \dfrac{\vert k \vert}{2}\widehat{\delta a}(k) e^{ikz_0} dk
\end{aligned}
}
\end{equation}
The remaining integral $\delta K_\mathrm{czm}^{a,5}$ yields:
\begin{equation}
\label{eq:Derivation_SIF_14}
\scalemath{0.9}{
\begin{aligned}
\delta K_\mathrm{czm}^{a,5} (z_0) = & \int_{u} \int_{z_1} \dfrac{\sqrt{2}}{\pi^{3/2}}\sigmac^0\omega_0^{3/2} u^{1/2} \fw(u) \frac{\omega_0 u}{\left[(z_0-z_1)^2+(\omega_0 u)^2\right]^2} \left[\delta a(z_0) - \delta a(z_1) \right] \,dz_1 du \\
= & \int_{u} \dfrac{\sqrt{2}}{\pi^{3/2}}\sigmac^0\omega_0^{1/2} \dfrac{\fw(u)}{u^{1/2}} \dfrac{1}{2\pi} \int_{k} dk \,\widehat{\delta a}(k) e^{ikz_0} \left[ \int_{z_1} \frac{\omega_0 u \left(1 - e^{ik(z_1-z_0)}\right)}{\left[(z_0-z_1)^2 + (\omega_0 u)^2\right]^2} dz_1 \right] du \\
= & \,\KIc^{0} \cdot \frac{1}{2\pi}\int_{k} \left[ \frac{1}{\Cw} \int_{u} \dfrac{\fw\left(u\right)}{2u^{3/2}} \left[\left(1+\kw_0 u\right)e^{-\kw_0 u}-1\right] du \right] \dfrac{\widehat{\delta a}(k)}{\omega_0} e^{ikz_0} dk \\
\underset{\mathrm{I.B.P/} u}{=} & \, -\KIc^{0} \cdot \frac{1}{2\pi}\int_{k} \left[\dfrac{1}{\Cw} \int_{u} \dfrac{\fw\left(u\right)}{u^{1/2}} e^{-\kw_0 u} du \right] \dfrac{\vert k \vert}{2} \widehat{\delta a}(k) e^{ikz_0} dk \\
& + \,\KIc^{0} \cdot \frac{1}{2\pi}\int_{k} \left[\dfrac{1}{\Cw} \int_{u} -\dfrac{\dfw\left(u\right)}{u^{1/2}} \left(1-e^{-\kw_0 u}\right) du \right] \dfrac{\widehat{\delta a}(k)}{\omega_0} e^{ikz_0} dk
\end{aligned}
}
\end{equation}
Regrouping Eqs.~\eqref{eq:Derivation_SIF_10}-\eqref{eq:Derivation_SIF_14}, one finds :
\begin{equation}
\label{eq:Derivation_SIF_15}
\scalemath{0.9}{
\begin{aligned}
\dfrac{\delta K_\mathrm{czm}^{a} (z_0)}{\KIc^{0} } = \, \frac{1}{2\pi}\int_{k} -\dfrac{\vert k \vert}{2} \widehat{\delta a}(k) e^{ikz_0} dk + \,\frac{1}{2\pi}\int_{k} \left[\dfrac{1}{\Cw} \int_{u} -\dfrac{\dfw\left(u\right)}{u^{1/2}} \left(1-e^{-\kw_0 u}\right) du \right] \dfrac{\widehat{\delta a}(k)}{\omega_0} e^{ikz_0} dk
\end{aligned}
}
\end{equation}
One finally gets Eq.~\eqref{eq:Cohesive_SIF_Variations} collecting the terms in $\delta \sigma_\mathrm{c}$, $\delta \omega$, and $\delta a$ from Eqs.\eqref{eq:Derivation_SIF_08}, \eqref{eq:Derivation_SIF_09} and \eqref{eq:Derivation_SIF_15}.

\modif{Note that Eq.~\eqref{eq:Cohesive_SIF_Variations} can be expressed in terms of variations of strength $\delta \sigma_\mathrm{c}$ and critical crack opening $\delta \deltac$ with respect to their average value $\sigmac^0$ and $\deltac^0$. From Eq.~\eqref{eq:CohesiveStress_ProcessZoneSize}, one has:}
\begin{equation}
\modif{\omega_0 = \alpha \dfrac{\mu}{\sigmac^0} \deltac^0 \text{ and } \dfrac{\delta \omega}{\omega_0} = \dfrac{\delta \deltac}{\deltac^0} -  \dfrac{\delta \sigmac}{\sigmac^0}}
\end{equation}
\modif{So that Eq.~\eqref{eq:Cohesive_SIF_Variations} boils down to:}
\begin{equation}
\label{eq:PropagationCriterion_Dc}
\scalemath{0.7}{
\modif{\KI^0\left[ 1 + \left(\dfrac{1}{\KI^0} \dfrac{\partial \KI^0}{\partial a} -\dfrac{\vert k\vert}{2}\right) \widehat{\delta a}\left(k\right)\right] = \KIc^0\left[1+\left(\dfrac{\hat{\mathcal{A}}\left(\kw_0\right)}{\omega_0} - \dfrac{\vert k\vert}{2}\right) \widehat{\delta a}\left(k\right) + \left[\hat{\Sigma}\left(\kw_0\right) - \dfrac{\hat{\Omega}\left(\kw_0\right)}{2}\right] \dfrac{\widehat{\delta \sigmac}\left(k\right)}{\sigmac^0} + \dfrac{\hat{\Omega}\left(\kw_0\right)}{2} \dfrac{\widehat{\delta \deltac}\left(k\right)}{\deltac^0}\right]}
}
\end{equation}
\modif{Note that, as $\hat{\Omega}\left(\kw_0\right)$ decreases more strongly than $\hat{\Sigma}\left(\kw_0\right)$ when $\kw_0 \rightarrow +\infty$ (see \ref{app:Asymptotic_Weakenings}), heterogeneities of strength $\sigma$ with constant critical opening $\deltac$ have a similar effect to that with constant process zone size described in Section~\ref{subsec:FrontDeformations_Strength}. Conversely, heterogeneities of critical opening $\deltac$ with constant strength $\sigma$ have a similar effect to that of process zone size described in Section~\ref{subsec:FrontDeformations_ProcessZoneSize}.}

%When $\omega_0 > 0$: 
%\begin{align}
%\label{eq:Cohesive_SIF_Variations_bis}
%\dfrac{\delta K_\mathrm{czm}}{\KIc^0}(z) = & -\dfrac{1}{2\pi} \mathrm{PV} \int_{-\infty}^{+\infty} \dfrac{\delta a\left(z\right) - \delta a(z')}{(z-z')^2} dz' \nonumber\\
% & + \dfrac{1}{\Cw}\int_{0}^{+\infty} 2\dfw(v) v^{1/2} \left[\dfrac{1}{2\pi} \int_{-\infty}^{+\infty} \dfrac{\delta a\left(z\right) - \delta a(z')}{(z-z')^2+\omega_0^2 v^2} dz'\right] dv \\
% & + \dfrac{1}{\Cw}\int_{0}^{+\infty} \fw(v)v^{1/2} \left[\dfrac{1}{2\pi} \int_{-\infty}^{+\infty} \dfrac{2\omega_0}{(z-z')^2+\omega_0^2 v^2} \dfrac{\delta \sigmac\left(z'\right)}{\sigmac^0} dz'\right] dv \nonumber\\
% & - \dfrac{1}{\Cw}\int_{0}^{+\infty} \dfw(v)v^{3/2} \left[\dfrac{1}{2\pi} \int_{-\infty}^{+\infty} \dfrac{2\omega_0}{(z-z')^2+\omega_0^2 v^2} \dfrac{\delta \omega\left(z'\right)}{\omega_0} dz'\right] dv \nonumber
%\end{align}

 \section{Expression and asymptotic behavior of $\hat{\mathcal{A}}\left(\vert k\vert \omega_0\right)$, $\hat{\Sigma}\left(\vert k\vert \omega_0\right)$, $\hat{\Omega}\left(\vert k\vert \omega_0\right)$ for different types of weakening}
\label{app:Asymptotic_Weakenings}

In this work, we consider four different cohesive laws. The first one is the linear distance-weakening law of Fig.~\ref{fig:FrontStability_Cohesive} for which the cohesive stress decays linearly from $\sigmac^0$ to $0$ along a distance $\omega_0$. The second one is the cohesive law of \cite{dugdale_yielding_1960} and \cite{barenblatt_processzone_1962} for which the cohesive stress is constant to $\sigmac^0$ along a distance $\omega_0$, and $0$ elsewhere. The third one is an exponentially distance-weakening law for which the cohesive stress decays as $\sigmac^0 e^{-x/\omega_0}$ behind the crack tip, and never reaches $0$. The last one is a linear traction-separation law for which the stress decay linearly from $\sigmac^0$ to $0$ with the local opening $\delta$ up to a critical value $\deltac$.  In this Appendix, we derive the expressions and the asymptotic behaviors for the three cohesive pre-factors $\hat{\mathcal{A}}$, $\hat{\Sigma}$ and $\hat{\Omega}$ of Eq.~\eqref{eq:Fourier_Cohesive_Prefactors} that control the front deformations. The first three cohesive laws yield analytical expressions for $\hat{\mathcal{A}}$, $\hat{\Sigma}$ and $\hat{\Omega}$, while they are computed numerically for the traction-separation law.

\subsection{Linear-distance weakening}
\label{app:Asymptotic_LinearDistance}

\begin{figure}[!h]
\centering \includegraphics[width=\textwidth]{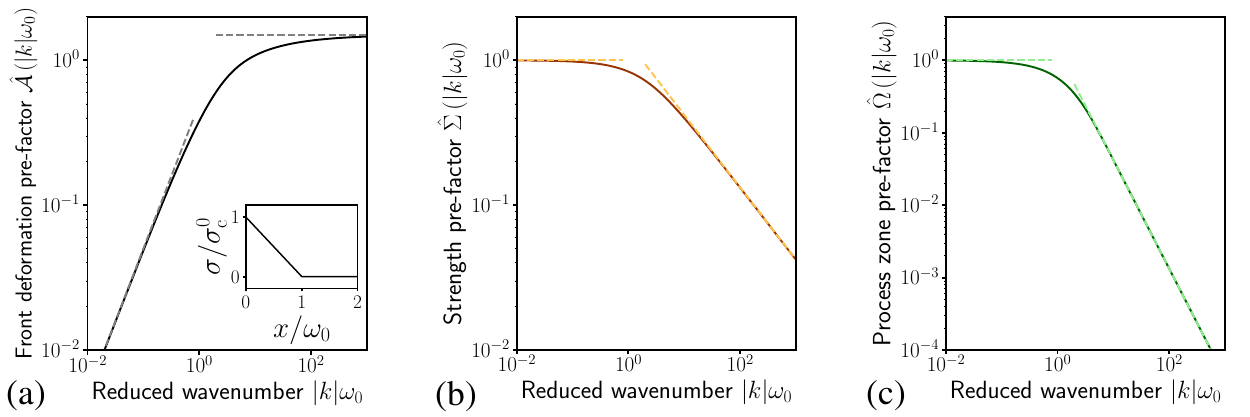}
\caption{Evolution of the cohesive pre-factors for a linear distance-weakening cohesive law: (a) the front deformations pre-factor $\hat{\mathcal{A}}(\kw_0)$ scales as $\propto \kw_0/2$ when $\kw_0 \rightarrow 0$ (dashed gray line on the left) and saturates to $\mathcal{A}_w^\infty$ when $\kw_0 \rightarrow +\infty$ (horizontal dashed gray line on the right). (b) The strength variation pre-factor $\hat{\Sigma}(\kw_0)$ goes to $1$ in the brittle limit $\kw_0 \rightarrow 0$ and decays as $(\kw_0)^{-1/2}$ when $\kw_0 \rightarrow +\infty$. (c) The process zone size variations pre-factor $\hat{\Omega}(\kw_0)$ goes to $1$ when $\kw_0 \rightarrow 0$ and decays as $(\kw_0)^{-3/2}$ when $\kw_0 \rightarrow +\infty$.}
\label{fig:AsymptoticBehavior_LinearDistance}
\end{figure}

The first one is the linear distance-weakening law of Fig.~\ref{fig:FrontStability_Cohesive} for which the cohesive stress decay linearly from $\sigmac^0$ to $0$ along a distance $\omega_0$:
\begin{equation}
\label{eq:Weakening_LinearDistance}
\fw(x) = \max\left(1-x, 0\right)
\end{equation}
In that case, Eq.~\eqref{eq:Fourier_Cohesive_Prefactors} yields:
\begin{equation}
\label{eq:Functionals_LinearDistance}
\scalemath{0.9}{
\begin{cases}
\hat{\mathcal{A}}\left(\kw_0\right) & = \dfrac{3}{2} - \dfrac{3\sqrt{\pi}}{4}\dfrac{\mathrm{erf}(\sqrt{\kw_0})}{\sqrt{\kw_0}} \\
\hat{\Sigma}\left(\kw_0\right) & = \dfrac{3\sqrt{\pi}}{4}\left(\kw_0-\frac{1}{2}\right)\dfrac{\mathrm{erf}(\sqrt{\kw_0})}{(\kw_0)^{3/2}} + \dfrac{3}{4}\dfrac{e^{-\kw_0}}{\kw_0} \\
\hat{\Omega}\left(\kw_0\right) & = \dfrac{3\sqrt{\pi}}{4}\dfrac{\mathrm{erf}(\sqrt{\kw_0})}{(\kw_0)^{3/2}} - \dfrac{3}{2}\dfrac{e^{-\kw_0}}{\kw_0}
\end{cases}
\Rightarrow
\begin{aligned}
\hat{\mathcal{A}}\left(\kw_0\right) & \underset{\kw_0 \rightarrow +\infty}{\longrightarrow} \dfrac{3}{2} \\
\hat{\Sigma}\left(\kw_0\right) & \underset{\kw_0 \rightarrow +\infty}{\sim} \dfrac{3\sqrt{\pi}}{4} (\kw_0)^{-1/2} \\
\hat{\Omega}\left(\kw_0\right) & \underset{\kw_0 \rightarrow +\infty}{\sim} \dfrac{3\sqrt{\pi}}{4} (\kw_0)^{-3/2}
\end{aligned}
}
\end{equation}
The evolution of $\hat{\mathcal{A}}$, $\hat{\Sigma}$ and $\hat{\Omega}$ with $\kw_0$ is given in Fig.~\ref{fig:AsymptoticBehavior_LinearDistance} for the linear distance-weakening law.

\subsection{Dugdale-Barenblatt distance weakening}
\label{app:Asymptotic_DugdaleBarenblatt}

\begin{figure}[!h]
\centering \includegraphics[width=\textwidth]{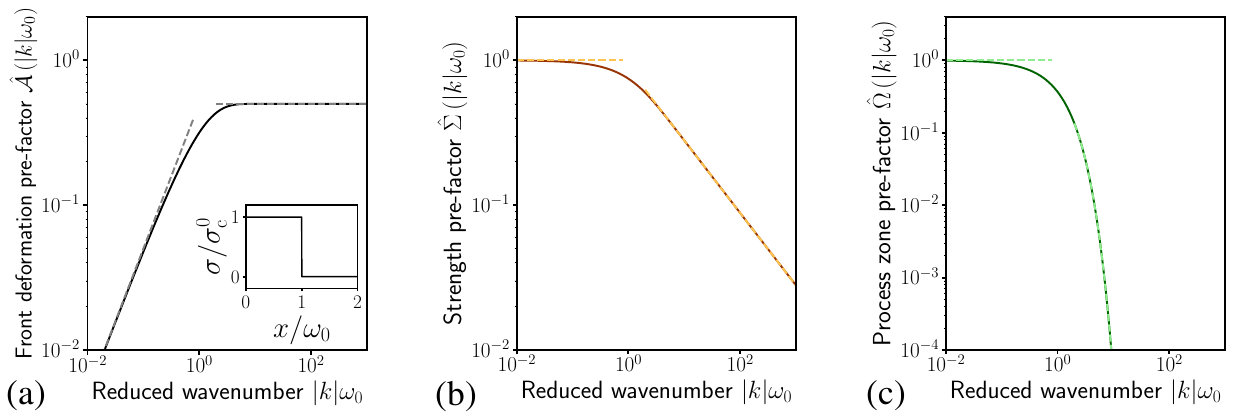}
\caption{Evolution of the cohesive pre-factors for a Dugdale-Barenblatt cohesive law: (a) the front deformations pre-factor $\hat{\mathcal{A}}(\kw_0)$ scales as $\propto \kw_0/2$ when $\kw_0 \rightarrow 0$ (dashed gray line on the left) and saturates to $\mathcal{A}_w^\infty$ when $\kw_0 \rightarrow +\infty$ (horizontal dashed gray line on the right). (b) The strength variation pre-factor $\hat{\Sigma}(\kw_0)$ goes to $1$ in the brittle limit $\kw_0 \rightarrow 0$ and decays as $(\kw_0)^{-1/2}$ when $\kw_0 \rightarrow +\infty$. (c) The process zone size variations pre-factor $\hat{\Omega}(\kw_0)$ goes to $1$ when $\kw_0 \rightarrow 0$ and decays as $e^{-\kw_0}$ when $\kw_0 \rightarrow +\infty$.}
\label{fig:AsymptoticBehavior_DugdaleBarenblatt}
\end{figure}

The second one is the cohesive law of \cite{dugdale_yielding_1960} and \cite{barenblatt_processzone_1962} for which the cohesive stress is constant to $\sigmac^0$ along a distance $\omega_0$, and $0$ elsewhere:
\begin{equation}
\label{eq:Weakening_DugdaleBarenblatt}
\fw(x) = \chi_{[0,1]}(x)
\end{equation}
where $\chi_\mathcal{E}$ is the indicator function of the ensemble $\mathcal{E}$. Eq.~\eqref{eq:Fourier_Cohesive_Prefactors} yields:
\begin{equation}
\label{eq:Functionals_DugdaleBarenblatt}
\begin{cases}
\hat{\mathcal{A}}\left(\kw_0\right) & = \dfrac{1-e^{-\kw_0}}{2} \\
\hat{\Sigma}\left(\kw_0\right) & = \dfrac{\sqrt{\pi}}{2}\dfrac{\mathrm{erf}(\sqrt{\kw_0})}{\sqrt{\kw_0}} \\
\hat{\Omega}\left(\kw_0\right) & = e^{-\kw_0}
\end{cases}
\Rightarrow
\begin{aligned}
\hat{\mathcal{A}}\left(\kw_0\right) & \underset{\kw_0 \rightarrow +\infty}{\longrightarrow} \dfrac{1}{2} \\
\hat{\Sigma}\left(\kw_0\right) & \underset{\kw_0 \rightarrow +\infty}{\sim} \dfrac{\sqrt{\pi}}{2} (\kw_0)^{-1/2} \\
\hat{\Omega}\left(\kw_0\right) & \underset{\kw_0 \rightarrow +\infty}{\sim} e^{-\kw_0}
\end{aligned}
\end{equation}
The evolution of $\hat{\mathcal{A}}$, $\hat{\Sigma}$ and $\hat{\Omega}$ with $\kw_0$ is given in Fig.~\ref{fig:AsymptoticBehavior_DugdaleBarenblatt} for the Dugdale-Barenblatt weakening law.

\subsection{Exponential-distance weakening}
\label{app:Asymptotic_ExponentialDistance}

\begin{figure}[!h]
\centering \includegraphics[width=\textwidth]{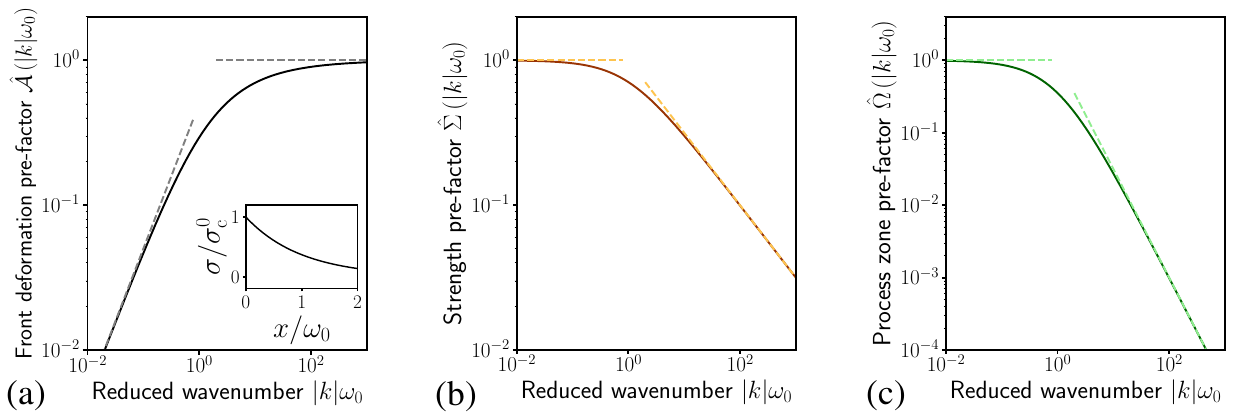}
\caption{Evolution of the cohesive pre-factors for an exponential distance-weakening cohesive law: (a) the front deformations pre-factor $\hat{\mathcal{A}}(\kw_0)$ scales as $\propto \kw_0/2$ when $\kw_0 \rightarrow 0$ (dashed gray line on the left) and saturates to $\mathcal{A}_w^\infty$ when $\kw_0 \rightarrow +\infty$ (horizontal dashed gray line on the right). (b) The strength variation pre-factor $\hat{\Sigma}(\kw_0)$ goes to $1$ in the brittle limit $\kw_0 \rightarrow 0$ and decays as $(\kw_0)^{-1/2}$ when $\kw_0 \rightarrow +\infty$. (c) The process zone size variations pre-factor $\hat{\Omega}(\kw_0)$ goes to $1$ when $\kw_0 \rightarrow 0$ and decays as $(\kw_0)^{-3/2}$ when $\kw_0 \rightarrow +\infty$.}
\label{fig:AsymptoticBehavior_ExponentialDistance}
\end{figure}

The third one is an exponentially distance-weakening law for which the cohesive stress decays as $\sigmac^0 e^{-x/\omega_0}$ behind the crack tip, and never reaches $0$:
\begin{equation}
\label{eq:Weakening_ExponentialDistance}
\fw(x) = e^{-x}
\end{equation}
In that case, Eq.~\eqref{eq:Fourier_Cohesive_Prefactors} yields:
\begin{equation}
\label{eq:Functionals_ExponentialDistance}
\begin{cases}
\hat{\mathcal{A}}\left(\kw_0\right) & = 1 - \dfrac{1}{(1+\kw_0)^{1/2}} \\
\hat{\Sigma}\left(\kw_0\right) & = \dfrac{1}{(1+\kw_0)^{1/2}} \\
\hat{\Omega}\left(\kw_0\right) & = \dfrac{1}{(1+\kw_0)^{3/2}}
\end{cases}
\Rightarrow
\begin{aligned}
\hat{\mathcal{A}}\left(\kw_0\right) & \underset{\kw_0 \rightarrow +\infty}{\longrightarrow} 1 \\
\hat{\Sigma}\left(\kw_0\right) & \underset{\kw_0 \rightarrow +\infty}{\sim} (\kw_0)^{-1/2} \\
\hat{\Omega}\left(\kw_0\right) & \underset{\kw_0 \rightarrow +\infty}{\sim} (\kw_0)^{-3/2}
\end{aligned}
\end{equation}
The evolution of $\hat{\mathcal{A}}$, $\hat{\Sigma}$ and $\hat{\Omega}$ with $\kw_0$ is given in Fig.~\ref{fig:AsymptoticBehavior_ExponentialDistance} for the exponential distance-weakening law.

\subsection{Linear-slip weakening}
\label{app:Asymptotic_SlipWeakening}

The fourth and last one is a linear traction-separation law for which the stress decay linearly from $\sigmac^0$ to $0$ with the local opening $\delta$ up to a critical value $\deltac$:
\begin{equation}
\label{eq:Weakening_SlipWeakening}
\fw(x) = \max\left(1-\delta(x)/\deltac, 0\right)
\end{equation}
This type of weakening is widely used in numerical simulations of cohesive fracture. One cannot find an analytical formula in that case, but $\fw$ can be computed numerically using the method of \cite{viesca_numerical_2018} that builds on a Gauss–Chebyshev quadrature. The resulting $\fw$ is plotted in Fig.~\ref{fig:NatureWeakening_Stability}a.\\

\begin{figure}[!h]
\centering \includegraphics[width=\textwidth]{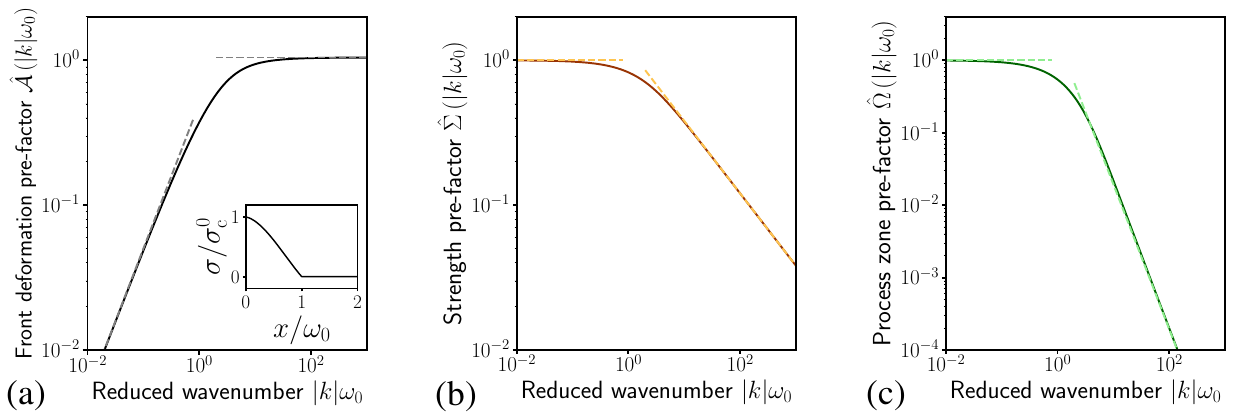}
\caption{Evolution of the cohesive pre-factors for a linear traction-separation law: (a) the front deformations pre-factor $\hat{\mathcal{A}}(\kw_0)$ scales as $\propto \kw_0/2$ when $\kw_0 \rightarrow 0$ (dashed gray line on the left) and saturates to $\mathcal{A}_w^\infty$ when $\kw_0 \rightarrow +\infty$ (horizontal dashed gray line on the right). (b) The strength variation pre-factor $\hat{\Sigma}(\kw_0)$ goes to $1$ in the brittle limit $\kw_0 \rightarrow 0$ and decays as $(\kw_0)^{-1/2}$ when $\kw_0 \rightarrow +\infty$. (c) The process zone size variations pre-factor $\hat{\Omega}(\kw_0)$ goes to $1$ when $\kw_0 \rightarrow 0$ and decays as $(\kw_0)^{-2}$ when $\kw_0 \rightarrow +\infty$.}
\label{fig:AsymptoticBehavior_SlipWeakening}
\end{figure}

The finiteness of the stress at the crack tip implies that $\fw(x) \underset{x \rightarrow 0}{\sim} 1-a_\mathrm{w}x^{3/2}$ where $a_w$ is a cohesive constant related to the weakening shape \citep{rice_mechanics_1980}. This very peculiar behavior leads a change in the asymptotic behavior of $\hat{\Omega}$ when $\kw_0 \rightarrow +\infty$:
\begin{equation}
\label{eq:Fourier_Cohesive_Asymptotics_infinity_Omega_Slip}
\hat{\Omega}\left(\kw_0\right) \underset{\kw_0 \rightarrow +\infty}{\sim} \dfrac{\Omega_w^\infty}{(\kw_0)^{2}} \text{ if } \fw(x) \underset{x \rightarrow 0}{\sim} 1-a_\mathrm{w}x^{3/2} \text{, with } \Omega_w^\infty = \dfrac{2a_\mathrm{w}}{\Cw}
\end{equation}
The behavior of $\hat{\mathcal{A}}$ and $\hat{\Sigma}$ is left unchanged. For the linear traction-separation law, one finds:
\begin{equation}
\label{eq:Asymptotics_SlipWeakening}
\begin{cases}
\hat{\mathcal{A}}\left(\kw_0\right) & \underset{\kw_0 \rightarrow +\infty}{\simeq} 1.045 \\
\hat{\Sigma}\left(\kw_0\right) & \underset{\kw_0 \rightarrow +\infty}{\simeq} 1.209\, (\kw_0)^{-1/2} \\
\hat{\Omega}\left(\kw_0\right) & \underset{\kw_0 \rightarrow +\infty}{\simeq} 1.946\, (\kw_0)^{-2}
\end{cases}
\end{equation}
The evolution of $\hat{\mathcal{A}}$, $\hat{\Sigma}$ and $\hat{\Omega}$ with $\kw_0$ is given for the linear traction-separation cohesive law in Fig.~\ref{fig:AsymptoticBehavior_SlipWeakening}.

\section{Influence of the inclusion spacing}
\label{app:InclusionSpacing}

In this Appendix, we explore the influence of the inclusion spacing on the deformations of a crack front, interacting with heterogeneities of strength and process zone size.

\subsection{Heterogeneities of strength and process zone size}
\label{app:InclusionSpacing_Heterogeneities}

\begin{figure}[!h]
\centering \includegraphics[width=\textwidth]{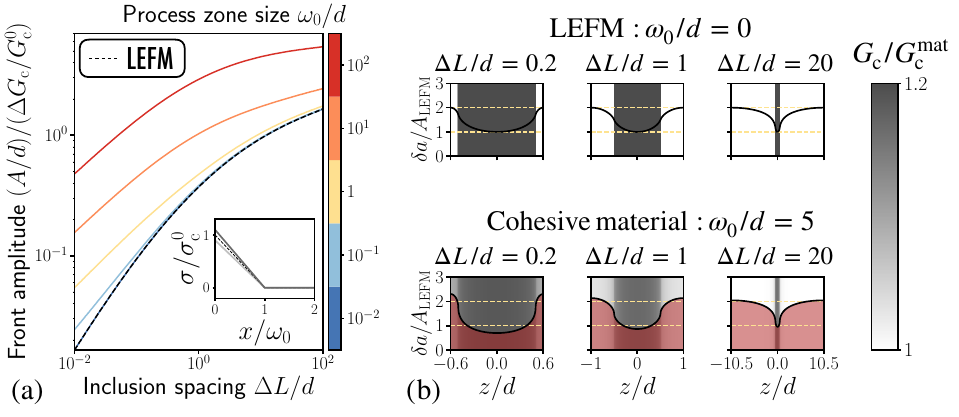}
\caption{\modif{(a) Influence of the inclusion spacing $\Delta L$ and the process zone size $\omega_0$ on the amplitude $A$ of the front deformations for a periodic array of obstacles of width $d$; Inset: the obstacles (in solid dark gray line) are stronger than the matrix $\sigma_\mathrm{c}^\mathrm{obs} = 1.1 \sigma_\mathrm{c}^\mathrm{mat}$ (in solid light gray line), but they are equally brittle $\omega^\mathrm{obs} = \omega^\mathrm{mat}$, so that $G_\mathrm{c}^\mathrm{obs} \simeq 1.2 G_\mathrm{c}^\mathrm{mat}$. (b) The deformation amplitude is always larger than in the LEFM case (in yellow dashed line) and increases with the average process zone size $\omega_0$ (in red).}}
\label{fig:FrontDeformation_Strength_InclusionSpacing}
\end{figure}

\begin{figure}[!h]
\centering \includegraphics[width=\textwidth]{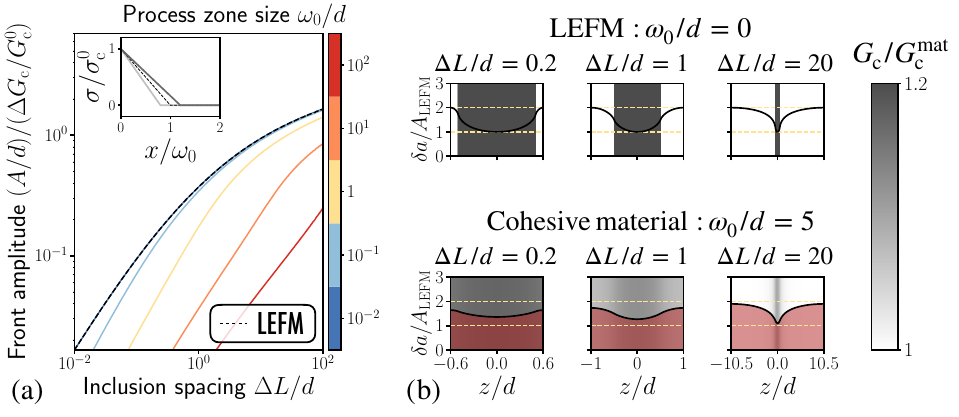}
\caption{\modif{(a) Influence of the inclusion spacing $\Delta L$ and the process zone size $\omega_0$ on the amplitude $A$ of the front deformations for a periodic array of obstacles of width $d$; Inset: the obstacles (in solid dark gray line) are more ductile than the matrix $\omega^\mathrm{obs} = 1.2 \omega^\mathrm{mat}$ (in solid light gray line), but they are equally strong $\sigmac^\mathrm{obs} = \sigmac^\mathrm{mat}$, so that $\Gc^\mathrm{obs} \simeq 1.2 \Gc^\mathrm{mat}$. (b) The deformation amplitude is always smaller than in the LEFM case (in yellow dashed line) and decreases with the average process zone size $\omega_0$ (in red).}}
\label{fig:FrontDeformation_ProcessZoneSize_InclusionSpacing}
\end{figure}

\modif{We consider the situations of Sections~\ref{subsec:FrontDeformations_Strength} and~\ref{subsec:FrontDeformations_ProcessZoneSize}. The crack front interacts with heterogeneities of (i) varying strength but uniform process zone size, or (ii) uniform strength but varying process zone size. In the former case, an increase of the average process zone size $\omega_0$ was associated with an increase in the front amplitude. In the latter, it was linked to a decrease in amplitude. We observe in Figs.~\ref{fig:FrontDeformation_Strength_InclusionSpacing} and~\ref{fig:FrontDeformation_ProcessZoneSize_InclusionSpacing}  that the two different behaviors are left unchanged by the inclusion spacing.}

\subsection{Limit case of the single defect in an infinite matrix}
\label{app:InclusionSpacing_SingleDefect}

\begin{figure}[!h]
\centering \includegraphics[width=\textwidth]{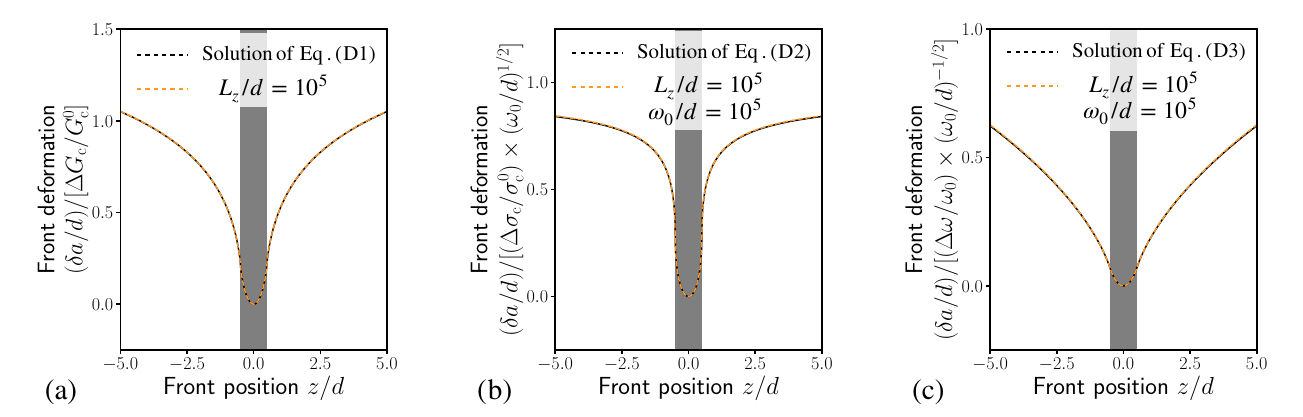}
\caption{\modif{Deformation profiles of a crack front pinned by a single tough heterogeneity ($\Delta L/d \rightarrow +\infty$): (a) In the perfectly brittle limit $\omega_0/d \rightarrow +\infty$, the deformations of the front follow Eq.~\eqref{eq:FrontDeformation_Brittle_SingleDefect} of \cite{chopin_crack_2011}. In the fully cohesive limit $\omega_0/d \rightarrow +\infty$, (b) the deformations of the front by heterogeneities of strength follow Eq.~\eqref{eq:FrontDeformation_Strength_SingleDefect} and increase as $\propto (\omega_0/d)^{1/2}$, while (c) that induced by heterogeneities of process zone size follow Eq.~\eqref{eq:FrontDeformation_ProcessZoneSize_SingleDefect} and vanish as $\propto (\omega_0/d)^{-1/2}$. The analytical solution (in solid black line) are compared to numerical solutions (in dashed orange line) computed from Eq.~\eqref{eq:PropagationCriterion}.}}
\label{fig:FrontDeformation_SingleDefect}
\end{figure}

\modif{A general expression of the front deformations $\delta a$ with the position $z$ cannot be found explicitly.  Yet, it is possible to derive it in some specific cases. In the limit of a single obstacle ($\Delta L \gg d$) embedded in a rather brittle medium ($\omega_0 \ll d$), one retrieves the solution of  \citep{chopin_crack_2011}:}
\begin{equation}
\label{eq:FrontDeformation_Brittle_SingleDefect}
\modif{\left(\delta a\left(z\right) - \delta a \left(0\right)\right)/d = \dfrac{1}{2\pi} \dfrac{\Delta \Gc}{\Gc^0}\left[\left(1+\dfrac{2z}{d}\right)\mathrm{ln}\left\lvert 1+\dfrac{2z}{d}\right\rvert + \left(1-\dfrac{2z}{d}\right)\mathrm{ln}\left\lvert 1-\dfrac{2z}{d}\right\rvert\right]}
\end{equation}
\modif{where $d$ is the obstacle width, $\Delta \Gc$ is the contrast in fracture energy and $\Gc^0$ its average value. The evolution of the front deformations $\delta a$ with the position is plotted in Fig.~\ref{fig:FrontDeformation_SingleDefect}a.}\\

\modif{On the contrary, it is also possible to derive an analytical expression of the front deformation when the same obstacle lies in a very ductile medium ($\omega_0 \gg d$). For heterogeneities of strength, one finds:}
\begin{equation}
\label{eq:FrontDeformation_Strength_SingleDefect}
\modif{\left(\delta a\left(z\right) - \delta a \left(0\right)\right)/d = \dfrac{2}{\sqrt{\pi}} \dfrac{\Sigma_w^\infty}{\mathcal{A}_w^\infty} \dfrac{\Delta \sigmac}{\sigmac^0} \left(\dfrac{\omega_0}{d}\right)^{1/2} \left[2 - \sqrt{1+\dfrac{2\abs{z}}{d}} + \mathrm{sgn}\left(\abs{z}-\dfrac{d}{2}\right) \sqrt{\left\lvert 1-\dfrac{2\abs{z}}{d}\right\rvert}\right]}
\end{equation}
\modif{where $d$ is the obstacle width, $\omega_0$ is the average process zone size, $\Delta \sigmac$ is the contrast in strength and $\sigmac^0$ its average value. $\mathcal{A}_w^\infty$ and $\Sigma_w^\infty$ relate to the cohesive law and are given in Eqs.~\eqref{eq:Fourier_Cohesive_Asymptotics_infinity_A} and \eqref{eq:Fourier_Cohesive_Asymptotics_infinity_Sigma}. The evolution of the front deformations $\delta a$ with the position is plotted in Fig.~\ref{fig:FrontDeformation_SingleDefect}b, and successfully compared to the numerical solution of Eq.~\eqref{eq:FrontDeformation_Cohesive_Strength} for $L_z/d=10^5$ and $\omega_0/d=10^5$. Eq.~\eqref{eq:FrontDeformation_Strength_SingleDefect} validates the linear dependence in $(\Delta \sigmac/\sigmac^0)$ and that in $(\omega_0/d)^{1/2}$ that was mentioned in Section~\ref{subsec:FrontDeformations_Strength}. Moreover, we observe that the deformation localizes at the edges of the obstacle, where $\delta a$ is non-differentiable. It may challenge the assumptions lying under our first-order theory as we assumed $\partial \delta a/\partial z \ll 1$, justifying future derivation of a second-order theory for the quasi-static cohesive front deformations to rationalize the experimental results of \cite{chopin_crack_2011}.}\\

\modif{For heterogeneities of process zone size, one finds:}
\begin{equation}
\label{eq:FrontDeformation_ProcessZoneSize_SingleDefect}
\scalemath{0.95}{
\modif{\left(\delta a\left(z\right) - \delta a \left(0\right)\right)/d = \dfrac{1}{6\sqrt{\pi}} \dfrac{\Omega_w^\infty}{\mathcal{A}_w^\infty} \dfrac{\Delta \omega}{\omega_0} \left(\dfrac{\omega_0}{d}\right)^{-1/2} \left[-2 + \left(1+\dfrac{2\abs{z}}{d}\right)^{3/2} - \mathrm{sgn}\left(\abs{z}-\dfrac{d}{2}\right) \left\lvert 1-\dfrac{2\abs{z}}{d}\right\rvert^{3/2}\right]}
}
\end{equation}
\modif{where $d$ is the obstacle width, $\Delta \omega$ is the contrast in process zone size and $\omega_0$ its average value. $\mathcal{A}_w^\infty$ and $\Omega_w^\infty$ relate to the cohesive law and are given in Eqs.~\eqref{eq:Fourier_Cohesive_Asymptotics_infinity_A} and \eqref{eq:Fourier_Cohesive_Asymptotics_infinity_Omega}. The evolution of the front deformations $\delta a$ with the position is plotted in Fig.~\ref{fig:FrontDeformation_SingleDefect}c. Eq.~\eqref{eq:FrontDeformation_ProcessZoneSize_SingleDefect} confirms the linear dependence in both $(\Delta \omega/\omega_0)$ and $(\omega_0/d)^{-1/2}$.}

\modif{Note that Eq.~\eqref{eq:FrontDeformation_ProcessZoneSize_SingleDefect} strongly depends on the weakening shape $\fw$, as it controls the asymptotic behavior of $\hat{\Omega}$. For example, one finds for the linear traction-separation law:}
\begin{equation}
\label{eq:FrontDeformation_ProcessZoneSize_SingleDefect_SlipWeakening}
\scalemath{1}{
\modif{\left(\delta a\left(z\right) - \delta a \left(0\right)\right)/d = 
\begin{cases}
\dfrac{\Omega_w^\infty}{4\mathcal{A}_w^\infty} \dfrac{\Delta \omega}{\omega_0} \left(\dfrac{\omega_0}{d}\right)^{-1} \left(\dfrac{z}{d}\right)^2 & \text{ if } \abs{z} \leq d/2 \\
\dfrac{\Omega_w^\infty}{4\mathcal{A}_w^\infty} \dfrac{\Delta \omega}{\omega_0} \left(\dfrac{\omega_0}{d}\right)^{-1} \left(\dfrac{\abs{z}}{d} -\dfrac{1}{4} \right) & \text{ if } \abs{z} > d/2
\end{cases}}
}
\end{equation}

\section{Influence of the nature of weakening}
\label{app:NatureWeakening}

In this Appendix, we explore the influence of the shape $\fw$ of the cohesive law on the stability of crack fronts and their interaction with periodic arrays of tough obstacles. We consider the four weakening shapes studied in \ref{app:Asymptotic_Weakenings}: the reference linear distance-weakening, a Dugdale-Barenblatt weakening, an exponential distance-weakening, and a more conventional linear traction-separation cohesive law.

\subsection{Influence of the nature of weakening on the crack front stability}
\label{app:NatureWeakening_Stability}

We saw in Eqs.~\eqref{eq:Stability_Cohesive_CriticalLengthScale} and \eqref{eq:Stability_Cohesive_CriticalWavelength} that the stability of a perturbed crack strongly depends on the evolution of $\hat{\mathcal{A}}$ and its asymptotic behavior as $\kw_0 \rightarrow +\infty$. We thus expect that the stability of crack fronts somehow depends on the spatial distribution $\fw$ of cohesive stress in the crack wake.

\begin{figure}[!h]
\centering \includegraphics[width=\textwidth]{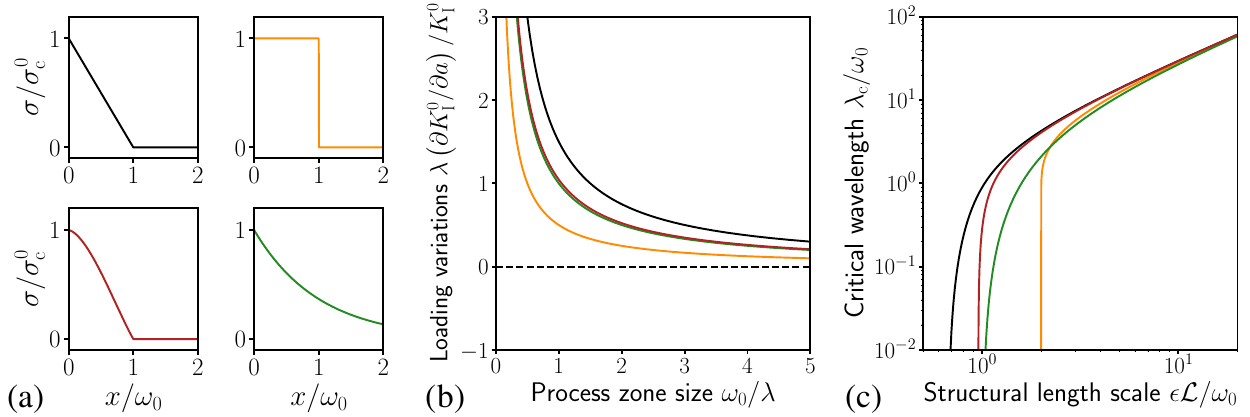}
\caption{Influence of the nature of weakening $f_w$ on the stability of a sinusoidal crack front: (a) linear distance-weakening (in black line), Dugdale-Barenblatt (in orange line), linear traction-separation (in red line), and exponential distance-weakening (in green line) cohesive laws are considered. (b) The frontier delimiting the region of conditional (Regime III) to unconditional instability (Regime II) is controlled by the evolution of $\hat{\mathcal{A}}$ which is set by $f_w$. (c) It also controls the value of the critical wavelength $\lambda_\mathrm{c}$ above which modal perturbations are unstable, and the value of the critical structural length scale $\mathcal{L}_\mathrm{c}$ below which perturbations of any wavelengths are unstable.}
\label{fig:NatureWeakening_Stability}
\end{figure}

We observe in Fig.~\ref{fig:NatureWeakening_Stability} that all four cohesive laws lead to a similar stability behavior. The main difference is a shift of the frontier separating Regimes II and III, as it is controlled by the asymptotic value $\mathcal{A}_w^\infty$ of the operator $\hat{\mathcal{A}}$. Namely, the shift from the two regimes occurs for smaller process zone size $\omega_0$ for the Dugdale-Barenblatt cohesive law than for the linear distance-weakening law. The cases of exponentially distance-weakening and linear traction-separation laws are found in between.

\subsection{Influence of the nature of weakening on the crack front deformations}
\label{app:NatureWeakening_FrontDeformations}

We address next the question of the influence of the nature of weakening, characterized by the weakening shape $\fw$, on the front deformations.\\

\begin{figure}[!h]
\centering \includegraphics[width=\textwidth]{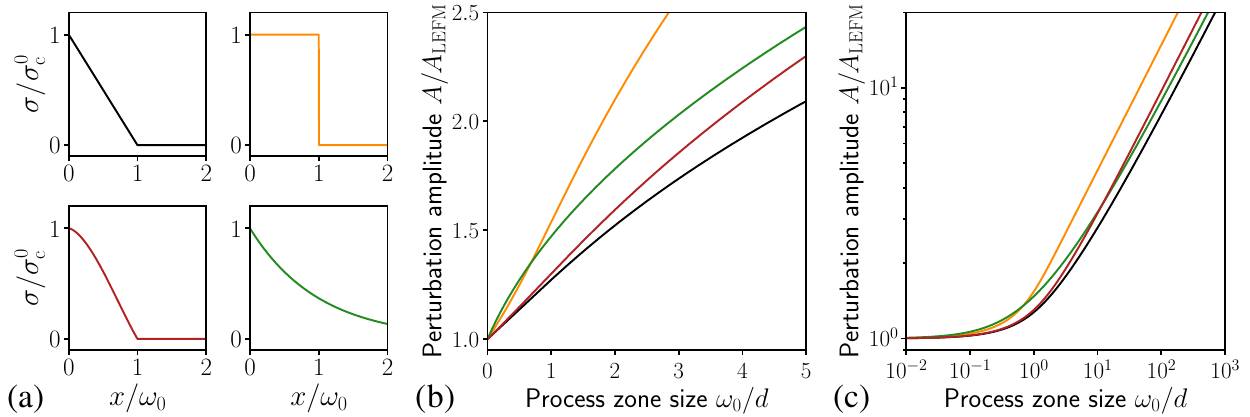}
\caption{Influence of the nature of weakening $f_w$ on the front deformations associated with spatial variations of strength: (a) linear distance-weakening (in black line), Dugdale-Barenblatt (in orange line), linear traction-separation (in red line), and exponential distance-weakening (in green line) cohesive laws are considered. (b) The perturbation amplitude always increases with the average process zone size, but at a rate controlled by the weakening $\fw$. (c) It scales as $\propto (\omega_0/d)^{1/2}$ when $\omega_0/d \rightarrow +\infty$ no matter how the material weakens.}
\label{fig:NatureWeakening_Strength}
\end{figure}

We first observe in Fig.~\ref{fig:NatureWeakening_Strength} that the influence of the weakening shape on the front deformation is rather weak for heterogeneities of strength. The deformations are overall larger for a Dugdale-Barenblatt law, and lower for a linear distance-weakening law. The other two laws are found in-between. They all display a similar scaling $A/d \propto (\omega_0/d)^{1/2}$ for large process zone sizes $\omega_0/d \rightarrow +\infty$, as the asymptotic behavior of $\hat{\Sigma}$ of Eq.~\eqref{eq:Fourier_Cohesive_Asymptotics_infinity_Sigma} depends very weakly on $\fw$.\\

\begin{figure}[!h]
\centering \includegraphics[width=\textwidth]{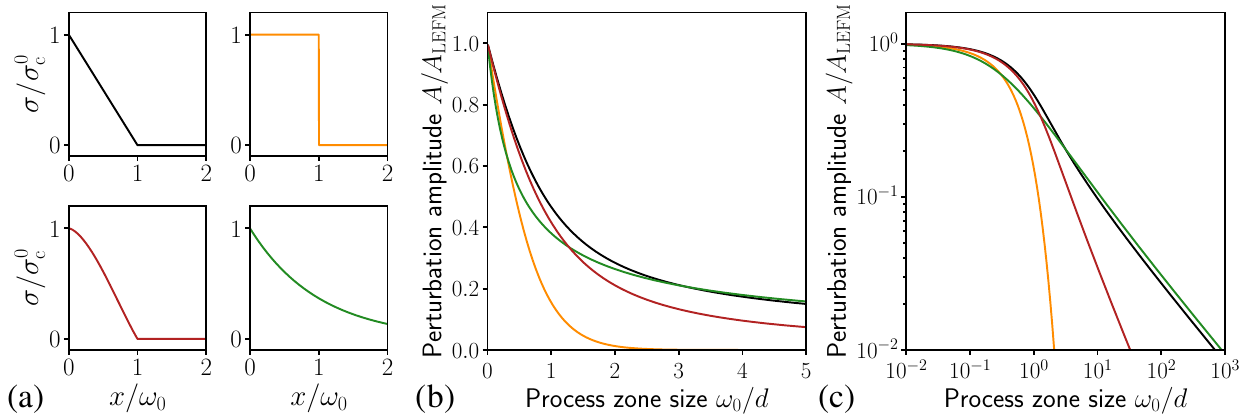}
\caption{Influence of the nature of weakening $f_w$ on the front deformation emerging from spatial variations of process zone size: (a) linear distance-weakening (in black line), Dugdale-Barenblatt (in orange line), linear traction-separation (in red line), and exponential distance-weakening (in green line) cohesive laws are considered. (b) The perturbation amplitude always decreases with the average process zone size $\omega_0$ but at a rate controlled by the weakening $f_w$. (c) Its scaling as when $\omega_0/d \rightarrow +\infty$ depends on how the material weakens. For distance-weakening laws, it scales as $\propto (\omega_0/d)^{-1/2}$, while it decreases as $\propto (\omega_0/d)^{-1}$ for traction-separation laws. The decrease is even stronger for the Dugdale-Barenblatt cohesive law $\propto (\omega_0/d) e^{-\omega_0/d}$.}
\label{fig:NatureWeakening_ProcessZoneSize}
\end{figure}

The case of heterogeneities of process zone size is more interesting, as the asymptotic behavior of $\hat{\Omega}$ changes significantly with the definition of $\fw$. Indeed, the decrease in magnitude of the effective fluctuations of fracture energy is sharper for traction-separation laws $A/d \propto (\omega_0/d)^{-1}$ than for distance-weakening ones $A/d \propto (\omega_0/d)^{-1/2}$ (as long as $\dfw(0) \neq 0$). Consequently, the amplitude of front deformations vanishes at a much higher rate as the average process zone size $\omega_0$ gets larger than the obstacle width $d$ (see Fig.~\ref{fig:NatureWeakening_ProcessZoneSize}). This effect is even stronger for the Dugdale-Barenblatt law, for which $A/d \propto (\omega_0/d) \times e^{-\omega_0/d}$. 

\bibliographystyle{elsarticle-harv} 
%\bibliography{JMPS_CohesiveFrontDeformations}

\begin{thebibliography}{45}
\expandafter\ifx\csname natexlab\endcsname\relax\def\natexlab#1{#1}\fi
\providecommand{\url}[1]{\texttt{#1}}
\providecommand{\href}[2]{#2}
\providecommand{\path}[1]{#1}
\providecommand{\DOIprefix}{doi:}
\providecommand{\ArXivprefix}{arXiv:}
\providecommand{\URLprefix}{URL: }
\providecommand{\Pubmedprefix}{pmid:}
\providecommand{\doi}[1]{\href{http://dx.doi.org/#1}{\path{#1}}}
\providecommand{\Pubmed}[1]{\href{pmid:#1}{\path{#1}}}
\providecommand{\bibinfo}[2]{#2}
\ifx\xfnm\relax \def\xfnm[#1]{\unskip,\space#1}\fi
%Type = Incollection
\bibitem[{Barenblatt(1962)}]{barenblatt_processzone_1962}
\bibinfo{author}{Barenblatt, G.I.}, \bibinfo{year}{1962}.
\newblock \bibinfo{title}{The mathematical theory of equilibrium cracks in
  brittle fracture}, in: \bibinfo{editor}{Dryden, H.L.}, \bibinfo{editor}{von
  K{\'a}rm{\'a}n, T.}, \bibinfo{editor}{Kuerti, G.}, \bibinfo{editor}{van~den
  Dungen, F.H.}, \bibinfo{editor}{Howarth, L.} (Eds.),
  \bibinfo{booktitle}{Advances in Applied Mechanics}.
  \bibinfo{publisher}{Elsevier}. volume~\bibinfo{volume}{7}, pp.
  \bibinfo{pages}{55--129}.
%Type = Article
\bibitem[{Barés et~al.(2018)Barés, Dubois, Hattali, Dalmas and
  Bonamy}]{bares_aftershock_2018}
\bibinfo{author}{Barés, J.}, \bibinfo{author}{Dubois, A.},
  \bibinfo{author}{Hattali, L.}, \bibinfo{author}{Dalmas, D.},
  \bibinfo{author}{Bonamy, D.}, \bibinfo{year}{2018}.
\newblock \bibinfo{title}{Aftershock sequences and seismic-like organization of
  acoustic events produced by a single propagating crack}.
\newblock \bibinfo{journal}{Nature Communications} \bibinfo{volume}{9},
  \bibinfo{pages}{1253}.
\newblock \URLprefix \url{https://www.nature.com/articles/s41467-018-03559-4},
  \DOIprefix\doi{10.1038/s41467-018-03559-4}.
%Type = Article
\bibitem[{Bonamy and Bouchaud(2011)}]{bonamy_failure_2011}
\bibinfo{author}{Bonamy, D.}, \bibinfo{author}{Bouchaud, E.},
  \bibinfo{year}{2011}.
\newblock \bibinfo{title}{Failure of heterogeneous materials: A dynamic phase
  transition?}
\newblock \bibinfo{journal}{Physics Reports} \bibinfo{volume}{498},
  \bibinfo{pages}{1--44}.
\newblock \URLprefix
  \url{http://www.sciencedirect.com/science/article/pii/S0370157310002115},
  \DOIprefix\doi{10.1016/j.physrep.2010.07.006}.
%Type = Article
\bibitem[{Bower and Ortiz(1991)}]{bower_bridging_1991}
\bibinfo{author}{Bower, A.}, \bibinfo{author}{Ortiz, M.}, \bibinfo{year}{1991}.
\newblock \bibinfo{title}{A three-dimensional analysis of crack trapping and
  bridging by tough particles}.
\newblock \bibinfo{journal}{Journal of the Mechanics and Physics of Solids}
  \bibinfo{volume}{39}, \bibinfo{pages}{815--858}.
\newblock \URLprefix
  \url{http://www.sciencedirect.com/science/article/pii/002250969190026K},
  \DOIprefix\doi{10.1016/0022-5096(91)90026-K}.
%Type = Article
\bibitem[{Bueckner(1987)}]{bueckner_weight_1987}
\bibinfo{author}{Bueckner, H.}, \bibinfo{year}{1987}.
\newblock \bibinfo{title}{Weight functions and fundamental fields for the
  penny-shaped and the half-plane crack in three-space}.
\newblock \bibinfo{journal}{International Journal of Solids and Structures}
  \bibinfo{volume}{23}, \bibinfo{pages}{57--93}.
\newblock \URLprefix
  \url{http://www.sciencedirect.com/science/article/pii/0020768387900321},
  \DOIprefix\doi{10.1016/0020-7683(87)90032-1}.
%Type = Article
\bibitem[{Chopin et~al.(2018)Chopin, Bhaskar, Jog and
  Ponson}]{chopin_depinning_2018}
\bibinfo{author}{Chopin, J.}, \bibinfo{author}{Bhaskar, A.},
  \bibinfo{author}{Jog, A.}, \bibinfo{author}{Ponson, L.},
  \bibinfo{year}{2018}.
\newblock \bibinfo{title}{Depinning dynamics of crack fronts}.
\newblock \bibinfo{journal}{Physical Review Letters} \bibinfo{volume}{121},
  \bibinfo{pages}{235501}.
\newblock \URLprefix
  \url{https://link.aps.org/doi/10.1103/PhysRevLett.121.235501},
  \DOIprefix\doi{10.1103/PhysRevLett.121.235501}.
%Type = Article
\bibitem[{Chopin et~al.(2011)Chopin, Prevost, Boudaoud and
  Adda-Bedia}]{chopin_crack_2011}
\bibinfo{author}{Chopin, J.}, \bibinfo{author}{Prevost, A.},
  \bibinfo{author}{Boudaoud, A.}, \bibinfo{author}{Adda-Bedia, M.},
  \bibinfo{year}{2011}.
\newblock \bibinfo{title}{Crack front dynamics across a single heterogeneity}.
\newblock \bibinfo{journal}{Physical Review Letters} \bibinfo{volume}{107},
  \bibinfo{pages}{144301}.
\newblock \URLprefix
  \url{https://link.aps.org/doi/10.1103/PhysRevLett.107.144301},
  \DOIprefix\doi{10.1103/PhysRevLett.107.144301}.
%Type = Article
\bibitem[{Creton and Ciccotti(2016)}]{creton_fracture_2016}
\bibinfo{author}{Creton, C.}, \bibinfo{author}{Ciccotti, M.},
  \bibinfo{year}{2016}.
\newblock \bibinfo{title}{Fracture and adhesion of soft materials: a review}.
\newblock \bibinfo{journal}{Reports on Progress in Physics}
  \bibinfo{volume}{79}, \bibinfo{pages}{046601}.
\newblock \URLprefix \url{https://doi.org/10.1088/0034-4885/79/4/046601},
  \DOIprefix\doi{10.1088/0034-4885/79/4/046601}.
%Type = Article
\bibitem[{Dalmas et~al.(2009)Dalmas, Barthel and
  Vandembroucq}]{dalmas_pinning_2009}
\bibinfo{author}{Dalmas, D.}, \bibinfo{author}{Barthel, E.},
  \bibinfo{author}{Vandembroucq, D.}, \bibinfo{year}{2009}.
\newblock \bibinfo{title}{Crack front pinning by design in planar heterogeneous
  interfaces}.
\newblock \bibinfo{journal}{Journal of the Mechanics and Physics of Solids}
  \bibinfo{volume}{57}, \bibinfo{pages}{446--457}.
\newblock \URLprefix
  \url{http://www.sciencedirect.com/science/article/pii/S002250960800210X},
  \DOIprefix\doi{10.1016/j.jmps.2008.11.012}.
%Type = Article
\bibitem[{Delaplace et~al.(1999)Delaplace, Schmittbuhl and
  Maloy}]{delaplace_high_1999}
\bibinfo{author}{Delaplace, A.}, \bibinfo{author}{Schmittbuhl, J.},
  \bibinfo{author}{Maloy, K.}, \bibinfo{year}{1999}.
\newblock \bibinfo{title}{High resolution description of a crack front in a
  heterogeneous plexiglas block}.
\newblock \bibinfo{journal}{Physical Review E} \bibinfo{volume}{60},
  \bibinfo{pages}{1337--1343}.
\newblock \URLprefix \url{https://link.aps.org/doi/10.1103/PhysRevE.60.1337},
  \DOIprefix\doi{10.1103/PhysRevE.60.1337}.
%Type = Article
\bibitem[{D{\'e}mery et~al.(2014)D{\'e}mery, Rosso and
  Ponson}]{demery_microstructural_2014}
\bibinfo{author}{D{\'e}mery, V.}, \bibinfo{author}{Rosso, A.},
  \bibinfo{author}{Ponson, L.}, \bibinfo{year}{2014}.
\newblock \bibinfo{title}{From microstructural features to effective toughness
  in disordered brittle solids}.
\newblock \bibinfo{journal}{{EPL} (Europhysics Letters)} \bibinfo{volume}{105},
  \bibinfo{pages}{34003}.
\newblock \URLprefix \url{http://stacks.iop.org/0295-5075/105/i=3/a=34003},
  \DOIprefix\doi{10.1209/0295-5075/105/34003}.
%Type = Article
\bibitem[{Dugdale(1960)}]{dugdale_yielding_1960}
\bibinfo{author}{Dugdale, D.S.}, \bibinfo{year}{1960}.
\newblock \bibinfo{title}{Yielding of steel sheets containing slits}.
\newblock \bibinfo{journal}{Journal of the Mechanics and Physics of Solids}
  \bibinfo{volume}{8}, \bibinfo{pages}{100--104}.
\newblock \URLprefix
  \url{http://www.sciencedirect.com/science/article/pii/0022509660900132},
  \DOIprefix\doi{10.1016/0022-5096(60)90013-2}.
%Type = Article
\bibitem[{Favier et~al.(2006)Favier, Lazarus and
  Leblond}]{favier_coplanar_2006}
\bibinfo{author}{Favier, E.}, \bibinfo{author}{Lazarus, V.},
  \bibinfo{author}{Leblond, J.}, \bibinfo{year}{2006}.
\newblock \bibinfo{title}{Coplanar propagation paths of 3d cracks in infinite
  bodies loaded in shear}.
\newblock \bibinfo{journal}{International Journal of Solids and Structures}
  \bibinfo{volume}{43}, \bibinfo{pages}{2091--2109}.
\newblock \URLprefix
  \url{http://www.sciencedirect.com/science/article/pii/S0020768305003689},
  \DOIprefix\doi{10.1016/j.ijsolstr.2005.06.041}.
%Type = Article
\bibitem[{Gao and Rice(1986)}]{gao_shear_1986}
\bibinfo{author}{Gao, H.}, \bibinfo{author}{Rice, J.}, \bibinfo{year}{1986}.
\newblock \bibinfo{title}{Shear stress intensity factors for a planar crack
  with slightly curved front}.
\newblock \bibinfo{journal}{Journal of Applied Mechanics} \bibinfo{volume}{53},
  \bibinfo{pages}{774--778}.
\newblock \URLprefix \url{http://dx.doi.org/10.1115/1.3171857},
  \DOIprefix\doi{10.1115/1.3171857}.
%Type = Article
\bibitem[{Gao and Rice(1989)}]{gao_trapping_1989}
\bibinfo{author}{Gao, H.}, \bibinfo{author}{Rice, J.}, \bibinfo{year}{1989}.
\newblock \bibinfo{title}{A first-order perturbation analysis of crack trapping
  by arrays of obstacles}.
\newblock \bibinfo{journal}{Journal of Applied Mechanics} \bibinfo{volume}{56},
  \bibinfo{pages}{828--836}.
\newblock \URLprefix \url{http://dx.doi.org/10.1115/1.3176178},
  \DOIprefix\doi{10.1115/1.3176178}.
%Type = Article
\bibitem[{Geubelle and Rice(1995)}]{geubelle_spectral_1995}
\bibinfo{author}{Geubelle, P.H.}, \bibinfo{author}{Rice, J.R.},
  \bibinfo{year}{1995}.
\newblock \bibinfo{title}{A spectral method for three-dimensional elastodynamic
  fracture problems}.
\newblock \bibinfo{journal}{Journal of the Mechanics and Physics of Solids}
  \bibinfo{volume}{43}, \bibinfo{pages}{1791--1824}.
\newblock \URLprefix
  \url{http://www.sciencedirect.com/science/article/pii/002250969500043I},
  \DOIprefix\doi{10.1016/0022-5096(95)00043-I}.
%Type = Book
\bibitem[{Gradshteyn and Ryzhik(2014)}]{gradshteyn_integrals_2014}
\bibinfo{author}{Gradshteyn, I.S.}, \bibinfo{author}{Ryzhik, I.M.},
  \bibinfo{year}{2014}.
\newblock \bibinfo{title}{Table of integrals, series, and products}.
\newblock \bibinfo{edition}{Seventh} ed., \bibinfo{publisher}{Elsevier/Academic
  Press, Amsterdam}.
%Type = Article
\bibitem[{Griffith(1921)}]{griffith_phenomena_1921}
\bibinfo{author}{Griffith, A.}, \bibinfo{year}{1921}.
\newblock \bibinfo{title}{The phenomena of rupture and flow in solids}.
\newblock \bibinfo{journal}{Phil. Trans. R. Soc. Lond. A}
  \bibinfo{volume}{221}.
\newblock \URLprefix
  \url{http://rsta.royalsocietypublishing.org/content/221/582-593/163},
  \DOIprefix\doi{10.1098/rsta.1921.0006}.
%Type = Incollection
\bibitem[{Irwin(1958)}]{irwin_fracture_1958}
\bibinfo{author}{Irwin, G.}, \bibinfo{year}{1958}.
\newblock \bibinfo{title}{Fracture}, in: \bibinfo{editor}{Fl{\"u}gge, S.}
  (Ed.), \bibinfo{booktitle}{Elasticity and Plasticity / Elastizit{\"a}t und
  Plastizitt{\"a}t}. \bibinfo{publisher}{Springer Berlin Heidelberg}. Handbuch
  der Physik / Encyclopedia of Physics, pp. \bibinfo{pages}{551--590}.
\newblock \URLprefix \url{https://doi.org/10.1007/978-3-642-45887-3\_5}.
%Type = Article
\bibitem[{Kolvin et~al.(2017)Kolvin, Fineberg and
  Adda-Bedia}]{kolvin_nonlinear_2017}
\bibinfo{author}{Kolvin, I.}, \bibinfo{author}{Fineberg, J.},
  \bibinfo{author}{Adda-Bedia, M.}, \bibinfo{year}{2017}.
\newblock \bibinfo{title}{Nonlinear focusing in dynamic crack fronts and the
  microbranching transition}.
\newblock \bibinfo{journal}{Physical Review Letters} \bibinfo{volume}{119},
  \bibinfo{pages}{215505}.
\newblock \URLprefix
  \url{https://link.aps.org/doi/10.1103/PhysRevLett.119.215505},
  \DOIprefix\doi{10.1103/PhysRevLett.119.215505}.
%Type = Article
\bibitem[{Larkin and Ovchinnikov(1979)}]{larkin_pinning_1979}
\bibinfo{author}{Larkin, A.}, \bibinfo{author}{Ovchinnikov, Y.},
  \bibinfo{year}{1979}.
\newblock \bibinfo{title}{Pinning in type {II} superconductors}.
\newblock \bibinfo{journal}{Journal of Low Temperature Physics}
  \bibinfo{volume}{34}, \bibinfo{pages}{409--428}.
\newblock \URLprefix \url{https://doi.org/10.1007/BF00117160},
  \DOIprefix\doi{10.1007/BF00117160}.
%Type = Article
\bibitem[{Lazarus(2011)}]{lazarus_review_2011}
\bibinfo{author}{Lazarus, V.}, \bibinfo{year}{2011}.
\newblock \bibinfo{title}{Perturbation approaches of a planar crack in linear
  elastic fracture mechanics: A review}.
\newblock \bibinfo{journal}{Journal of the Mechanics and Physics of Solids}
  \bibinfo{volume}{59}, \bibinfo{pages}{121--144}.
\newblock \URLprefix
  \url{http://www.sciencedirect.com/science/article/pii/S0022509610002462},
  \DOIprefix\doi{10.1016/j.jmps.2010.12.006}.
%Type = Article
\bibitem[{Lebihain(2021)}]{lebihain_towards_2021}
\bibinfo{author}{Lebihain, M.}, \bibinfo{year}{2021}.
\newblock \bibinfo{title}{Towards brittle materials with tailored fracture
  properties: the decisive influence of the material disorder and its
  microstructure}.
\newblock \bibinfo{journal}{International Journal of Fracture} \URLprefix
  \url{https://doi.org/10.1007/s10704-021-00538-7},
  \DOIprefix\doi{10.1007/s10704-021-00538-7}.
%Type = Article
\bibitem[{Leblond et~al.(2012)Leblond, Patinet, Frelat and
  Lazarus}]{leblond_second_2012}
\bibinfo{author}{Leblond, J.}, \bibinfo{author}{Patinet, S.},
  \bibinfo{author}{Frelat, J.}, \bibinfo{author}{Lazarus, V.},
  \bibinfo{year}{2012}.
\newblock \bibinfo{title}{Second-order coplanar perturbation of a semi-infinite
  crack in an infinite body}.
\newblock \bibinfo{journal}{Engineering Fracture Mechanics}
  \bibinfo{volume}{90}, \bibinfo{pages}{129--142}.
\newblock \URLprefix
  \url{http://www.sciencedirect.com/science/article/pii/S0013794412000963},
  \DOIprefix\doi{10.1016/j.engfracmech.2012.03.002}.
%Type = Article
\bibitem[{Legrand et~al.(2011)Legrand, Patinet, Leblond, Frelat, Lazarus and
  Vandembroucq}]{legrand_coplanar_2011}
\bibinfo{author}{Legrand, L.}, \bibinfo{author}{Patinet, S.},
  \bibinfo{author}{Leblond, J.B.}, \bibinfo{author}{Frelat, J.},
  \bibinfo{author}{Lazarus, V.}, \bibinfo{author}{Vandembroucq, D.},
  \bibinfo{year}{2011}.
\newblock \bibinfo{title}{Coplanar perturbation of a crack lying on the
  mid-plane of a plate}.
\newblock \bibinfo{journal}{International Journal of Fracture}
  \bibinfo{volume}{170}, \bibinfo{pages}{67--82}.
\newblock \URLprefix \url{https://doi.org/10.1007/s10704-011-9603-0},
  \DOIprefix\doi{10.1007/s10704-011-9603-0}.
%Type = Article
\bibitem[{Morrissey and Rice(2000)}]{morrissey_perturbative_2000}
\bibinfo{author}{Morrissey, J.W.}, \bibinfo{author}{Rice, J.R.},
  \bibinfo{year}{2000}.
\newblock \bibinfo{title}{Perturbative simulations of crack front waves}.
\newblock \bibinfo{journal}{Journal of the Mechanics and Physics of Solids}
  \bibinfo{volume}{48}, \bibinfo{pages}{1229--1251}.
\newblock \URLprefix
  \url{http://www.sciencedirect.com/science/article/pii/S0022509699000691},
  \DOIprefix\doi{10.1016/S0022-5096(99)00069-1}.
%Type = Article
\bibitem[{Movchan et~al.(1998)Movchan, Gao and
  Willis}]{movchan_perturbations_1998}
\bibinfo{author}{Movchan, A.}, \bibinfo{author}{Gao, H.},
  \bibinfo{author}{Willis, J.}, \bibinfo{year}{1998}.
\newblock \bibinfo{title}{On perturbations of plane cracks}.
\newblock \bibinfo{journal}{International Journal of Solids and Structures}
  \bibinfo{volume}{35}, \bibinfo{pages}{3419--3453}.
\newblock \URLprefix
  \url{http://www.sciencedirect.com/science/article/pii/S002076839700231X},
  \DOIprefix\doi{10.1016/S0020-7683(97)00231-X}.
%Type = Article
\bibitem[{Måløy et~al.(2006)Måløy, Santucci, Schmittbuhl and
  Toussaint}]{maloy_local_2006}
\bibinfo{author}{Måløy, K.J.}, \bibinfo{author}{Santucci, S.},
  \bibinfo{author}{Schmittbuhl, J.}, \bibinfo{author}{Toussaint, R.},
  \bibinfo{year}{2006}.
\newblock \bibinfo{title}{Local waiting time fluctuations along a randomly
  pinned crack front}.
\newblock \bibinfo{journal}{Physical Review Letters} \bibinfo{volume}{96},
  \bibinfo{pages}{045501}.
\newblock \URLprefix
  \url{https://link.aps.org/doi/10.1103/PhysRevLett.96.045501},
  \DOIprefix\doi{10.1103/PhysRevLett.96.045501}.
%Type = Article
\bibitem[{Palmer et~al.(1973)Palmer, Rice and Hill}]{palmer_growth_1973}
\bibinfo{author}{Palmer, A.C.}, \bibinfo{author}{Rice, J.R.},
  \bibinfo{author}{Hill, R.}, \bibinfo{year}{1973}.
\newblock \bibinfo{title}{The growth of slip surfaces in the progressive
  failure of over-consolidated clay}.
\newblock \bibinfo{journal}{Proceedings of the Royal Society of London. A.
  Mathematical and Physical Sciences} \bibinfo{volume}{332},
  \bibinfo{pages}{527--548}.
\newblock \URLprefix
  \url{https://royalsocietypublishing.org/doi/10.1098/rspa.1973.0040},
  \DOIprefix\doi{10.1098/rspa.1973.0040}.
%Type = Article
\bibitem[{Patinet et~al.(2013a)Patinet, Alzate, Barthel, Dalmas, Vandembroucq
  and Lazarus}]{patinet_pinning_2013}
\bibinfo{author}{Patinet, S.}, \bibinfo{author}{Alzate, L.},
  \bibinfo{author}{Barthel, E.}, \bibinfo{author}{Dalmas, D.},
  \bibinfo{author}{Vandembroucq, D.}, \bibinfo{author}{Lazarus, V.},
  \bibinfo{year}{2013}a.
\newblock \bibinfo{title}{Finite size effects on crack front pinning at
  heterogeneous planar interfaces: Experimental, finite elements and
  perturbation approaches}.
\newblock \bibinfo{journal}{Journal of the Mechanics and Physics of Solids}
  \bibinfo{volume}{61}, \bibinfo{pages}{311--324}.
\newblock \URLprefix
  \url{http://www.sciencedirect.com/science/article/pii/S0022509612002335},
  \DOIprefix\doi{10.1016/j.jmps.2012.10.012}.
%Type = Article
\bibitem[{Patinet et~al.(2013b)Patinet, Vandembroucq and
  Roux}]{patinet_quantitative_2013}
\bibinfo{author}{Patinet, S.}, \bibinfo{author}{Vandembroucq, D.},
  \bibinfo{author}{Roux, S.}, \bibinfo{year}{2013}b.
\newblock \bibinfo{title}{Quantitative prediction of effective toughness at
  random heterogeneous interfaces}.
\newblock \bibinfo{journal}{Physical Review Letters} \bibinfo{volume}{110},
  \bibinfo{pages}{165507}.
\newblock \URLprefix
  \url{https://link.aps.org/doi/10.1103/PhysRevLett.110.165507},
  \DOIprefix\doi{10.1103/PhysRevLett.110.165507}.
%Type = Article
\bibitem[{Poliakov et~al.(2002)Poliakov, Dmowska and
  Rice}]{poliakov_dynamic_2002}
\bibinfo{author}{Poliakov, A.N.B.}, \bibinfo{author}{Dmowska, R.},
  \bibinfo{author}{Rice, J.R.}, \bibinfo{year}{2002}.
\newblock \bibinfo{title}{Dynamic shear rupture interactions with fault bends
  and off-axis secondary faulting}.
\newblock \bibinfo{journal}{Journal of Geophysical Research: Solid Earth}
  \bibinfo{volume}{107}, \bibinfo{pages}{ESE 6--1--ESE 6--18}.
\newblock \URLprefix
  \url{https://agupubs.onlinelibrary.wiley.com/doi/abs/10.1029/2001JB000572},
  \DOIprefix\doi{10.1029/2001JB000572}.
%Type = Article
\bibitem[{Ponson and Bonamy(2010)}]{ponson_crack_2010}
\bibinfo{author}{Ponson, L.}, \bibinfo{author}{Bonamy, D.},
  \bibinfo{year}{2010}.
\newblock \bibinfo{title}{Crack propagation in brittle heterogeneous solids:
  Material disorder and crack dynamics}.
\newblock \bibinfo{journal}{International Journal of Fracture}
  \bibinfo{volume}{162}, \bibinfo{pages}{21--31}.
\newblock \URLprefix \url{https://doi.org/10.1007/s10704-010-9481-x},
  \DOIprefix\doi{10.1007/s10704-010-9481-x}.
%Type = Article
\bibitem[{Rice(1985)}]{rice_first-order_1985}
\bibinfo{author}{Rice, J.}, \bibinfo{year}{1985}.
\newblock \bibinfo{title}{First-order variation in elastic fields due to
  variation in location of a planar crack front}.
\newblock \bibinfo{journal}{Journal of Applied Mechanics} \bibinfo{volume}{52},
  \bibinfo{pages}{571--579}.
\newblock \URLprefix \url{http://dx.doi.org/10.1115/1.3169103},
  \DOIprefix\doi{10.1115/1.3169103}.
%Type = Inproceedings
\bibitem[{Rice(1980)}]{rice_mechanics_1980}
\bibinfo{author}{Rice, J.R.}, \bibinfo{year}{1980}.
\newblock \bibinfo{title}{The mechanics of earthquake rupture}, in:
  \bibinfo{booktitle}{in Physics of the Earth's Interior, edited by A.M.
  Dziewonski and E. Boschi}, pp. \bibinfo{pages}{555--649}.
%Type = Article
\bibitem[{Rice(1989)}]{rice_weight_1989}
\bibinfo{author}{Rice, J.R.}, \bibinfo{year}{1989}.
\newblock \bibinfo{title}{Weight function theory for three-dimensional elastic
  crack analysis}.
\newblock \bibinfo{journal}{Fracture Mechanics: Perspectives and Directions
  (Twentieth Symposium)} \URLprefix
  \url{http://www.astm.org/DIGITAL\_LIBRARY/STP/PAGES/STP18819S.htm},
  \DOIprefix\doi{10.1520/STP18819S}.
%Type = Misc
\bibitem[{Roch et~al.(2022)Roch, Lebihain and Molinari}]{roch_dynamic_2022}
\bibinfo{author}{Roch, T.}, \bibinfo{author}{Lebihain, M.},
  \bibinfo{author}{Molinari, J.F.}, \bibinfo{year}{2022}.
\newblock \bibinfo{title}{Dynamic crack front deformations in cohesive
  materials}.
\newblock \URLprefix \url{http://arxiv.org/abs/2206.04588},
  \DOIprefix\doi{10.48550/arXiv.2206.04588}. \bibinfo{note}{number:
  arXiv:2206.04588 arXiv:2206.04588 [cond-mat]}.
%Type = Article
\bibitem[{Roux et~al.(2003)Roux, Vandembroucq and Hild}]{roux_effective_2003}
\bibinfo{author}{Roux, S.}, \bibinfo{author}{Vandembroucq, D.},
  \bibinfo{author}{Hild, F.}, \bibinfo{year}{2003}.
\newblock \bibinfo{title}{Effective toughness of heterogeneous brittle
  materials}.
\newblock \bibinfo{journal}{European Journal of Mechanics - A/Solids}
  \bibinfo{volume}{22}, \bibinfo{pages}{743--749}.
\newblock \URLprefix
  \url{http://www.sciencedirect.com/science/article/pii/S0997753803000780},
  \DOIprefix\doi{10.1016/S0997-7538(03)00078-0}.
%Type = Article
\bibitem[{Sevillano et~al.(2007)Sevillano, González and
  Martínez-Esnaola}]{sevillano_roughness_2007}
\bibinfo{author}{Sevillano, J.G.}, \bibinfo{author}{González, D.},
  \bibinfo{author}{Martínez-Esnaola, J.M.}, \bibinfo{year}{2007}.
\newblock \bibinfo{title}{Roughness of a mode i in-plane crack front
  propagating along a heterogeneous cohesive interface}.
\newblock \bibinfo{journal}{Journal of Computer-Aided Materials Design}
  \bibinfo{volume}{14}, \bibinfo{pages}{15--24}.
\newblock \URLprefix \url{https://doi.org/10.1007/s10820-007-9086-5},
  \DOIprefix\doi{10.1007/s10820-007-9086-5}.
%Type = Article
\bibitem[{Vasoya et~al.(2016a)Vasoya, Lazarus and
  Ponson}]{vasoya_fingering_2016}
\bibinfo{author}{Vasoya, M.}, \bibinfo{author}{Lazarus, V.},
  \bibinfo{author}{Ponson, L.}, \bibinfo{year}{2016}a.
\newblock \bibinfo{title}{Bridging micro to macroscale fracture properties in
  highly heterogeneous brittle solids: weak pinning versus fingering}.
\newblock \bibinfo{journal}{Journal of the Mechanics and Physics of Solids}
  \bibinfo{volume}{95}, \bibinfo{pages}{755--773}.
\newblock \URLprefix
  \url{http://www.sciencedirect.com/science/article/pii/S0022509615303604},
  \DOIprefix\doi{10.1016/j.jmps.2016.04.022}.
%Type = Article
\bibitem[{Vasoya et~al.(2013)Vasoya, Leblond and Ponson}]{vasoya_second_2013}
\bibinfo{author}{Vasoya, M.}, \bibinfo{author}{Leblond, J.},
  \bibinfo{author}{Ponson, L.}, \bibinfo{year}{2013}.
\newblock \bibinfo{title}{A geometrically nonlinear analysis of coplanar crack
  propagation in some heterogeneous medium}.
\newblock \bibinfo{journal}{International Journal of Solids and Structures}
  \bibinfo{volume}{50}, \bibinfo{pages}{371--378}.
\newblock \URLprefix
  \url{http://www.sciencedirect.com/science/article/pii/S0020768312004118},
  \DOIprefix\doi{10.1016/j.ijsolstr.2012.10.001}.
%Type = Article
\bibitem[{Vasoya et~al.(2016b)Vasoya, Unni, Leblond, Lazarus and
  Ponson}]{vasoya_experimental_2016}
\bibinfo{author}{Vasoya, M.}, \bibinfo{author}{Unni, A.},
  \bibinfo{author}{Leblond, J.}, \bibinfo{author}{Lazarus, V.},
  \bibinfo{author}{Ponson, L.}, \bibinfo{year}{2016}b.
\newblock \bibinfo{title}{Finite size and geometrical non-linear effects during
  crack pinning by heterogeneities: An analytical and experimental study}.
\newblock \bibinfo{journal}{Journal of the Mechanics and Physics of Solids}
  \bibinfo{volume}{89}, \bibinfo{pages}{211--230}.
\newblock \URLprefix
  \url{http://www.sciencedirect.com/science/article/pii/S0022509615303884},
  \DOIprefix\doi{10.1016/j.jmps.2015.12.023}.
%Type = Article
\bibitem[{Viesca and Garagash(2018)}]{viesca_numerical_2018}
\bibinfo{author}{Viesca, R.C.}, \bibinfo{author}{Garagash, D.I.},
  \bibinfo{year}{2018}.
\newblock \bibinfo{title}{Numerical methods for coupled fracture problems}.
\newblock \bibinfo{journal}{Journal of the Mechanics and Physics of Solids}
  \bibinfo{volume}{113}, \bibinfo{pages}{13--34}.
\newblock \URLprefix
  \url{https://www.sciencedirect.com/science/article/pii/S0022509617306853},
  \DOIprefix\doi{10.1016/j.jmps.2018.01.008}.
%Type = Article
\bibitem[{Xia et~al.(2012)Xia, Ponson, Ravichandran and
  Bhattacharya}]{xia_toughening_2012}
\bibinfo{author}{Xia, S.}, \bibinfo{author}{Ponson, L.},
  \bibinfo{author}{Ravichandran, G.}, \bibinfo{author}{Bhattacharya, K.},
  \bibinfo{year}{2012}.
\newblock \bibinfo{title}{Toughening and asymmetry in peeling of heterogeneous
  adhesives}.
\newblock \bibinfo{journal}{Physical Review Letters} \bibinfo{volume}{108},
  \bibinfo{pages}{196101}.
\newblock \URLprefix
  \url{https://link.aps.org/doi/10.1103/PhysRevLett.108.196101}.
%Type = Article
\bibitem[{Xia et~al.(2015)Xia, Ponson, Ravichandran and
  Bhattacharya}]{xia_adhesion_2015}
\bibinfo{author}{Xia, S.M.}, \bibinfo{author}{Ponson, L.},
  \bibinfo{author}{Ravichandran, G.}, \bibinfo{author}{Bhattacharya, K.},
  \bibinfo{year}{2015}.
\newblock \bibinfo{title}{Adhesion of heterogeneous thin films - ii: Adhesive
  heterogeneity}.
\newblock \bibinfo{journal}{Journal of the Mechanics and Physics of Solids}
  \bibinfo{volume}{83}, \bibinfo{pages}{88--103}.
\newblock \URLprefix
  \url{http://www.sciencedirect.com/science/article/pii/S0022509615001593},
  \DOIprefix\doi{10.1016/j.jmps.2015.06.010}.

\end{thebibliography}

\end{document}